%% file: tnoqs-final.tex
\documentclass[letterpaper,notitlepage,onecolumn,nofootinbib]{revtex4-1}

\usepackage{amsmath,amsfonts}
\usepackage{graphicx}  % needed for figures
\usepackage{dcolumn}   % needed for some tables
\usepackage{bm}        % for math
\usepackage{amssymb}   % for math
\usepackage{verbatim}
\usepackage{enumerate}
\usepackage[caption=false]{subfig}

\usepackage{graphicx}

\usepackage{verbatim}
\usepackage{array}
\usepackage{hyperref}
\usepackage{epsfig}

\hypersetup{
	pdfnewwindow=true, 
	colorlinks=true, 
	linkcolor=blue, 
	citecolor=magenta, 
	urlcolor=cyan,
pdftitle={A Graphical Calculus for Open Quantum Systems}, 
pdfauthor={Wood},
  pdfsubject={Completely Positive Maps},
  pdfkeywords={Open Quantum Systems, Quantum Operations, Completely Positive Maps, Choi Jamiolkowski Isomorphism, Tensor Network States, 
  Penrose graphical notation, graphical calculus, Quantum Theory, Matrix Product States}
}

\newcommand{\bra}[1]{\mbox{$\left\langle #1\right|$}}
\newcommand{\ket}[1]{\mbox{$\left|#1\right\rangle$}}
\newcommand{\braket}[2]{\mbox{$\langle#1|#2\rangle$}}
\newcommand{\ketbra}[2]{\mbox{$\left|#1\right\rangle\!\!\left\langle #2\right|$}}

\newcommand{\dket}[1]{\mbox{$\left|\left.#1\right\rangle\!\right\rangle$}}
\newcommand{\dbra}[1]{\mbox{$\left\langle\!\left\langle #1\right.\right|$}}
\newcommand{\dbradket}[2]{\mbox{$\langle\!\langle #1|#2\rangle\!\rangle$}}
\newcommand{\dketdbra}[2]{\mbox{$|#1\rangle\!\rangle\!\!\langle\!\langle #2|$}}

\DeclareMathOperator{\Tr}{Tr}

\newcommand{\Eqref}[1]{Eq.~\eqref{#1}}

\newcommand{\be}{\begin{equation}}
\newcommand{\ee}{\end{equation}}
\newcommand{\bea}{\begin{eqnarray}}
\newcommand{\eea}{\end{eqnarray}}

\newcommand{\figeq}[1]{\begin{equation}\begin{split}#1\end{split}\end{equation}}

\usepackage{color}
\definecolor{dred}{rgb}{.8,0.2,.2}
\definecolor{ddred}{rgb}{.8,0.5,.5}
\definecolor{dblue}{rgb}{.2,0.2,.8}
\newtheorem{theorem}{Theorem}

\newtheorem{proposition}[theorem]{Proposition}

\def\1#1{{\bf #1}}
\def\2#1{{\cal #1}}
\def\3#1{{\sl #1}}
\def\4#1{{\tt #1}}
\def\5#1{{\sf #1}}
\def\6#1{{\mathfrak #1}}
\def\7#1{{\mathbb #1}}

\newcommand{\C}{{\mathbb{C}}}

\newcommand{\id}{\mbox{$1 \hspace{-1.0mm} {\bf l}$}}

\newcommand{\Lx}[1]{{L(\mathcal{#1})}}
\newcommand{\Cx}[1]{{C(\mathcal{#1})}}
\newcommand{\Tx}[1]{{T(\mathcal{#1})}}

\newcommand{\XX}{\mathcal{X}\otimes\mathcal{X}}
\newcommand{\XY}{\mathcal{X}\otimes\mathcal{Y}}
\newcommand{\YX}{\mathcal{Y}\otimes\mathcal{X}}
\newcommand{\YY}{\mathcal{Y}\otimes\mathcal{Y}}
\newcommand{\XZ}{\mathcal{X}\otimes\mathcal{Z}}

\newcommand{\LXY}{L(\mathcal{X}\otimes\mathcal{Y})}

\numberwithin{equation}{section}

\graphicspath{{Figures/}}
\begin{document}

%==============================================================
%==============================================================
%		TITLE INFO
%==============================================================
%==============================================================
\title{Tensor networks and graphical calculus for open quantum systems} 
\date{May 7, 2015}

\author{Christopher J. Wood}
\email{christopher.j.wood@uwaterloo.ca}
\affiliation{Institute for Quantum Computing, Waterloo, ON, Canada}
\affiliation{Department of Physics and Astronomy, University of Waterloo, Waterloo, ON, Canada} 

\author{Jacob D. Biamonte}
\affiliation{Centre for Quantum Technologies, National University of Singapore, Singapore}
\affiliation{ISI Foundation, Torino, TO, Italy}
\affiliation{Institute for Quantum Computing, Waterloo, ON, Canada}

\author{David G. Cory}
\affiliation{Institute for Quantum Computing, Waterloo, ON, Canada}
\affiliation{Perimeter Institute for Theoretical Physics, Waterloo, ON, Canada}
\affiliation{Department of Chemistry, University of Waterloo, Waterloo, ON, Canada}

%==============================================================
%==============================================================
%		ABSTRACT
%==============================================================
%==============================================================\
\begin{abstract}

We describe a graphical calculus for completely positive maps and in doing so review the theory of open quantum systems and other fundamental primitives of quantum information theory using the language of tensor networks. In particular we demonstrate the construction of tensor networks to pictographically represent the Liouville-superoperator, Choi-matrix, process-matrix, Kraus, and system-environment representations for the evolution of quantum states, review how these representations interrelate, and illustrate how graphical manipulations of the tensor networks may be used to concisely transform between them.
To further demonstrate the utility of the presented graphical calculus we include several examples where we provide arguably simpler graphical proofs of several useful quantities in quantum information theory including the composition and contraction of multipartite channels, a condition for whether an arbitrary bipartite state may be used for ancilla assisted process tomography, and the derivation of expressions for the average gate fidelity and entanglement fidelity of a channel in terms of each of the different representations of the channel.

\end{abstract}

\maketitle

\input{Sections/tnoqs-1-intro}
\input{Sections/tnoqs-2-tn}
\input{Sections/tnoqs-3-cpmaps}
\input{Sections/tnoqs-4-trans}
\input{Sections/tnoqs-5-examples}
\input{Sections/tnoqs-6-conc}

\begin{appendix}
\input{Sections/tnoqs-7-appendix}
\end{appendix}

\end{document}

%% file: Sections/tnoqs-1-intro.tex
%===============================================================
%===============================================================
%		INTRODUCTION
%===============================================================
%===============================================================
\section{Introduction}

A complete description of the evolution of quantum systems is an important tool in quantum information processing (QIP). In contrast to closed quantum systems, for open quantum systems the evolution need no longer be unitary. In general the evolution of an open quantum system is called a \emph{quantum operation} or \emph{quantum channel} which, for a discrete time interval, is described mathematically by a \emph{completely positive map} (CP-map)~\cite{Nielsen2000}. 
In the context of QIP, a quantum channel is a completely positive linear map acting on the density operators which describe a physical systems state. A map $\2 E$ is \emph{positive} if and only if it preserves the positivity of an operators spectrum, and \emph{completely positive} if and only if it further satisfies the condition that the composite map $\2 I\otimes \2 E$ is positive, where $\2 I$ is the identity map on a space of density operators with dimension greater than or equal to the dimension of the space on which $\2 E$ acts. We will consider the case when $\2 E$ is also \emph{trace preserving}, which is to say $\Tr[\2 E(\rho)]=\Tr[\rho]$ for all density operators $\rho$. Since quantum systems are described by density operators, positive operators $\rho$ with unit trace, the requirements that $\2 E$ be complete-positive and trace-preserving ensures that the output state of the map will always be a valid density operator. Such maps are called completely positive trace preserving maps (CPTP-maps). 

% FIGURE - CP REPS
\begin{figure*}
\includegraphics[width=0.7\textwidth]{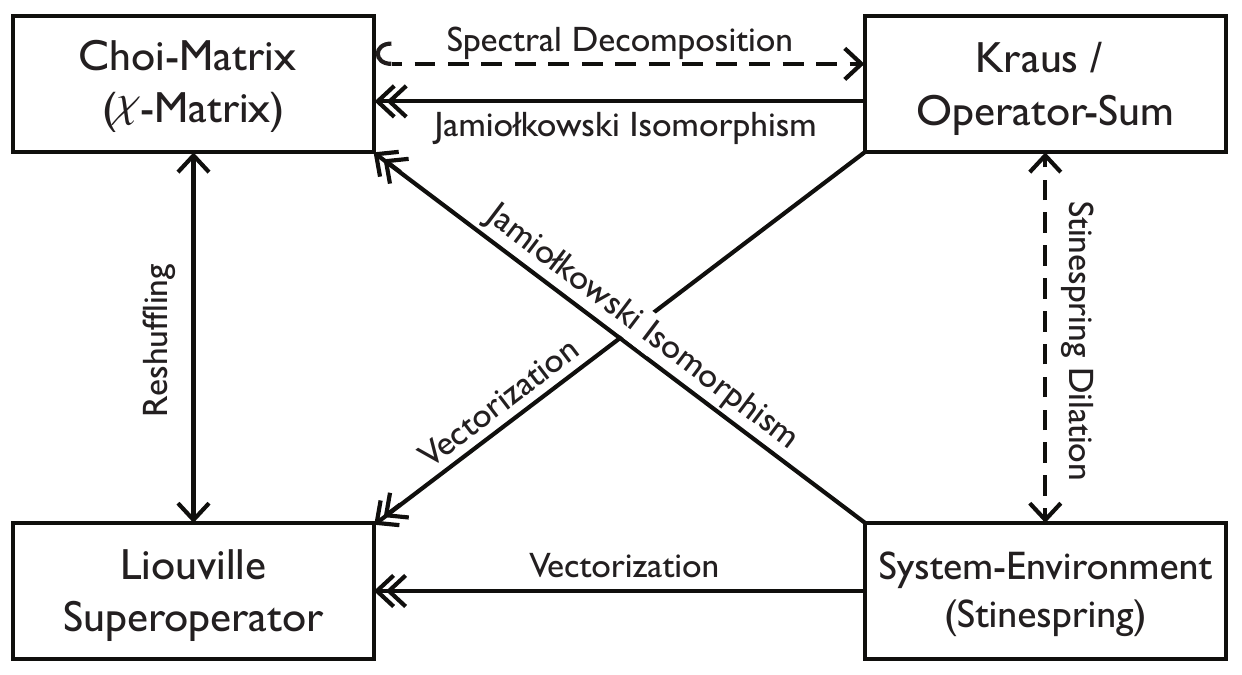}
\caption{The main mathematical representations for completely positive maps and how one may transform between them. Solid arrows represent linear operations which we prove can be done by ``wire bending'' transformations in our graphical calculus. Dashed arrows represent non-linear transformations. Reshuffling and Stinespring dilation are bijective transformations, vectorization and the Jamio{\l}kowski isomorphism are surjective transformations, and the spectral decomposition is an injective transformation.}
\label{fig:cpreps}
\end{figure*}

There are numerous representations for completely-positive maps~\cite{Jordan1961,Choi1975,Kraus1983}, and although well understood, the transformations between these representations is often cumbersome and tedious. Thus one goal of this paper is to present and summarize graphical calculus methods that facilitate an intuitive unification and interoperability between these representations, the outcome of which is depicted in Fig.~\ref{fig:cpreps}. In doing so we provide a compact review of the properties and transformations of CP-maps.

% GRAPHICAL CALCULUS
Graphical calculi have been used to great benefit in several areas of modern physics with the most prolific example being the use of Feynman diagrams to calculate scattering amplitudes in quantum field theories~\cite{Baez2009}. In the context of QIP there has been recent interest in employing graphical techniques more general than standard quantum circuits, with two popular approaches being based on \emph{tensor networks} and \emph{category theory}. The approach presented here casts the theory of open quantum systems into the framework of tensor networks, which comes equipped with a graphical means to represent and reason about the contraction of sequences of tensors~\cite{Penrose1971}.
% TENSOR NETWORKS
The use of tensor networks dates back to earlier work by Penrose, who's graphical notation is a useful starting point~\cite{Penrose1971}. They have been used as computational tools for simulating certain many-body quantum systems efficiently~\cite{Vidal2008,Evenbly2009,Verstraete2008}, as a tool for manipulating tensor networks~\cite{Biamonte2010} and to generalize quantum circuits~\cite{Bergholm2010}. Although it is straightforward to translate equations into so-called tensor string diagrams, a missing piece has been a graphical calculus for open systems theory which provided new results, and hence enhanced the potential for diagrammatic reasoning. 

% MONOIDAL CATEGORIES
The category theoretic approaches for QIP are based on so-called \emph{dagger-compact monoidal categories}
%~\footnote{The term dagger-compact monoidal category is due to Selinger~\cite{Selinger2007}. Abramsky and Coecke originally introduced this under the name strongly compact closed category.} 
which were used by Abramsky and Coecke to abstractly describe quantum teleportation~\cite{Abramsky2004}.
 This approach was then extended to include CP-maps by Selinger in the \emph{CPM-construction}~\cite{Selinger2007}. The graphical language of these approaches are built upon well established graphical calculi for \emph{compact closed categories}~\cite{Kelly1980} and \emph{symmetric monoidal categories}~\cite{Joyal1991}. A key result is that Selinger's calculus for CP-maps is \emph{complete} for finite-dimensional Hilbert spaces~\cite{Selinger2011}. This means that any identity which can be represented graphically is true if and only if it is is algebraically true, which is important for diagrammatic proofs. Subsequent work based on these constructions has been used to graphically depict quantum protocols \cite{Coecke2009a,Boixo2011} and Bayesian inference~\cite{Coecke2011}; and for the axiomatic formulation of quantum theory~\cite{Coecke2011a}. For a review of graphical calculi for monoidal categories see~\cite{Selinger2009}.  Other alternative graphical approaches have also been used in the axiomatic formational of quantum theory~\cite{Chiribella2011,Hardy2011}.

 % CP-MAP CALC
There have been at least two graphical calculi previously presented for CP-maps: Selinger's aforementioned category theoretic approach, and a graph-theory approach by Collins and Nechita which was used to compute ensemble averages of random quantum states~\cite{Collins2009,Collins2010}. Selinger's CPM-construction bears some similarities to our approach, however there are important and practical differences between the two. The CPM-construction is most closely related to our superoperator representation in the row-vectorization convention which we present in Section~\ref{sec:sop}. 
In the presented graphical calculus we tailor the tensor string diagrams of Penrose to unify several mathematical representations used in open quantum systems and to transform freely between them. In accomplishing this, we express our graphical tensor calculus in the Dirac notation familiar in QIP instead of the abstract index notation used by Penrose, or the category theoretic notation used by others. This provides a toolset which can be used for the manipulation, visual representation, and contraction of quantum circuits and general open quantum system equations. 

To demonstrate the utility of the presented graphical calculus we provide examples where we use these tools to derive several common quantities used in quantum information theory. We emphasize that these are not new results, but rather it is the application of the graphical methods that we are aiming to highlight. In particular we demonstrate the construction of composite channels involving the composition of several subsystem channels, and also the contraction of multipartite channels to an effective subsystem channel.
Another example we explore is  \emph{ancilla assisted process tomography} (AAPT) which is a general method of experimentally characterizing an unknown CPTP map $\2E$. This is done by preparing a joint state $\rho_{AS}$ across the system of interest and an ancilla, subjecting this state to the channel $\2 I\otimes \2E$, where $\2I$ is the identity channel on the ancilla, and then performing state tomography to determine the resulting output state. For an appropriately chosen initial state $\rho_{AS}$ a complete description of $\2E$ can be recovered from the measured output state. A necessary and sufficient condition for the recovery of $\2E$ is that the input state state $\rho_{AS}$ has maximally Schmidt-number~\cite{White2003,DAriano2003}. We present an equivalent though simpler to check condition on $\rho_{AS}$, and a method of reconstructing the channel $\2E$, in terms of the reshuffling transformation between Choi-matrices and superoperators. 
The final examples demonstrating graphical proof techniques relate to fidelity measures on quantum channels. In particular we present graphical proofs for the \emph{average gate fidelity} and \emph{entanglement fidelity} of a channel $\2E$ in terms of the Choi-matrix, superoperator, Kraus, $\chi$-matrix, and Stinespring representations of $\2E$. These expressions have previously been presented in the literature~\cite{Schumarcher1996,Horodecki1999,Nielsen2002,Emerson2005,Fletcher2007,Johnston2011}, however the graphical proof is arguably more succinct.

The present paper has five sections, in Section~\ref{sec:tensor} we introduce the elements of our graphical calculus with a particular emphasis on bipartite systems and vectorization of linear operators that are important in the subsequent sections. In Section~\ref{sec:cpmaps} we introduce the  mathematical representations of CP-maps listed in Fig.~\ref{fig:cpreps}, and their corresponding graphical representation. In Section~\ref{sec:trans} we describe how one may transform between any of the representations of CP-maps as shown by the arrows in Fig.~\ref{fig:cpreps}. In Section~~\ref{sec:ex} we give several more advanced examples to further demonstrate the utility of the graphical calculus. 

In what follows we will use the notation that $\2 X, \2 Y, \2 Z$ are finite-dimensional complex Hilbert spaces,  $L(\2X,\2Y)$ is the space of bounded linear operators $A: \2 X\rightarrow \2 Y$ (with $L(\2X)\equiv L(\2X,\2X)$), $T(\2X,\2Y)$ is the space of operator maps $\2 E:L(\2X)\rightarrow L(\2Y)$  (with $T(\2X)\equiv T(\2X,\2X)$), and $C(\2X,\2Y)$ is the space of operator maps $\2 E$ which are CP. 

%% file: Sections/tnoqs-2-tn.tex
%==============================================================
%==============================================================
%		TENSOR NETWORKS
%==============================================================
%==============================================================
\section{Tensor Networks}
\label{sec:tensor}

Tensors can be thought of as indexed multi-dimensional arrays of complex numbers with respect to a fixed standard basis. The number of indices is called the \emph{order} of a tensor, and the concurrent evaluation of all indices returns a complex number. For example, consider the Hilbert space $\2 X\cong \C^d$, where as is typical in QIP, we choose our standard basis to be the \emph{computational basis} $\{\ket{i}\,:\, i=0,...,d-1\}$. Then in Dirac notation a vector $\ket{v}\in \2 X$ is a 1st-order tensor which can be expressed in terms of its tenor components $v_i:=\braket{i}{v}$ with respect to the standard basis as $\ket{v}=\sum_{i=0}^{d-1} v_i\ket{i}$. Similarly one can represent linear operators on this Hilbert space, $A\in L(\2 X)$,  as 2nd-order tensors with components $A_{ij}:= \bra{i}A\ket{j}$ as $A = \sum_{i,j=0}^{d-1} A_{ij} \ketbra{i}{j}$.  

Hence, in Dirac notation the number of indices of a tensors components are what we refer to as the order of the tensor. Vectors $\ket{v}\in \2 X$ refer to tensors which only have \emph{ket} ``$\ket{i}$'' basis elements, vectors in the dual vector space $\bra{u}\in \2 X^\dagger$ refer to those with only \emph{bras} ``$\bra{i}$'', and linear operators on $A\in \2 L(\2 X)$ refer to tensors with a mixture of kets and bras in their component decomposition.
 
\begin{figure}[h]
   \vspace{-1em}
 \centering
 	\subfloat[Vector $\ket{v}\in\2 X$]{
		\label{sfig:ket}\includegraphics[width=0.1\textwidth]{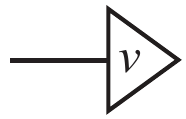}}
		\hspace{0.5em}
  	\subfloat[Dual-vector $\bra{v}\in \2 X^\dagger$]{
		\label{sfig:bra}\includegraphics[width=0.1\textwidth]{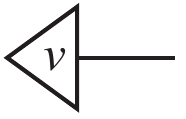}}
		\hspace{0.5em}
   	\subfloat[Linear Operator $A\in \2 L(\2 X)$]{
		\label{sfig:mat}\includegraphics[width=0.14\textwidth]{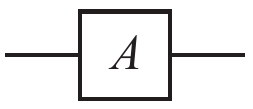}}
		\hspace{0.5em}
   	\subfloat[Scalar $\lambda\in\C$]{
		\label{sfig:scalar}\includegraphics[width=0.07\textwidth]{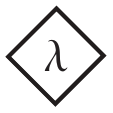}}
		\hspace{1em}
	\subfloat[Vector $\ket{v}\in \bigotimes_{i=1}^n \2 X_i$]{
		\label{sfig:gket}\includegraphics[width=0.1\textwidth]{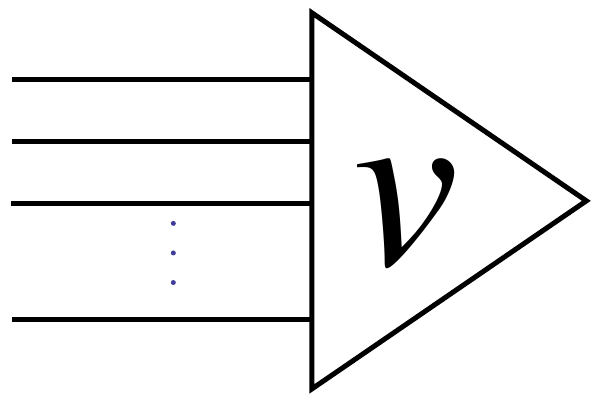}}
		\hspace{1em}
  	\subfloat[Dual-vector $\bra{v}\in \bigotimes_{i=1}^n \2 X^\dagger_i$]{
		\label{sfig:gbra}\includegraphics[width=0.1\textwidth]{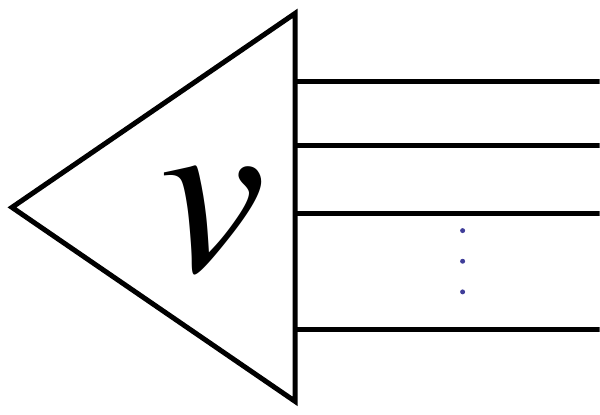}}
		\hspace{1em}
   	\subfloat[Linear operator $A:\2 \bigotimes_{i=1}^n \2 X_i \rightarrow \bigotimes_{j=1}^m \2 X_j$]{
		\label{sfig:gmat}\includegraphics[width=0.17\textwidth]{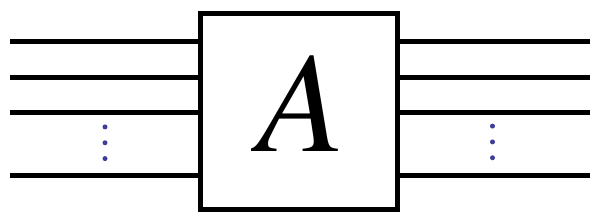}}
 \caption{Graphical depiction of elementary tensors. We represent vectors (states) and dual-vectors (effects) as triangles, linear operators as boxes, and scalars as diamonds, with each index of the tensor depicted as an open wire on the diagram. The orientation of the wires determines the type of tensor, in our convention the open end of the wires point to the left for vectors, right for dual-vectors, and both left and right for linear operators.}
  \label{fig:ketbramat}
 \end{figure}

The idea of representing states, operators and maps (etc.) diagrammatically dates back to several works by Penrose and is often referred to as \textit{Penrose graphical notation} or \textit{string diagrams}.
We adopt Penrose's notation of representing states (vectors) and effects (dual-vectors) as triangles, linear operators as boxes, and scalars as diamonds, as illustrated in Fig.~\ref{fig:ketbramat}. Here each index corresponds to an open wire on the diagram and so we may define higher order tensors with increasingly more wires. The number of wires is then the order of the tensor, with each wire acting on a separate vector space $\2 X_j$.  
% FIGURE - BASIC TENSORS
 
We also insist that the orientation of these wires, rather than the number of wires, specifies whether they represent multi-partite vectors, dual-vectors, or linear operators. We have a freedom in choosing our orientation for the tensors, top-to-bottom, bottom-to-top, left-to-right or right-to-left. In this paper we will choose the \emph{right-to-left} convention (the opposite of most orthodox quantum circuits) so that the graphical representation will most closely match the underlying equations. Thus we use the terms \emph{vector}, \emph{dual-vector} and \emph{linear operator} to refer to tensors of any order, not just 1st-order and 2nd-order, based on the orientation of their wires as follows:
% TENSOR ORIENTATION
\begin{enumerate} 
\item\emph{Vectors} $\ket{v}\in \bigotimes_{j=1}^n \2 X_j$ are tensors with $n\ge 1$ wires oriented to the left. 
\item\emph{Vectors in the dual space} $\bra{v}\in\bigotimes_{j=1}^n \2 X^\dagger_j$ are tensors with $n\ge 1$ wires are oriented to the right.
\item\emph{Linear operators} $A:\2 \bigotimes_{i=1}^n \2 X_i \rightarrow \bigotimes_{j=1}^m \2 X_j$ are tensors which have $n\ge 1$ wires going to the right and  $m\ge 1$ wires to the left. 
\item Tensors with no open wires are \emph{scalars} $\lambda\in\C$.
\end{enumerate}
The graphical depictions of the these tensors are also illustrated in Fig.~\ref{fig:ketbramat}. In the present paper, we will generally be interested in the case where each wire indexes $\2X_j\cong\C^d$ for fixed dimension $d$, though one may generalize most of what follows to situations where the dimensions of each wire are not equal. Note that we represent scalars as tensors with no open wires, this could either be a contracted tensor $\lambda=\braket{v}{u}$, or a multiplicative factor $\lambda$ acting on the tensor $A$ as $\lambda A$.

 % WIRE BENDING
The mathematical rules of tensor network theory assert that the wires of tensors may be manipulated, with each manipulation corresponding to a specific contraction or transformation. We now introduce some tools which we have tailored for manipulations common in open quantum systems. Transposition of 1st-order vectors and dual-vectors, and 2nd-order linear operators is represented by a \emph{bending} of a tensors wires as follows: 
  % FIGURE - TRANSPOSITION
\figeq{
\begin{tabular}{c|c|c}
\includegraphics[height=4.5em]{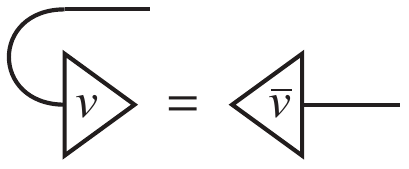} \quad\quad&\quad\quad
\includegraphics[height=4.5em]{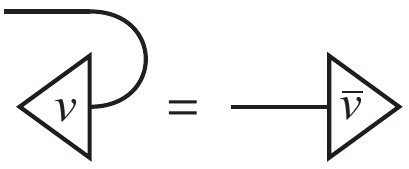} \quad\quad&\quad\quad
\includegraphics[height=4.5em]{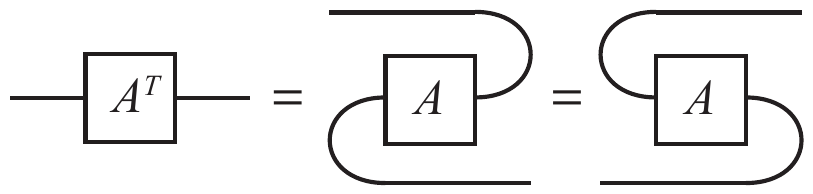} \\
\footnotesize{(a) Vector transposition:} \quad&\quad
\footnotesize{(b) Dual-vector transposition:} \quad&\quad
\footnotesize{(c) Linear operator transposition}\\
\footnotesize{$\ket{v}^T = \bra{\overline{v}}$} \quad&\quad
\footnotesize{$\bra{v}^T=\ket{\overline{v}}$} \quad&\quad
\end{tabular}
\label{fig:transpose}
}

Complex conjugation of a tensor's coefficients however is depicted by a bar over the tensor label in the diagram: 
  % FIGURE - CONJUGATION
\figeq{
\begin{tabular}{c|c|c}
\includegraphics[height=3.5em]{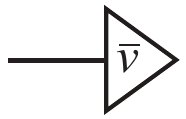} \quad&\quad
\includegraphics[height=3.5em]{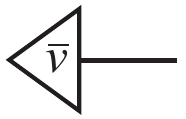} \quad&\quad
\includegraphics[height=3.5em]{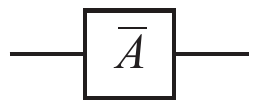} \\
\footnotesize{(a) Complex conjugation of $\ket{v}$} \quad&\quad
\footnotesize{(b) Complex conjugation of $\bra{v}$} \quad&\quad
\footnotesize{(c) Complex conjugation of $A$}
\end{tabular}
\label{fig:conj}
}

Hence we may represent the transformation of a vector to its dual vector, or the hermitian conjugation of a linear operator as the combination of these two operations:
% FIGURE - ADJOINT
\figeq{
\begin{tabular}{c|c|c}
\includegraphics[height=4.7em]{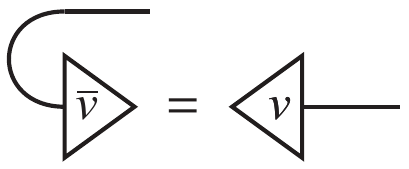} \quad&\quad
\includegraphics[height=4.7em]{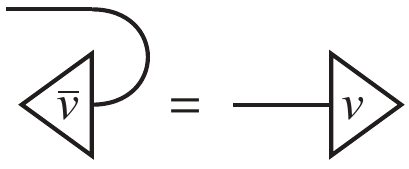} \quad&\quad
\includegraphics[height=4.7em]{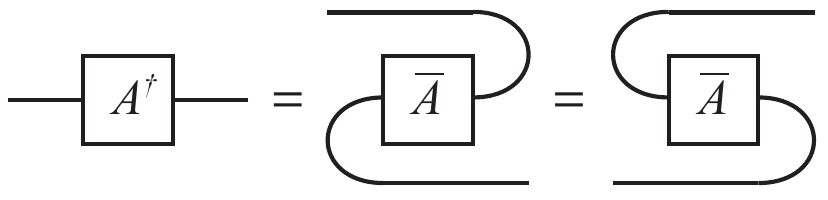} \\
\footnotesize{(a) Hermitian conjugation} \quad&\quad
\footnotesize{(b) Hermitian conjugation} \quad&\quad
\footnotesize{(c) Hermitian Conjugation of an operator $A$}\\
\footnotesize{of a vector : $\ket{v}^\dagger=\bra{v}$} \quad&\quad
\footnotesize{of a dual-vector : $\bra{v}^\dagger=\ket{v}$} \quad&\quad
\end{tabular}
\label{fig:adjoint}
}
We stress that under this convention a vector $\ket{v}=\sum_i v_i \ket{i}$ and its hermitian conjugate dual-vector $\bra{v}=\sum_i \overline{v}_i\bra{i}$ are represented as shown in Fig.~\ref{sfig:ket} and~\ref{sfig:bra} respectively.
 
% TENSOR CONTRACTION
Tensor contraction is represented by joining the wires corresponding to the indices to be contracted. In the case of matrix multiplication $A\cdot B$ is represented by connecting the corresponding wires of the tensors representing the matrices:
  % FIGURE -MULTIPLICATION
\figeq{\includegraphics[height=3em]{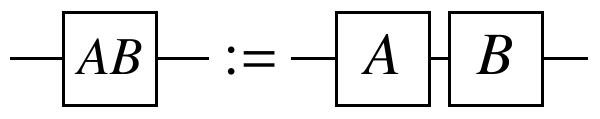}
  \label{fig:matmult}}

To form multi-partite tensors we denote the tensor product of two tensors $A\otimes B$ by the vertical juxtaposition of their tensor networks:
  % FIGURE - TENSOR PRODUCT
\figeq{\includegraphics[height=5em]{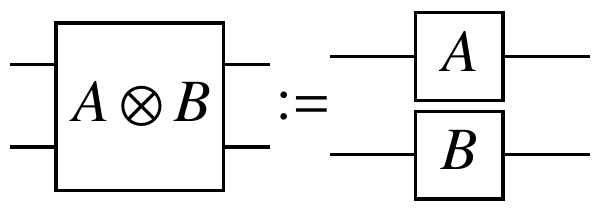}
  \label{fig:tensprod}} 
The trace, $\Tr[A]$, of an operator $A$ is depicted by connecting the corresponding left and right wires of a linear operator:
  % FIGURE -TRACE
\figeq{\includegraphics[height=5em]{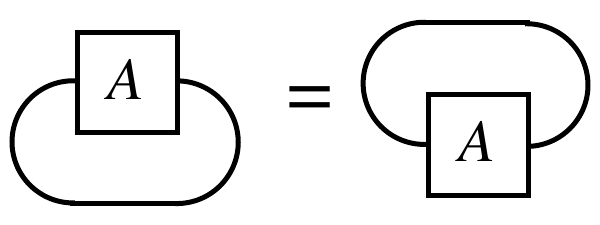}  \label{fig:trace}}
 
We represent summation in our variant of graphical calculus adapted to open systems by introducing shading or coloring of the tensors being summed over. We call this the color summation convention. Tensors corresponding to the same summation index will be shaded the same color, and we use different colors for different summation indexes. For example, consider the spectral decomposition of a normal operator $A$ with eigenvalues $\lambda_i$ and eigenvectors  $\ket{a_i}$. The graphical depiction of the spectral-decomposition $A=\sum_i \lambda_i\ket{a_i}$ using the color summation convention is given by:
% FIGURE - SPECTRAL DECOMP
\figeq{
 	\includegraphics[width=0.7\textwidth]{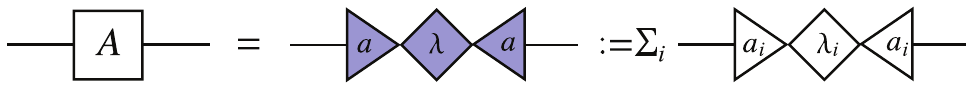}
	  \label{fig:spectral-sum}
}

% COLOUR CONVENTION
In our color summation convention, we will represent the sum over the standard basis as a shaded vector (or dual-vector) tensor with an empty label. This is demonstrated for the graphical resolution of the identity $\id = \sum_{i=0}^{d-1} \ketbra{i}{i}$ as follows:
% FIGURE -  IDENTITY
\figeq{\includegraphics[height=3em]{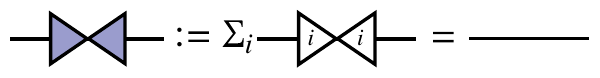}  \label{fig:identity-comp}}

The unnormalized maximally entangled Bell-state $\ket{\Phi^+}=\sum_{i=0}^{d-1} \ket{i}\otimes\ket{i}\in\XX$ is represented graphically as the curve:
% FIGURE -  BELL STATE
\figeq{\includegraphics[height=5.5em]{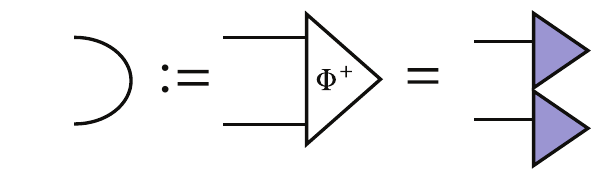}\label{fig:bellstate}}
Similarly the unnormalized Bell-effect $\bra{\Phi^+}$ is represented as: 
% FIGURE -  BELL EFFECT
\figeq{\includegraphics[height=5.5em]{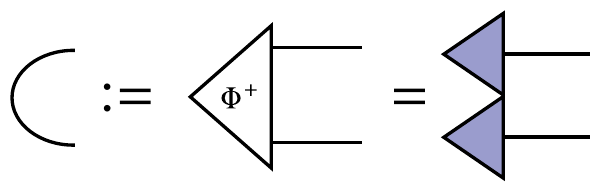}\label{fig:belldual}}
As will be shown in Section~\ref{sec:vec}, our choice of graphical notation for $\ket{\Phi^+}$ is due to its equivalence to the vectorization of the identity operator.

% DEF - Snake equation
Using the graphical definition for $\ket{\Phi^+}$ we can compose the unnormalized Bell-state and its dual to form an identity element~\cite{Penrose1971}. This is known as the \emph{snake equation} or \emph{zig-zag equation} and is given by: 
% FIGURE -  SNAKE EQUATION
\figeq{\includegraphics[height=4.5em]{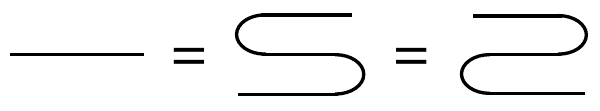}  \label{fig:snake}}
The snake equations have several uses and provide an equivalence class of diagrams. Anytime we have a curved wire with two bends we can ``pull the wire" to straighten it out into an identity. Anytime we bend a wire, transforming between say a bra and a ket, we can bend the wire to transform back again.

By combining the snake-equation with the wire-bending operation for transposition, we find that ``sliding'' a linear operator around an unnormalized Bell-state is also equivalent to transposition of the operator: 
% FIGURE - TRANSPOSE
\figeq{\includegraphics[width=0.25\textwidth]{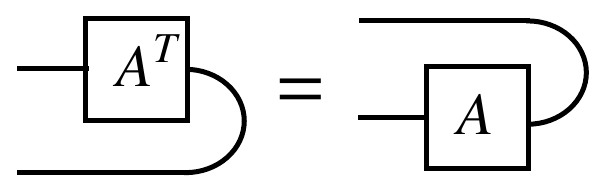}  \label{fig:slide-trans}}
Note that due to the orientation of the wires this graphical representation of the operator $A$ is actually a vector. This is called the vectorization of a matrix and we discuss this in more detail in Section~\ref{sec:vec}.

% SWAP
Another important operation is the graphical SWAP which exchanges the position of two Hilbert spaces in a composite system. Let $\2 
X$ and $\2 Y$ be complex Hilbert spaces of dimensions $d_1$ and $d_2$ respectively, then the SWAP operation is the map
\begin{eqnarray} 
\mbox{SWAP}: \XY &\rightarrow& \YX\\
\mbox{SWAP}: \ket{x}\otimes\ket{y}&\mapsto& \ket{y}\otimes\ket{x},
\end{eqnarray}
for all $ \ket{x}\in \2 X,\ket{y}\in\2 Y$.

Given any two orthonormal basis $\{\ket{x_i}: i=0,\hdots, d_1-1\}$ and $\{\ket{y_j}: j=0,\hdots, d_2-1 \}$ for $\2 X$ and $\2 Y$ respectively, we can give an explicit construction for the SWAP operation as
\begin{equation}
\mbox{SWAP} =  \sum_{i_1=0}^{d_1-1}\sum_{j_2=0}^{d_2-1}\ketbra{y_j}{x_i}\otimes\ketbra{x_i}{y_j}.
\label{eqn:swap}
\end{equation}

The SWAP operation is represented graphically by two crossing wires as shown:
% FIGURE -  Swap 
\figeq{\includegraphics[width=0.35\textwidth]{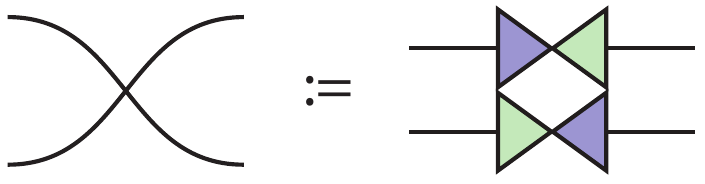}  \label{fig:swap}}
The basis decomposition in Eq.~\eqref{eqn:swap} is then an application of the resolution of the identity to each wire. In Section~\ref{sec:vec} we will see that the SWAP operation is the natural transformation between the row-stacking and column-stacking vectorization conventions.

 Using the above primitives one may represent any linear equation involving the composition and contraction of tensors by a tensor network diagram. In the reverse case, given a tensor network diagram one may always write down an equivalent equation by labelling each wire by an index, and then writing down the corresponding tensor components and summing over contracted indices. By manipulating the tensor diagrams and making use of the primitives introduced one may obtain equivalent expressions, however these manipulated forms may have a more convent equational form that is not inherently obvious from simply looking at the original equation. For explicit examples of writing down the equational form of a tensor diagram refer to the proofs in Appendix~\ref{app:tensor}.

%===================================================
%		BIPARTITE
%===================================================
\subsection{Bipartite Matrix Operations}
\label{sec:bipartite}

Bipartite matrices are used in several representations of CP-maps, and manipulations of these matrices will be important in the following discussion. Consider two complex Hilbert spaces $\2 X$, and $\2 Y$ with dimensions $d_x$ and $d_y$ respectively. The bipartite matrices we are interested in are then $d^2_x\times d^2_y$ matrices $M\in L(\2X\otimes\2Y)$ which we can represent as 4th-order tensors with tensor components
\begin{equation}
M_{m\mu,n\nu}:=\bra{m, \mu}M\ket{n,\nu}
\end{equation} 
where $m,n \in \{0,...,d_x-1\}$, $\mu,\nu \in \{0,...,d_y-1\}$ and $\ket{n,\nu}:=\ket{n}\otimes\ket{\nu}\in \2X\otimes\2Y$ is the tensor product of the standard bases for $\2 X$ and $\2 Y$. Graphically this is given by
% FIGURE - BIPARTITE TENSOR
\figeq{
\includegraphics[height=5em]{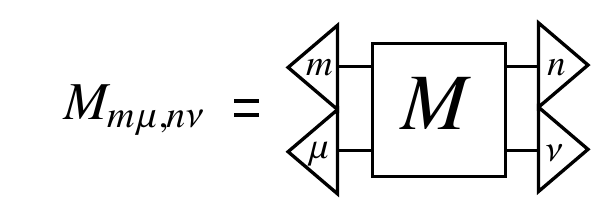}
 \label{fig:bipartite-m-4}
}
We can also express the matrix $M$ as a 2nd-order tensor in terms of the standard basis $\{\ket{\alpha} : \alpha = 0,\hdots,D-1\}$ for $\2X\otimes\2Y$ where $D=d_x d_y$. In this case $M$ has tensor components
\begin{equation}
M_{\alpha\beta}=\bra{\alpha}M\ket{\beta}
\end{equation}
This is represented graphically as
% FIGURE - BIPARTITE TENSOR
\figeq{\includegraphics[height=5em]{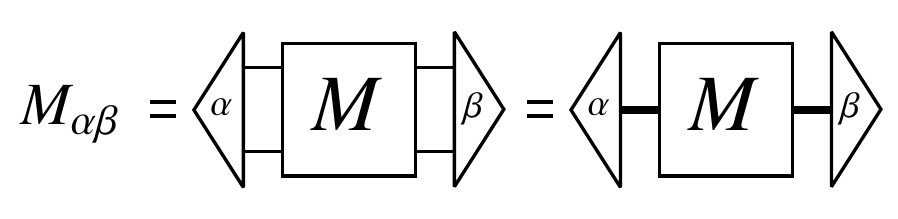}
 \label{fig:bipartite-m-2}
}
We can specify the equivalence between the tensor components $M_{\alpha\beta}$ and $M_{m\mu,n\nu}$ by making the assignment
\begin{eqnarray}
\alpha &=& d_y m + \mu\\
\beta &=& d_y n +\nu,
\end{eqnarray} 
where $d_y$ is the dimension of the Hilbert space $\2 Y$.

The bipartite matrix operations which are the most relevant for open quantum systems (see Fig.~\ref{fig:cpreps}) are the \emph{partial trace over $\2 X$}  ($\Tr_{\2 X}$) (and $\Tr_{\2 Y}$ over $\2 Y$), \emph{transposition} ($T$), \emph{bipartite-SWAP} ($S$), \emph{col-reshuffling} ($R_c$), and \emph{row-reshuffling} ($R_r$). The corresponding graphical manipulations are:
% FIGURE - BIPARTITE TENSOR
\figeq{
 \begin{tabular}{cccccc}
\includegraphics[width=0.13\textwidth]{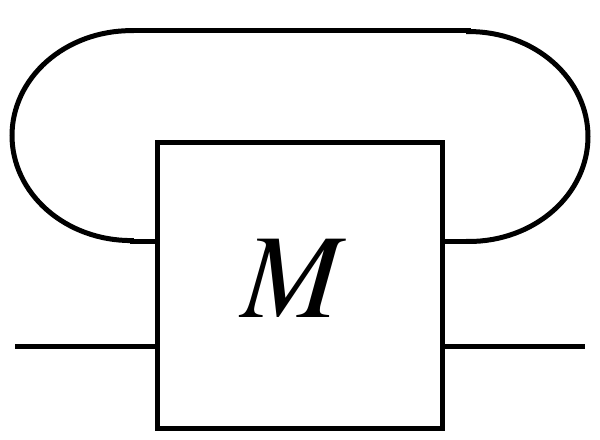}	\quad&\quad
\includegraphics[width=0.13\textwidth]{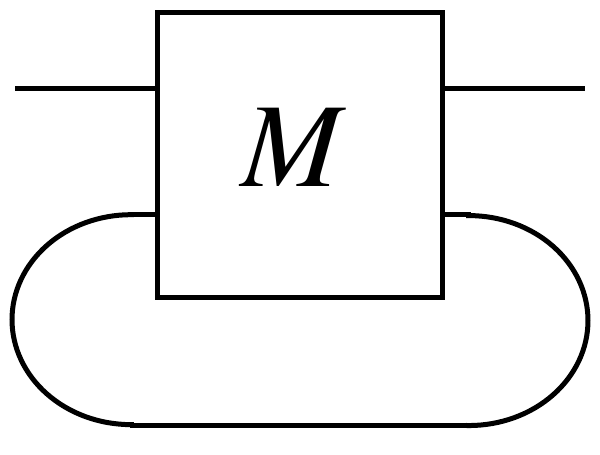}	\quad&\quad
\includegraphics[width=0.12\textwidth]{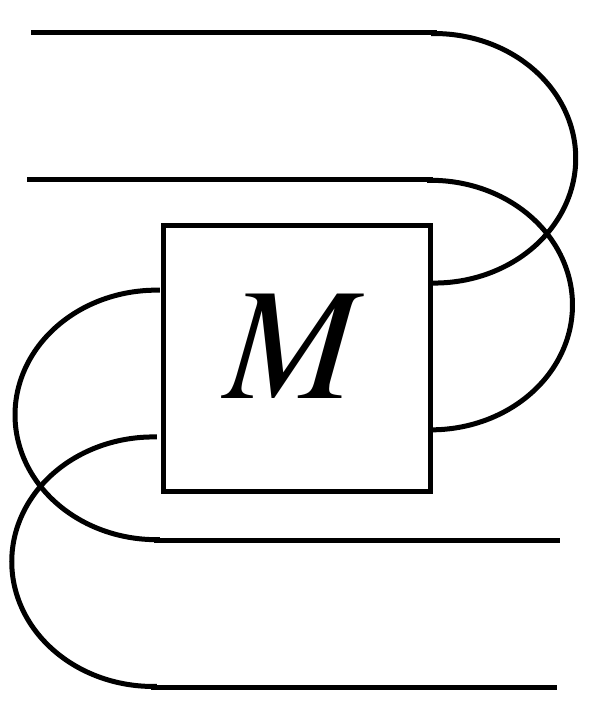}	\quad&\quad
\includegraphics[width=0.15\textwidth,trim=1.5cm 1.5cm 1.5cm 1.5cm,clip]{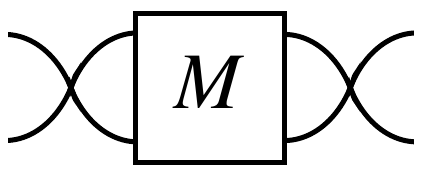}	\quad&\quad
\includegraphics[width=0.13\textwidth,trim=1.5cm 1.5cm 1.5cm 1.5cm,clip]{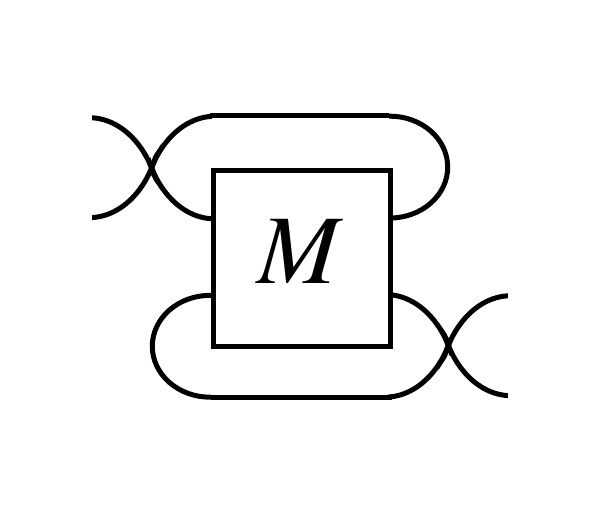}	\quad&\quad
\includegraphics[width=0.13\textwidth,trim=1.5cm 1.5cm 1.5cm 1.5cm,clip]{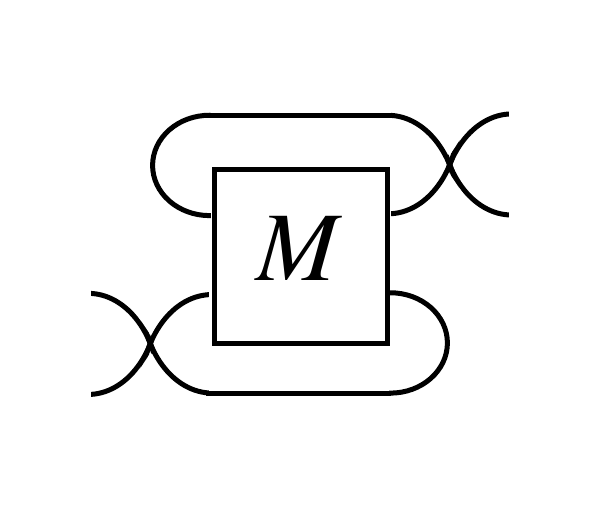}	\\
\footnotesize{(a) Partial Trace} \quad&\quad
\footnotesize{(b) Partial Trace} \quad&\quad
\footnotesize{(c) Transpose} \quad&\quad
\footnotesize{(d) Bipartite-Swap} \quad&\quad
\footnotesize{(e) Row-Reshuffle} \quad&\quad
\footnotesize{(f) Col-Reshuffle} 
\\
\footnotesize{$\Tr_{\2 X}[M]$} \quad&\quad
\footnotesize{$\Tr_{\2 Y}[M]$} \quad&\quad
\footnotesize{$M^T$} \quad&\quad
\footnotesize{$M^S$} \quad&\quad
\footnotesize{$M^{R_r}$} \quad&\quad
\footnotesize{$M^{R_c}$} 
 \end{tabular}
 \label{fig:bipartite}
}
 In terms of the tensor components of $M$ these operations are respectively given by:
\begin{center}
\begin{tabular}{lll}
Partial trace over $\2X$
\quad&\quad	$\Tr_{\2 X}: L(\2X\otimes\2Y)\rightarrow L(\2Y),$ 
\quad&\quad	$M_{m\mu,n\nu} \mapsto \sum_{m} M_{m\mu,m\nu}$
\\
Partial trace over $\2Y$
\quad&\quad	$\Tr_{\2 Y}: L(\2X\otimes\2Y)\rightarrow L(\2X)$
\quad&\quad	$M_{m\mu,n\nu} \mapsto \sum_{\mu} M_{m\mu,n\mu}$
\\
Tranpose
\quad&\quad	$T: L(\2X\otimes\2Y)\rightarrow L(\2X\otimes\2Y),$ 
\quad&\quad	$M_{m\mu,n\nu} \mapsto M_{n\nu,m\mu}$
\\
Bipartite-SWAP
\quad&\quad	$S: L(\2X\otimes\2Y)\rightarrow L(\2Y\otimes\2X),$ 
\quad&\quad	$M_{m\mu,n\nu} \mapsto M_{\mu m, \nu n}$
\\
Row-reshuffling
\quad&\quad	$R_r: L(\2X\otimes\2Y)\rightarrow L(\2Y\otimes\2Y,\2X\otimes\2X),$ 
\quad&\quad	$M_{m\mu,n\nu} \mapsto M_{m, n , \mu, \nu} $
\\
Col-reshuffling
\quad&\quad	$R_c: L(\2X\otimes\2Y)\rightarrow L(\2X\otimes\2X,\2Y\otimes\2Y),$ 
\quad&\quad	$M_{m\mu,n\nu} \mapsto M_{\nu \mu, n m}$
\end{tabular}
\end{center}
Note that we will generally use reshuffling $R$ to refer to col-reshuffling $R_c$. Similarly we can represent the partial transpose operation by only transposing the wires for $\2X$ (or $\2Y$), and the partial-SWAP operations by only swapping the left (or right) wires of $M$.

%===================================================
%		VECTORIZATION
%===================================================
\subsection{Vectorization of Matrices}
\label{sec:vec}

We now recall the concept of \emph{vectorization} which is a reshaping operation, transforming a $(m\times n)$-matrix into a $(1\times mn)$-vector~\cite{Horn1985}. This is necessary for the description of open quantum systems in the superoperator formalism, which we will consider in Section~\ref{sec:sop}. Vectorization can be done with using one of two standard conventions: \emph{column-stacking} (col-vec) or \emph{row-stacking} (row-vec). Consider two complex Hilbert spaces $\2 X\cong \C^m, \2 Y\cong \C^n$, and linear operators  $A\in L(\2X,\2Y)$ from $\2 X$ to $\2 Y$. Column and row vectorization are the mappings
\begin{eqnarray}
\mbox{col-vec: } \Lx{X,Y}&\rightarrow& \XY: \,\, A\mapsto\dket{A}_c\\
\mbox{row-vec: } \Lx{X,Y}&\rightarrow& \YX: \,\,A\mapsto\dket{A}_r
\end{eqnarray} 
respectively, where the operation col(row)-vec when applied to a matrix, outputs a vector with the columns (rows) of the matrix stacked on top of each other. Graphical representations for the row-vec and col-vec operations are found from bending a wire to the left either clockwise or counterclockwise respectively:
% FIGURE - VEC DEF
\figeq{
\begin{tabular}{c|c}
\includegraphics[height=4em]{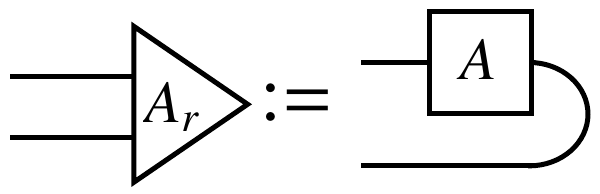} 
\quad\quad&\quad\quad
\includegraphics[height=4em]{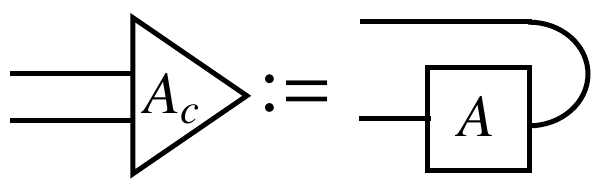} 
\\
\footnotesize{(a) Row-vec} &
\footnotesize{(b) Col-vec}
\end{tabular}
 \label{fig:vectorization}
}
Vectorized matrices in the col-vec and row-vec conventions are naturally equivalent under wire exchange (the SWAP operation)
% FIGURE - COL ROW SWAP
\figeq{\includegraphics[width=0.48\textwidth]{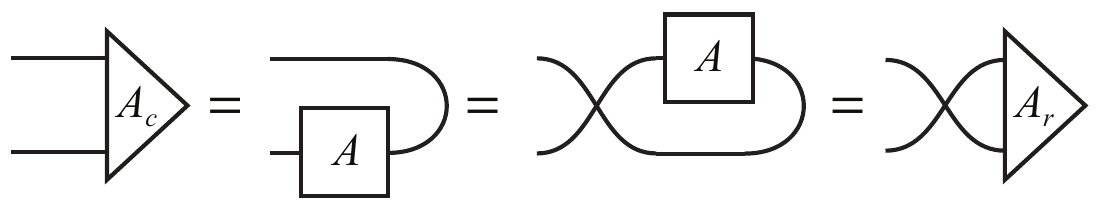} \label{fig:col-to-row}}

In particular we can see that the unnormalized Bell-state $\ket{\Phi^+}\in\2X\otimes\2X$ is in fact the vectorized identity operator $\id\in L(\2X)$
\begin{equation}
\ket{\Phi^+} = \dket{\id}_r = \dket{\id}_c.
\end{equation}

We may also define a vectorization operation with respect to an arbitrary operator basis for $L(\2X,\2Y)$. Let $\2 X\cong\C^{d_x}, \2 Y\cong \C^{d_y}$, and $\2Z \cong \C^D$ where $D=d_x d_y$. Vectorization with respect to an orthonormal operator basis $\{\sigma_\alpha:\alpha=0,...,D-1\}$ for $L(\2X,\2Y)$ is given by
\begin{equation}
	\sigma\mbox{-vec: }L(\2X,\2Y)\rightarrow \2Z :\,\,A\mapsto \dket{A}_\sigma.
\end{equation}
This operation extracts the coefficients of the basis elements returning the vector  
\begin{equation}
	\label{eqn:trace-rep}
	\dket{A}_\sigma := \sum_{\alpha=0}^{D-1} \Tr[\sigma^\dagger_\alpha A] \ket{\alpha}
\end{equation}
where $\{\ket{\alpha}: \alpha=0,...,D-1\}$ is the standard basis for $\2Z\cong\C^D$. This is depicted in our graphical calculus as 
% FIGURE - BASIS VEC
\figeq{\includegraphics[width=0.4\textwidth]{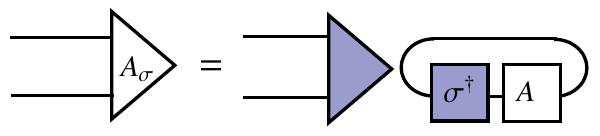} \label{fig:basis-vectorization}}
 
 To distinguish between these different conventions we use the notation $\dket{A}_x$ to denote the vectorization of a matrix $A$, were the subscript $x=c,r,\sigma$ labels which convention we use; either $c$ for col-vec, $r$ for row-vec, or $\sigma$ for an arbitrary operator basis.

For the case $\2 X\cong\2 Y\cong \C^d$, we can define row-vec and col-vec in terms terms of Eq.~\eqref{eqn:trace-rep} by taking our basis to be the elementary matrix basis $\{ E_{i,j}=\ketbra{i}{j} : i,j=0,...,d^2-1\}$, and making the assignment $\alpha=di+j$ and $\alpha=i+dj$ respectively. Hence we have
\begin{eqnarray}
\dket{A}_r &:=& \sum_{i,j=0}^{d-1} A_{ij}\, \ket{i}\otimes\ket{j}	\label{eqn:rowvec}\\ 
\dket{A}_c &:=& \sum_{i,j=0}^{d-1} A_{ij}\, \ket{j}\otimes\ket{i}	\label{eqn:colvec}.
\end{eqnarray} 
Using the definition of the unnormalized Bell-state $\ket{\Phi^+}$ and summing over $i$ and $j$ one can rewrite Eq.~\eqref{eqn:rowvec} and \eqref{eqn:colvec} as
\begin{eqnarray}
\dket{A}_r &=& (A \otimes \id) \ket{\Phi^+}\label{eqn:row-vec}\\
\dket{A}_c &=& (\id \otimes A) \ket{\Phi^+}\label{eqn:col-vec}
\end{eqnarray}
which are the equational versions of our graphical definition of row and col vectorization shown in \eqref{fig:vectorization}.
 
When working in the superoperator formalism for open quantum systems, it is sometimes convenient to transform between vectorization conventions in different bases. Given two orthonormal operator bases $\{\sigma_\alpha\}$ and $\{\omega_\alpha\}$ for $L(\2X,\2Y)$, the basis transformation operator 
\begin{equation}
T_{\sigma\rightarrow\omega}: \2Z\rightarrow\2Z: \dket{A}_\sigma\mapsto \dket{A}_\omega
\end{equation} 
transforms vectorized operators in the $\sigma$-vec convention to the $\omega$-vec convention. Graphically this is given by
% FIGURE - VEC BASIS CHANGE
\figeq{\includegraphics[width=0.3\textwidth]{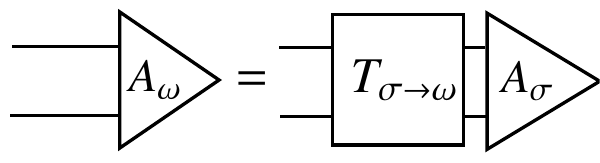} \label{fig:vec-basis-change}}
 
The basis transformation operator $T_{\sigma\rightarrow\omega}$ is given by the equivalent expressions
\begin{equation}
T_{\sigma\rightarrow\omega} 
	= \sum_\alpha \ket{\alpha}\dbra{\omega_\alpha}_\sigma 	
	= \sum_\alpha \dket{\sigma_\alpha}_\omega\bra{\alpha} 	\label{eqn:vec-change},
\end{equation}
and the corresponding graphical representations are: 
% FIGURE
\figeq{\includegraphics[width=0.55\textwidth]{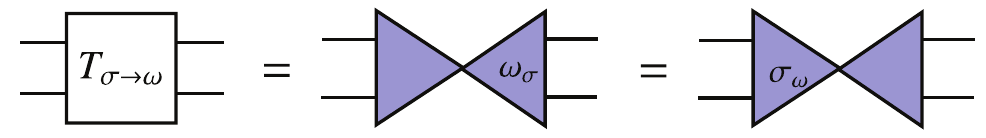}\label{fig:basis-change-op}}
The proofs of Eq.~\eqref{eqn:vec-change-1} and \eqref{eqn:vec-change-2} are found in Appendix~\ref{app:vec-proofs}.

For the remainder of this paper we will use the col-vec convention by default, and drop the vectorization label subscripts unless referring to a general $\sigma$-basis. The main transformation we will be interested in is then from col-vec to another arbitrary orthononormal operator basis $\{\sigma_\alpha\}$. Tensor networks for the change of basis $T_{c\rightarrow\sigma}$ and its inverse $T_{\sigma\rightarrow c}$ are
% FIGURE
\figeq{
 \begin{tabular}{c|c}
\includegraphics[width=0.35\textwidth]{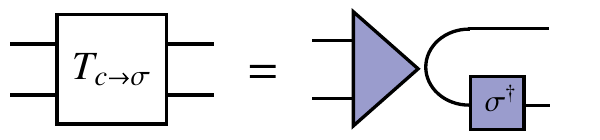}	
\quad\quad&\quad\quad
\includegraphics[width=0.35\textwidth]{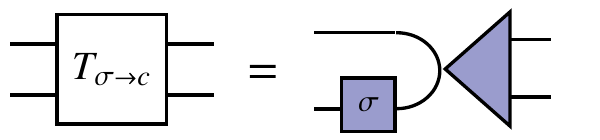}	
\\
\footnotesize{(a) Col-vec to $\sigma$-basis} &
\footnotesize{(b) Row-vec to $\sigma$-basis} 
\end{tabular}
 \label{fig:vec-basis-col}
}
In the case where one wants to convert to row-vec convention, as previously shown the transformation is given by
\begin{equation}
T_{c\rightarrow r}=T_{r\rightarrow c}= \mbox{SWAP}.
\end{equation}

One final important result that often arises when dealing with vectorized matrices is Roth's Lemma for the vectorization of the matrix product $ABC$~\cite{Horn1985}.  Given matrices $A,B,C\in L(\2X)$ we have
\begin{eqnarray}
\dket{ABC} &=& (C^T\otimes A) \dket{B}
\label{eqn:roth}
\end{eqnarray}
The graphical tensor network proof of this lemma is as follows:
% FIGURE
\figeq{
 	\includegraphics[width=0.3\textwidth]{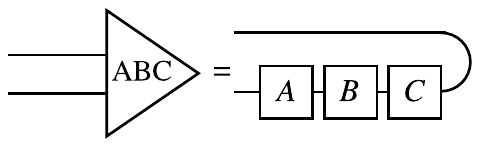}
  	\includegraphics[width=0.3\textwidth]{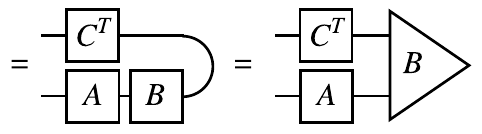}
	 \label{fig:vecabc}
}

%% file: Sections/tnoqs-3-cpmaps.tex
%==============================================================
%==============================================================
%		REPRESENTATIONS
%==============================================================
%==============================================================
\section{Representations of Completely positive maps}
\label{sec:cpmaps}

In this section we recall several common mathematical descriptions for completely-positive trace-preserving maps, and show how several key properties may be captured graphically using the diagrammatic notation we have introduced. The representations we will consider are the Kraus (or operator-sum) representation, the system-environment (or Stinespring) representation, the Liouville superoperator description based on vectorization of matrices, and the Choi-matrix or dynamical matrix description based on the Choi-Jamio{\l}kowski isomorphism. We will also describe the often used process matrix (or $\chi$-matrix) representation and show how this can be considered as a change of basis of the Choi-matrix. Following this, in Section~\ref{sec:trans} we will show how our framework enables one to freely transform between these representations as illustrated in Fig.~\ref{fig:cpreps}.

%===================================================
%		KRAUS
%===================================================
\subsection{Kraus / Operator-Sum Representation}
\label{sec:kraus}

The first representation of CPTP-maps we cast into our framework is the \emph{Kraus}~\cite{Kraus1983} or \emph{operator-sum}~\cite{Nielsen2000} representation. This representation is particularly useful in phenomenological models of noise in quantum systems. Kraus's theorem states that a linear map $\2 E\in T(\2X,\2Y)$ is CPTP if and only if it may be written in the form
\begin{equation}
\2 E(\rho)= \sum_{\alpha=1}^D K_\alpha \rho K_\alpha^\dagger
\label{eqn:kraus}
\end{equation}
where the Kraus operators $\{K_\alpha: \alpha=1,...,D\}$, $K_\alpha\in \Lx{X,Y}$, satisfy the completeness relation 
\begin{equation}
\sum_{\alpha=0}^D K_\alpha^\dagger K_\alpha = \id_{\2 X}.
\label{eqn:completeness}
\end{equation} 
The Kraus representation of $\2E$ in Eq.~\eqref{eqn:kraus} has the graphical representation 
% FIGURE - KRAUS
\figeq{\includegraphics[width=0.5\textwidth]{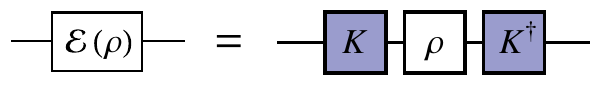}  \label{fig:kraus-evo}}

The maximum number of Kraus operators needed for a Kraus description of $\2 E$ is equal to the dimension of $L(\2X,\2Y)$. For the case where $\2 X\cong\2 Y\cong\C^d$ the maximum number of Kraus operators is $d^2$, and the minimum number case corresponds to unitary evolution where there is only a single Kraus operator.

It is important to note that the Kraus representation of $\2 E$ is not unique as there is unitary freedom in choosing the Kraus operators. We can give preference to a particular representation called the \emph{Canonical Kraus Representation}~\cite{Bengtsson2006} which is the unique set of Kraus operators satisfying the orthogonality relation $\Tr[K^\dagger_\alpha K_\beta]=\lambda_\alpha\delta_{\alpha\beta}$. The canonical Kraus representation will be important when transforming between representations in Section~\ref{sec:trans}.

%===================================================
%		 -ENVIRONMENT
%===================================================
\subsection{System-Environment / Stinespring Representation}
\label{sec:se}

The second representation of CPTP-maps we consider is the system-environment model~\cite{Nielsen2000}, which is typically considered the most physically intuitive description of open system evolution. This representation is closely related to (and sometimes referred to as) the Stinespring representation as it can be thought of as an application of the Stinespring dilation theorem~\cite{Stinespring1955}, which we also describe in this section. In this model, we consider a system of interest $\2 X$, called the \emph{principle system}, coupled to an additional system $\2 Z$ called the \emph{environment}. The composite system of the principle system and environment is then treated as a closed quantum system which evolves unitarily. We recover the reduced dynamics on the principle system by performing a partial trace over the environment. Suppose the initial state of our composite system is given by $\rho\otimes\tau \in L(\2X\otimes\2Z)$, where $\tau\in L(\2Z)$ is the initial state of the environment. The joint evolution is described by a unitary operator $U\in L(\2X\otimes\2Z)$ and the reduced evolution of the principle system's state $\rho$ is given by
\begin{equation}
\2 E(\rho) = \Tr_{\2 Z}[U(\rho\otimes\tau)U^\dagger]
\label{eqn:sys-env}
\end{equation}

For convenience we can assume that the environment starts in a pure state $\tau=\ketbra{v_0}{v_0}$, and in practice one only need consider the case where the Hilbert space describing the environment has at most dimension $d^2$ for $\2 X\cong\C^d$~\cite{Nielsen2000}. The system-environment representation of the CP-map $\2 E$ may then be represented graphically as 
% FIGURE - SE
\figeq{\includegraphics[width=0.42\textwidth]{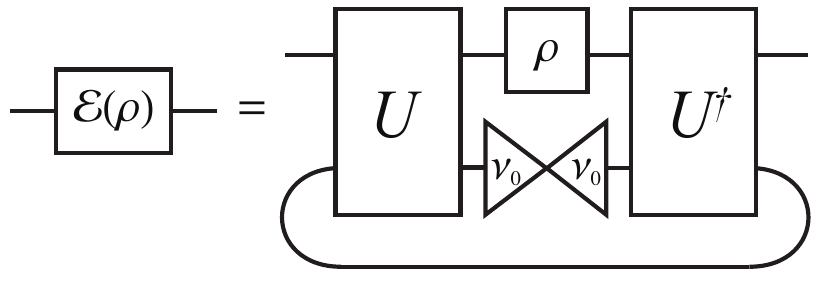} \label{fig:sys-env}}

The system-environment model is advantageous when modelling the environment as a physical system. However, care must be taken when ascribing physical reality to any particular model as the system-environment description is not unique. This is not surprising as many different physical interactions could give rise to the same reduced dynamics on the principle system. This freedom manifests in an ability to choose the initial state of the environment in the representation and then adjust the unitary operator accordingly. In practice, the system-environment model can be cumbersome for performing many calculations where the explicit dynamics of the environment system are irrelevant. The remaining descriptions, which we cast into diagrammatic form, may be more convenient in these contexts.

Note that the system-environment evolution for the most general case will be an isometry and this is captured in Stinespring's representation~\cite{Stinespring1955}. Stinespring's dilation theorem states that a CP-map $\2 E\in C(\2X,\2Y)$ can be written in the form
\begin{equation}
\2 E(\rho) = \Tr_Z\left[A\rho A^\dagger\right]
\label{eqn:stinespring}
\end{equation}
where $A\in L(\2X,\2Y\otimes\2Z)$ and the Hilbert space $\2 Z$ has dimension at most equal to $L(\2X,\2Y)$. Further, the map $\2 E$ is trace preserving if and only if $A^\dagger A=\id_{\2 X}$ ~\cite{Stinespring1955}.

In the case where $\2 Y\cong\2 X$, the Hilbert space $\XZ$ mapped into by the Stinespring operator $A$ is equivalent to the joint system-environment space in the system-environment representation. Hence one may move from the system-environment description to the Stinespring representation as follows:
% FIGURE - STINESPRING
\figeq{\includegraphics[width=0.45\textwidth]{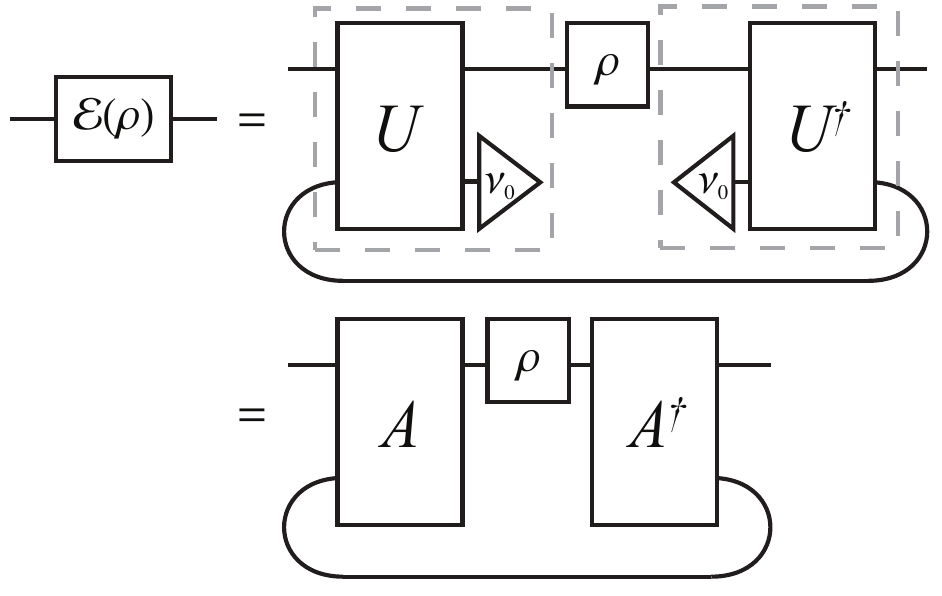}\label{fig:stinespring}}
where $\ket{v_0}\in\2 Z$ is the initial state of the environment, and we have defined the Stinespring operator
\begin{equation}
A = U\cdot(\id_{\2 X}\otimes\ket{v_0}),\label{eqn:stinespring-se}.
\end{equation}

This close relationship is why these two representations are often referred to by the same name, and as we will show in Section~\ref{sec:to-se}, it is straight forward to construct a Stinespring representation from the Kraus representation. However, generating a full description of the joint system-environment unitary operator $U$ from a Stinespring operator $A$ is cumbersome. It involves an algorithmic completion of the matrix elements in the unitary $U$ not contained within the subspace of the initial state of the environment~\cite{Bengtsson2006}. Since it usually suffices to define the action of $U$ when restricted to the initial state of the environment, which by Eq.~\eqref{eqn:stinespring-se} is the Stinepsring representation, this is often the only transformation one need consider.

A further important point is that the evolution of the principle system $\2 E(\rho)$ is guaranteed to be CP if and only if the initial state of the system and environment is separable; $\rho_{\2 X\2 Z}=\rho_{\2X}\otimes\rho_{\3 Z}$. In the case where the physical system is initially correlated with the environment, it is possible to have reduced dynamics which are non-completely positive~\cite{Weinstein2004,Carteret2008}, however such situations are beyond the scope of this paper.

%===================================================
%		SUPEROPERATOR
%===================================================
\subsection{Louiville-Superoperator Representation}
\label{sec:sop}

We now move to the \emph{linear superoperator} or \emph{Liouville} representation of a CP-map $\2 E\in C(\2X,\2Y)$. 
The superoperator representation is based on the vectorization of the density matrix $\rho \mapsto \dket{\rho}_\sigma$ with respect to some orthonormal operator basis $\{\sigma_\alpha : \alpha=0,...,d^2-1\}$ as introduced in Section~\ref{sec:vec}. Once we have chosen a vectorization basis (col-vec in our case) we define the superoperator for a map $\2 E\in T(\2X,\2Y)$ to be the linear map 
\begin{equation}
\2 S:\2X\otimes\2X\rightarrow\2Y\otimes\2Y: \dket{\rho}\mapsto \dket{\2 E(\rho)}	\label{eqn:sop}
\end{equation}
This is depicted graphically as
 %FIGURE - SUPEROP
 \figeq{\includegraphics[width=0.3\textwidth]{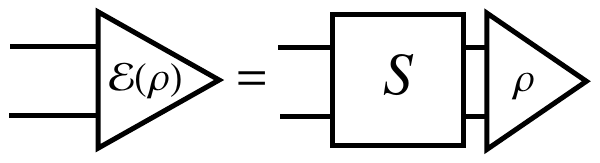}  \label{fig:superop}}
 
 In the col-vec basis we can express the evolution of a state $\rho$ in terms of tensor components of $\2 S$ as
\begin{eqnarray}
\2 E(\rho)_{mn} &=& \sum_{\mu\nu} \2S_{nm,\nu\mu} \rho_{\mu\nu}.
\end{eqnarray}

For the case where $\2 E \in T(\2X)$, it is sometimes  useful to change the basis of our superoperators from the col-vec basis to an orthonormal operator basis $\{\sigma_\alpha\}$ for $L(\2X)$. This is done using the basis transformation operator $T_{c\rightarrow\sigma}$ introduced in Section~\ref{sec:vec}. We have
\begin{eqnarray}
\2 S_\sigma 
	&=& T_{c\rightarrow\sigma}\cdot \2 S \cdot T_{c\rightarrow\sigma}^\dagger \label{eqn:sop-basis-change}\\
	&=& \sum_{\alpha\beta} \2 S_{\alpha\beta} \,\dket{\sigma_\alpha}\dbra{\sigma_\beta}.
\end{eqnarray}
where the subscript $\sigma$ indicates that $\2 S_\sigma$ is the superoperator in the $\sigma$-vec convention. The tensor networks for this transformation is given by
% FIGURE - BASIS CHANGE
\figeq{\includegraphics[width=0.4\textwidth]{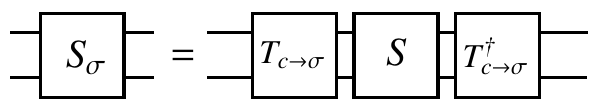}  \label{fig:sop-change-basis}} 
Note that for a general map $\2 E\in T(\2X,\2Y)$ we could do a similar construction but would need different bases for the initial and final Hilbert spaces $L(\2X)$ and $L(\2Y)$.

The structural properties the superoperator $\2 S$ must satisfy for the linear map $\2 E$ to be hermitian-preserving (HP), trace-preserving (TP), and completely positive (CP) are~\cite{Bengtsson2006}:
\begin{eqnarray}
% SOP - HP
\2 E \mbox{ is HP } 
	&\Longleftrightarrow&  \overline{\2 S}=\2 S^S \label{eqn:sop-hpres}\hspace{10em}\\
	&\Longleftrightarrow&
	% FIGURE
	\parbox[c]{1em}{\includegraphics[width=0.28\textwidth, trim= 0cm 1cm 0cm 1cm,clip]{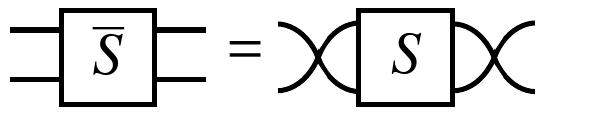}}\\
% SOP - TP
\2 E \mbox{ is TP } 
	&\Longleftrightarrow&\2 S_{mm,n\nu}=\delta_{n\nu}\\
	&\Longleftrightarrow&
	% FIGURE
	\parbox[c]{1em}{\includegraphics[width=0.2\textwidth]{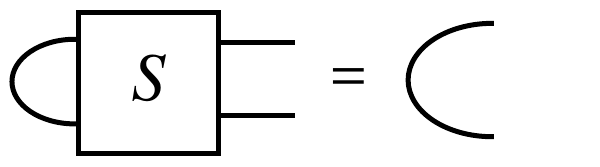}}\\
% SOP - CP
\2 E \mbox{ is CP } 
	&\Longleftrightarrow&\2 S_{\2 I\otimes \2 E}\dket{\rho_{AB}}\ge 0 ~\forall \rho_{AB}\ge0
\end{eqnarray}
Note that there is not a convenient structural criteria on the superoperator $\2 S$ which specifies if $\2 E$ is a CP-map. To test for positivity or complete positivity one generally uses the closely related \emph{Choi-matrix} representation.

Superoperators are convenient to use for many practical calculations. Unlike the system-environment model the superoperator $S$ is unique with respect to the choice of vectorization basis. Choosing an appropriate basis to express the superoperator in can often expose certain information about a quantum system. For example, if we want to model correlated noise for a mutli-partite system we can vectorize with respect to the mutli-qubit Pauli basis. Correlated noise would then manifest as non-zero entries in the superoperator corresponding to terms such as $\sigma_x\otimes\sigma_x$. 
We discus in more detail how this may be done in Section~\ref{sec:comp-sop}.

%===================================================
%		CHOI
%===================================================
\subsection{Choi-Matrix Representation}
\label{sec:choi}

The final representation shown in Fig.~\ref{fig:cpreps} is the \emph{Choi matrix}~\cite{Choi1975}, or dynamical matrix~\cite{Bengtsson2006}. This is an application of the Choi-Jamio{\l}kowski isomorphism which gives a bijection between linear maps and linear operators~\cite{Jamiolkowski1972}. Similarly to how vectorization mapped linear operators in $L(\2X,\2Y)$ to vectors in $\XY$ or $\YX$, the Choi-Jamio{\l}kowski isomorphism maps linear operators in $T(\2X,\2Y)$ to linear operators in $L(\2X\otimes\2Y)$ or $L(\2Y\otimes\2X)$. 
The two conventions are
\begin{eqnarray}
\mbox{col-}\Lambda&:& 
	T(\2X,\2Y)\rightarrow L(\2X\otimes\2Y):\,\, \2 E\mapsto \Lambda_c\\
\mbox{row-}\Lambda&:& 
	T(\2X,\2Y)\rightarrow L(\2Y\otimes\2X):\,\, \2 E\mapsto \Lambda_r.
\end{eqnarray}

For $\2 X\cong\C^d$, the explicit construction of the Choi-matrix is given by
\begin{eqnarray} 
	\label{eqn:col-choi}
	\Lambda_c &=& \sum_{i,j=0}^{d-1}
	 \ketbra{i}{j}\otimes\2 E(\ketbra{i}{j})\\
	\Lambda_r &=& \sum_{i,j=0}^{d-1}
	 \2 E(\ketbra{i}{j})\otimes\ketbra{i}{j}
\end{eqnarray}
where $\{\ket{i}:i=0,\hdots,d-1\}$ is an orthonormal basis for $\2 X$.

We call the two conventions col-$\Lambda$ and row-$\Lambda$ due to their relationship with the vectorization conventions introduced in Section~\ref{sec:vec}. The Choi-Jamio{\l}kowski isomorphism can also be thought of as having a map $\2 E\in T(\2X,\2Y)$ act on one half of an unnormalized Bell-state $\ket{\Phi^+}=\sum_i \ket{i}\otimes\ket{i} \in \2X\otimes\2X$, and hence these conventions corresponding to which half of the Bell state it acts on:
\begin{eqnarray} 
	\Lambda_c &=& (\2 I \otimes \2 E)\ketbra{\Phi^+}{\Phi^+}\label{eqn:jamiolkowski}\\
	\Lambda_r &=& (\2 E \otimes \2 I)\ketbra{\Phi^+}{\Phi^+}
\end{eqnarray}
where $\2 I\in T(\2X)$ is the identity map. In what follows we will use the col-$\Lambda$ convention and drop the subscript from $\Lambda_c$. We note that the alternative row-$\Lambda$ Choi-matrix is naturally obtained by applying the bipartite-SWAP operation to $\Lambda_c$. 

As will be considered in Section~\ref{sec:to-choi}, if the evolution of the CP map $\2 E$ is described by a Kraus representation $\{K_i\}$, then the Choi-Jamio{\l}kowski isomorphism states that we construct the Choi-matrix by acting on one half of a bell state with the Kraus map as shown:
\figeq{
\includegraphics[width=0.65\textwidth]{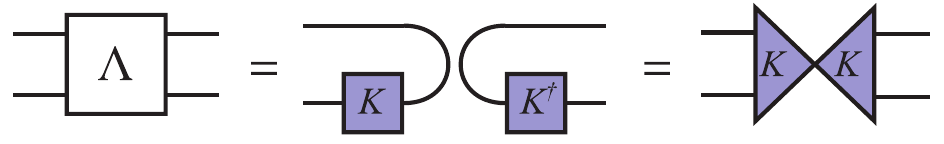}
\label{fig:choi-jamiolkowski}}
Note that in general any tensor network describing a linear map $\2 E$, not just the Kraus description, may be contracted with one-half of the maximally entangled state $\ketbra{\Phi^+}{\Phi^+}$ to construct the Choi-matrix.

With the Choi-Jamio{\l}kowski isomorphism defined, the evolution of a quantum state in terms of the Choi-matrix is then given by
\begin{eqnarray}
\2 E(\rho) &=& \Tr_{\2 X} \left[ (\rho^T\otimes \id_{\2 Y} )\Lambda \right]
\label{eqn:choi-evo}
\end{eqnarray}
or in terms of tensor components
\begin{eqnarray}
\2 E(\rho)_{mn}
		&=&\sum_{n,m} \Lambda_{\mu m,\nu n}\rho_{\mu\nu}.
\end{eqnarray}

%FIGURE - CHOI
The tensor network for Eq.~\eqref{eqn:choi-evo} is given by
\figeq{\includegraphics[width=0.45\textwidth]{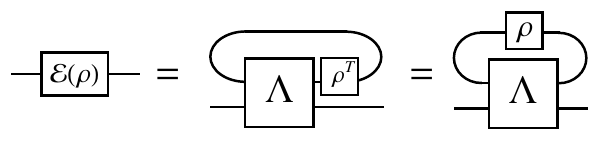}\label{fig:choi-evo}}
The graphical proof of \eqref{fig:choi-evo} for the case where $\2 E$ is described by a Kraus representation is as follows:
 %FIGURE - CHOI
 \figeq{\includegraphics[width=0.45\textwidth]{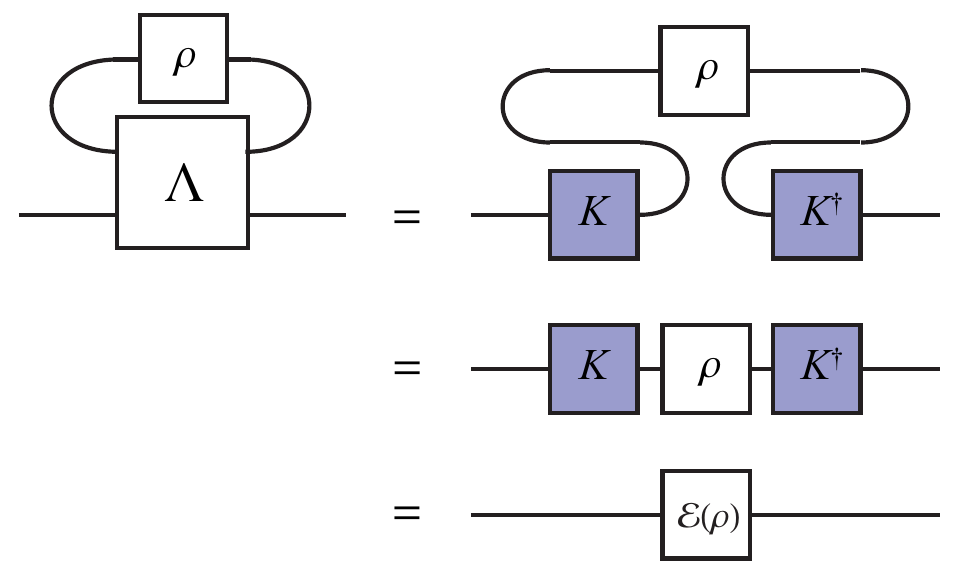}\label{fig:choi-evo-proof}}
 
The structural properties the Choi-matrix $\Lambda$ must satisfy for the linear map $\2 E$ to be hermitian-preserving (HP), trace-preserving (TP),  and completely positive (CP) are~\cite{Bengtsson2006}:
\begin{eqnarray}
% CHOI - HP
\2 E \mbox{ is HP } 
	&\Longleftrightarrow&  \Lambda^\dagger=\Lambda \label{eqn:choi-HP}\hspace{10em}\\
	&\Longleftrightarrow&
	% FIGURE
	\parbox[c]{1em}{\includegraphics[width=0.25\textwidth]{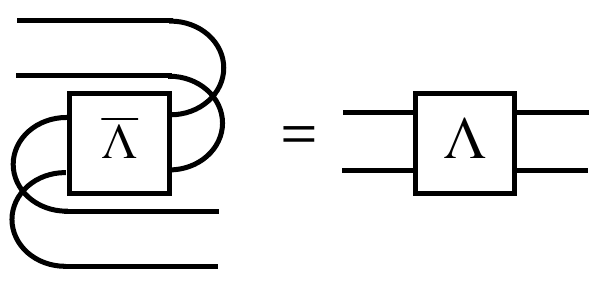}}\\
% CHOI - TP
\2 E \mbox{ is TP } 
	&\Longleftrightarrow& \Tr_{\2 Y}[\Lambda]=\id_{\2 X}\label{eqn:choi-TP}\\
	&\Longleftrightarrow&
	% FIGURE
	\parbox[c]{1em}{\includegraphics[width=0.25\textwidth]{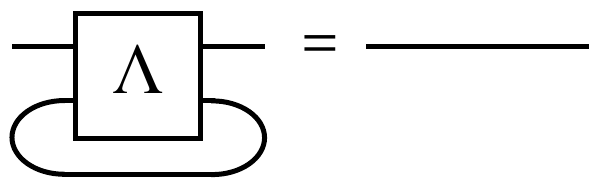}}\label{eqn:choi-tp}\\
% CHOI - CP
\2 E \mbox{ is CP } 
	&\Longleftrightarrow& \2 E \mbox{ is CP } \Longleftrightarrow \Lambda\ge 0.\label{eqn:choithm}
\end{eqnarray}

The Choi-matrix for a given map $\2 E$ is unique with respect to the isomorphism convention chosen. We will provide tensor networks to illustrate a close relationship to the superoperator formed with the corresponding vectorization convention in Section~\ref{sec:choi-sop}. The Choi-matrix finds practical utility as one can check the complete-positivity of the map $\2 E$ by computing the eigenvalues of $\Lambda$. It is also necessary to construct the Choi-matrix for a given superoperator to transform to the other representations. 

Due to the similarity of vectorization and the Choi-Jamio{\l}kowski isomorphism, one could then ask what happens if we vectorize in a different basis. This change of basis of the Choi-matrix is more commonly known as the $\chi$-matrix which we will discuss next. However, such a change of basis does not change the eigen-spectrum of a matrix, so the positivity criteria in Eq.~\eqref{eqn:choithm} holds for any basis.

Another desirable property of Choi matrices is that they can be directly determined for a given system experimentally by \emph{ancilla assisted process tomography (AAPT)}~\cite{DAriano2003,White2003}. This is an experimental realization of the Choi-Jamio{\l}kowski isomorphism which we discuss in detail in Section~\ref{sec:aapt}.

%===================================================
%		CHI MATRIX
%===================================================
\subsection{Process Matrix Representation}
\label{sec:chi}

As previously mentioned, one could consider a change of basis of the Choi-matrix analogous to that for the superoperator. The resulting operator is more commonly known as the \emph{$\chi$-matrix} or \emph{process matrix}~\cite{Nielsen2000}. Consider Hilbert spaces $\2 X\cong\C^{d_x}$, $\2 Y\cong\C^{d_y}$ and let $D=d_x d_y$, and $\2Z \cong \C^D$. If one chooses an orthonormal operator basis  $\{\sigma_\alpha: \alpha=0,...,D-1\}$ for $L(\2X,\2Y)$, then a CPTP map $\2 E\in C(\2X,\2Y)$ may be expressed in terms of a matrix $\chi\in L(\2Z)$ as
\begin{eqnarray}
\2 E(\rho)
&=&\sum_{\alpha,\beta=0}^{D-1} \chi_{\alpha\beta} \sigma_\alpha \rho \sigma_\beta^\dagger
\label{eqn:chi-evo}
\end{eqnarray}
where the process matrix $\chi$ is unique with respect to the choice of basis $\{\sigma_\alpha\}$. 

The process matrix with respect to an orthonormal operator basis $\{\sigma_\alpha\}$ is related to the Choi matrix by the change of basis
\begin{eqnarray}
\chi	&=& T_{c\rightarrow\sigma} \cdot \Lambda \cdot T^\dagger_{c\rightarrow\sigma}
\label{eqn:choi-to-chi}\\
\Rightarrow \Lambda&=& \sum_{\alpha,\beta} \chi_{\alpha\beta} \dketdbra{\sigma_\alpha}{\sigma_\beta}
\end{eqnarray}
where $T_{c\rightarrow\sigma}$ is the vectorization change of basis operator introduced in Section~\ref{sec:vec}. Thus evolution in terms of the $\chi$-matrix is analogous to our Choi evolution as shown below:
 %FIGURE - CHI EVO
 \figeq{\includegraphics[width=0.4\textwidth]{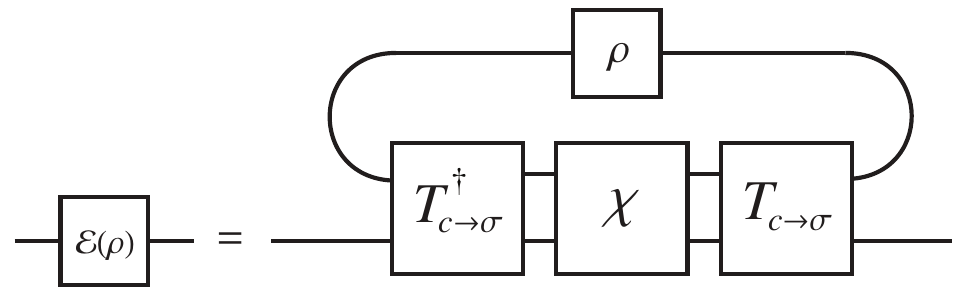}  \label{fig:chi-evo}}
 %FIGURE - CHI PROOFS
Starting with the expression for process matrix evolution in Eq.~\eqref{eqn:chi-evo}, the  graphical proof asserting the validity of Eq.~\eqref{eqn:choi-to-chi} is as follows
\figeq{\includegraphics[width=0.46\textwidth]{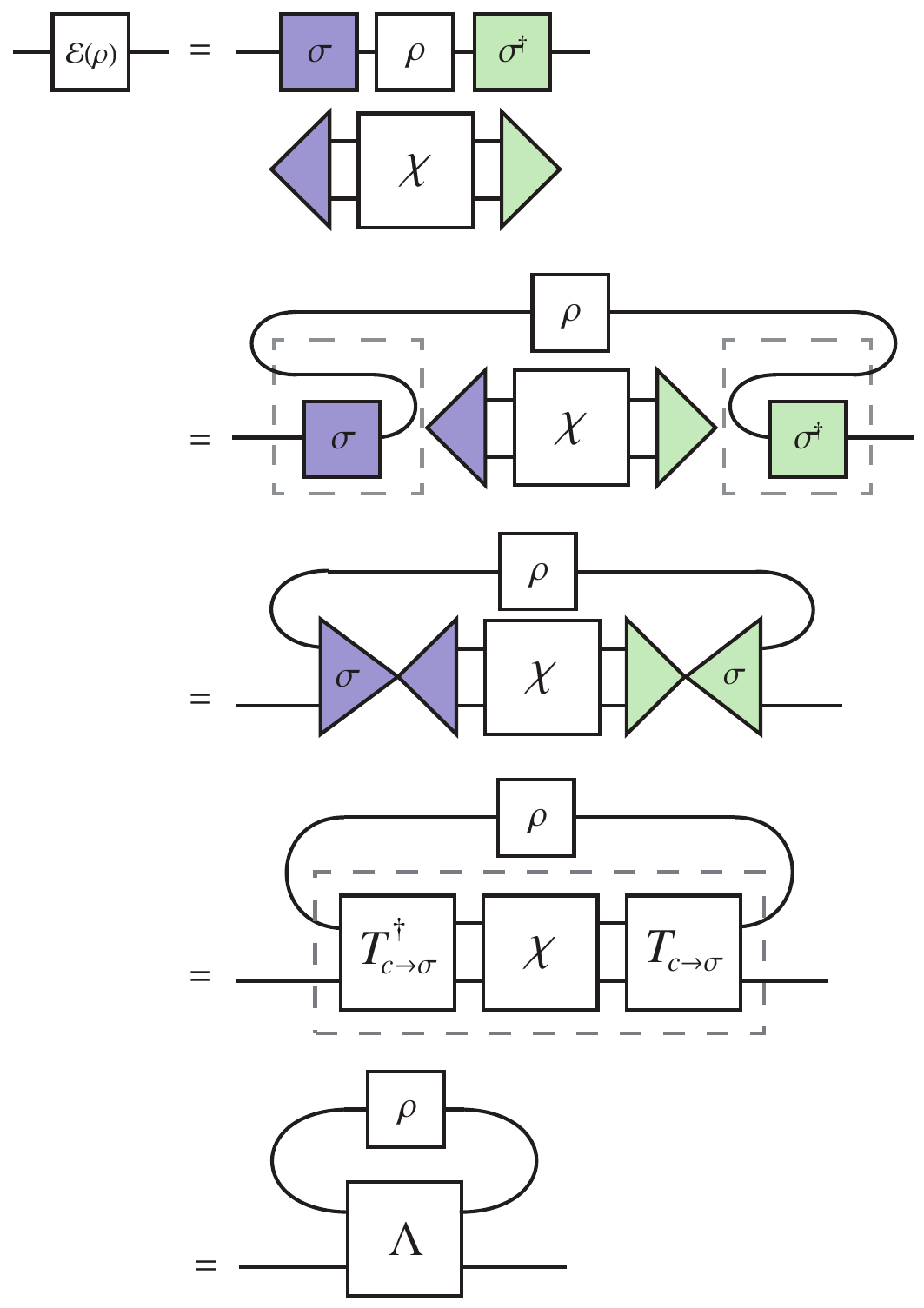}  \label{fig:choi-to-chi}}
We also see that if one forms the process matrix with respect to the col-vec basis $\sigma_\alpha = E_{j,i}$ where $\alpha = i+dj$ and $d$ is the dimension of $\2 H$, then we have $\chi=\Lambda$. 
 
Since the process matrix is a unitary transformation of the Choi-matrix, it shares the same structural conditions for hermitian preservation and complete-positivity as for the Choi-matrix given in Eq.~\eqref{eqn:choi-HP} and \eqref{eqn:choithm} respectively. The condition for it to be trace preserving may be written in terms of the matrix elements and basis however. These conditions are
\begin{eqnarray}
\2 E \mbox{ is TP} &\Longleftrightarrow&
\Tr_{\2 Y}\left[T^\dagger_{c\rightarrow\sigma}\chi T_{c\rightarrow\sigma}\right]=\id_{\2 X}\\
&\Longleftrightarrow& \sum_{\alpha,\beta} \chi_{\alpha,\beta} \sigma_{\alpha}^T \overline{\sigma}_\beta = \id_{\2 X}\\
\2 E \mbox{ is HP} &\Longleftrightarrow& \chi^\dagger=\chi\\
\2 E \mbox{ is CP} &\Longleftrightarrow& \chi\ge0.
\end{eqnarray}

To convert a process-matrix $\chi$ in a basis $\{\sigma_\alpha\}$ to another orthonormal operator basis $\{\omega_\alpha\}$, we may use the same change of basis transformation as used for the superoperator change of basis in Section~\ref{sec:sop}. That is
\begin{eqnarray}
\chi^\omega	&=& T_{\sigma\rightarrow\omega} \cdot \chi^\sigma \cdot T^\dagger_{\sigma\rightarrow\omega}\\
		&=& \sum_{\alpha\beta} \chi^\sigma_{\alpha\beta}\,\, \dket{\sigma_\alpha}_\omega\dbra{\sigma_\beta}_\omega
\end{eqnarray}
where the superscripts $\sigma,\omega$ denote the basis of the $\chi$-matries. This is illustrated as
%FIGURE - CHI BASIS CHANGE
\figeq{\includegraphics[width=0.35\textwidth]{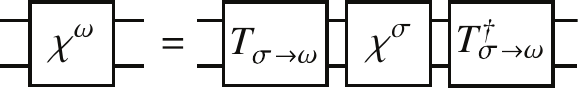}  \label{fig:chi-change-basis}}

%% file: Sections/tnoqs-4-trans.tex
%==============================================================
%==============================================================
%		TRANSFORMATIONS
%==============================================================
%==============================================================
\section{Transforming between representations}
\label{sec:trans}

We now proceed to the task of describing how one may transform between the representations of completely-positive trace-preserving maps depicted in Fig.~\ref{fig:cpreps}. In particular, the transformations depicted as solid arrows in Fig.~\ref{fig:cpreps} have succinct descriptions in the graphical calculus we introduced in Section~\ref{sec:tensor}. These transformations are based on the wire bending dualities for reshuffling, vectorization, and the Choi-Jamio{\l}kowski isomorphism. While the remaining transformations depicted as dashed lines are not based on dualities, but rather non-linear decompositions, or constructions, they also have diagrammatic representations in our graphical calculus for completely positive maps.

%===================================================
%		CHOI-SUPEROP
%===================================================
\subsection{Transformations between the Choi-matrix and superoperator representations}
\label{sec:choi-sop}

The Choi-matrix and superoperator are naturally equivalent under the reshuffling wire bending duality introduced in Section~\ref{sec:bipartite}.  In the col (row) convention we may transform between the two by applying the bipartite col (row)-reshuffling operation $R$ introduced in Section~\ref{sec:bipartite}. Let $\Lambda\in \LXY$ be the Choi-matrix, and $\2 S\in \Lx{\XX,\YY}$ be the superoperator, for a map $\2 E\in \Tx{X,Y}$. Then we have
\begin{eqnarray}
\Lambda &=& \2S^{R} \quad\quad\2 S = \Lambda^{R}
\end{eqnarray}
The tensor networks for these transformations using the col convention are
%FIGURE - CHOI <-> SOP
\figeq{
\begin{tabular}{c|c}
\includegraphics[width=0.3\textwidth]{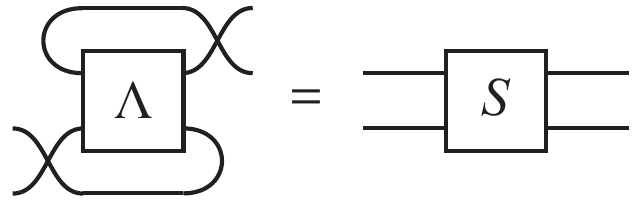}
\hspace{2em} & \hspace{2em}
\includegraphics[width=0.3\textwidth]{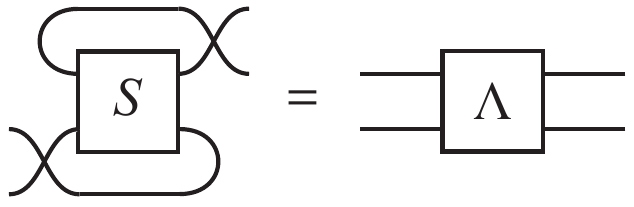}
\end{tabular}
\label{fig:choi-sop}
}
In terms of tensor components we have
\begin{eqnarray}
\Lambda_{mn,\mu\nu} 	&=& \2 S_{\nu n,\mu m}
\end{eqnarray}
where $m,n$ and $\mu,\nu$ index the standard bases for $\2 X$ and $\2 Y$ respectively. Graphical proofs of the relations $\Lambda^{R_c}=\2 S$ and $\2 S^{R_c}=\Lambda$ are given below
% FIGURE - CHOI SOP PROOF
\figeq{
\begin{tabular}{c|c}
\includegraphics[width=0.4\textwidth]{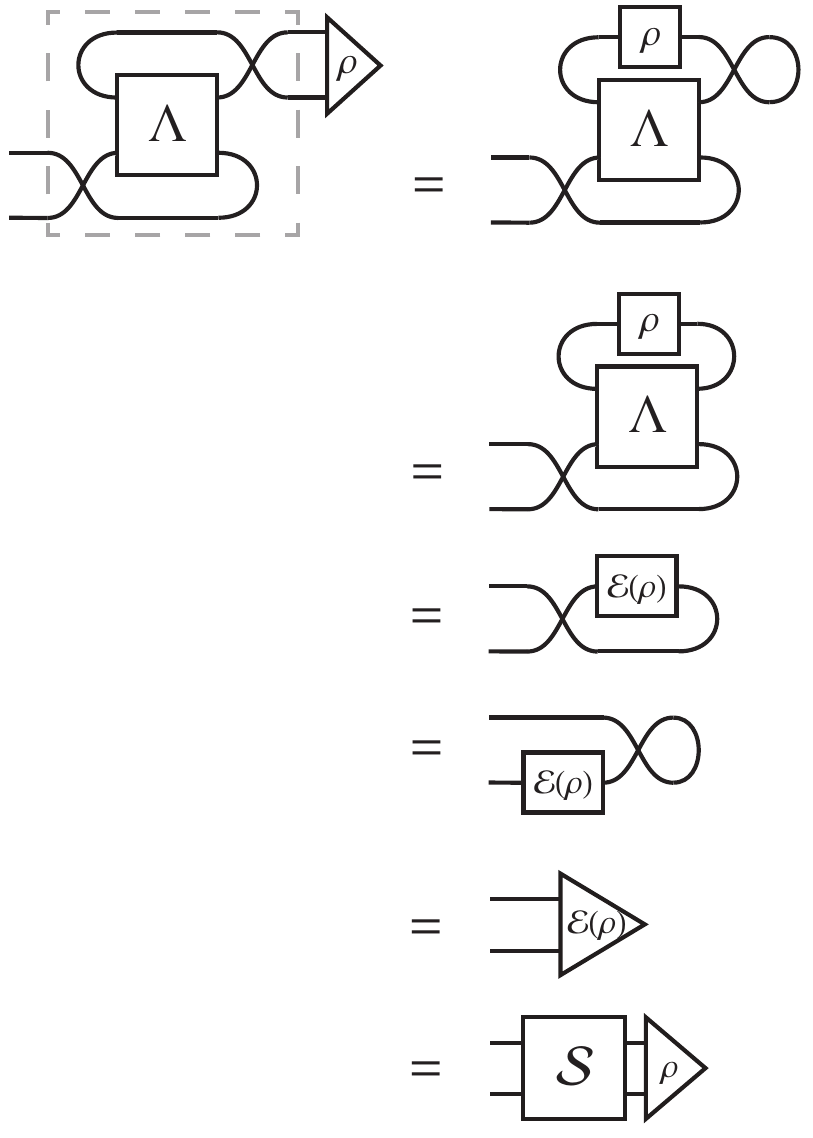} 
\quad\quad&\quad
\includegraphics[width=0.4\textwidth]{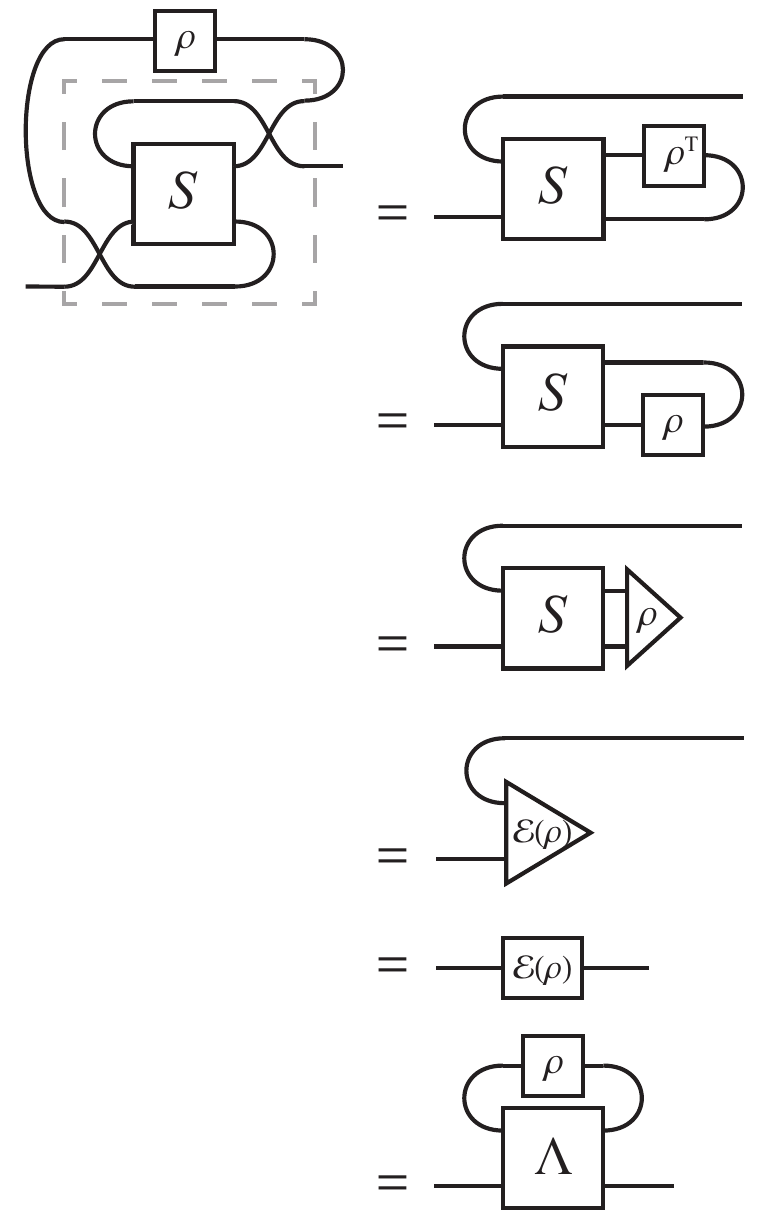}
\end{tabular}
 \label{fig:choi-sop-proof}
 }

To transfer between a $\chi$-matrix with respect to an arbitrary operator basis, and a superoperator with respect to an arbitrary vectorization basis, we must first convert both to col-vec (or row-vec) convention and then proceed by reshuffling.

Note that reshuffling is its own inverse, ie $(\Lambda^R)^R=\Lambda$, hence the solid bi-directional arrow connecting the Choi-matrix and superoperator representations in Fig.~\ref{fig:cpreps}. This is the only transformation between the representations we consider which is linear, bijective, and self-inverse.

%===================================================
%		TO SUPEROPERATOR
%===================================================
\subsection{Transformations to the superoperator representation}
\label{sec:to-sop}

Transformations to the superoperator from the Kraus and system-environment representations of a CP-map are also accomplished by a wire-bending duality, in this case vectorization. However, unlike the bijective equivalence of the Choi-matrix and superoperator under the reshuffling duality, the vectorization duality is only surjective.

If we start with a Kraus representation for a CPTP map $\2 E\in\Cx{X,Y}$ given by $\{K_\alpha: \alpha=0,...,D-1 \}$, with $K_\alpha\in \Lx{X,Y}$, we can construct the superoperator $\2 S \in \Lx{\XX,\YY}$ by 
\begin{eqnarray}
\2 S &=&\sum_{\alpha=0}^{D-1} \overline{K}_\alpha\otimes K_\alpha.
\label{eqn:kraus-to-choi}
\end{eqnarray}
The corresponding tensor network is  
% FIGURE - KRAUS TO SOP
\figeq{\includegraphics[width=0.3\textwidth]{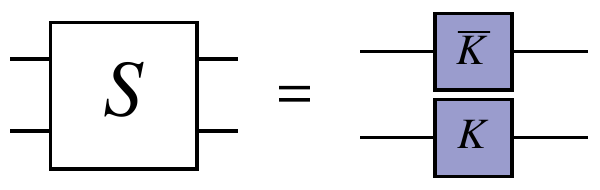} \label{fig:kraus-sop}}
and the graphical proof of this relationship follows directly from Roth's lemma:
% FIGURE - KRAUS TO SOP PROOF
\figeq{
\includegraphics[width=0.8\textwidth]{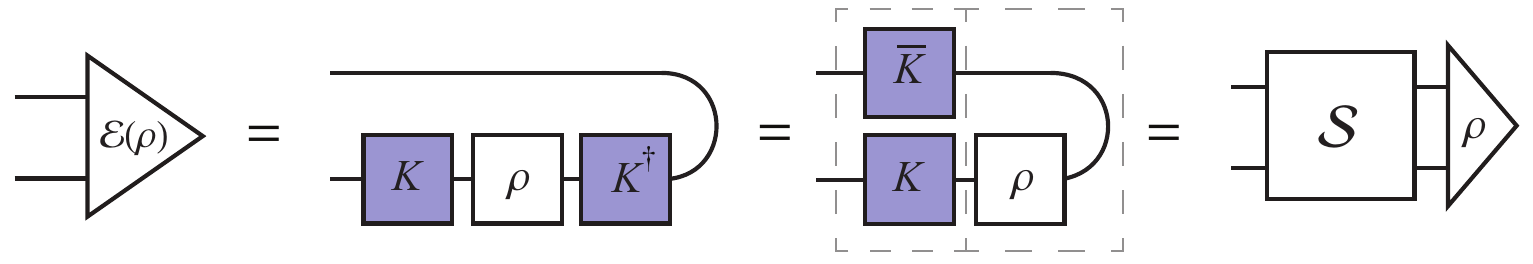}
  \label{fig:kraus-sop-proof}}

Starting with a system-environment (or Stinespring) representation of a map $\2 E\in\Cx{X,Y}$ with input and output system Hilbert spaces $\2 X\cong \C^{d_x}$ and $\2Y\cong \C_{d_y}$ respectively, and environment Hilbert space $\2 Z\cong \C^D$ with $1\le D \le d_x d_y$, we may construct the superoperator for this map from the joint system-environment unitary $U$ and initial environment state $\ket{v_0}$ by
\begin{eqnarray}
\2 S &=& \sum_{\alpha} \bra{\alpha}\overline{U}\ket{v_0}\otimes \bra{\alpha}U\ket{v_0}, \label{eqn:se-to-sop}
\end{eqnarray}
where $\{\ket{\alpha}:\alpha=0,...,D-1\}$ is an orthonormal basis for $\2 Z$. The corresponding tensor network is 
\figeq{\includegraphics[width=0.3\textwidth]{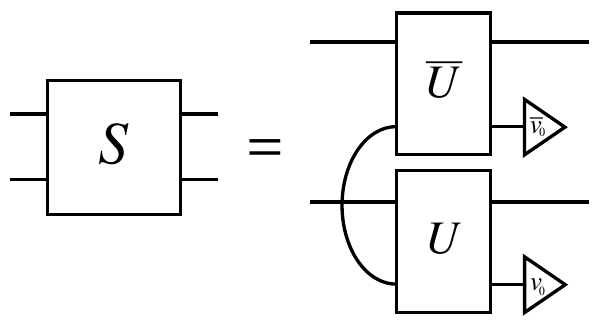}
  \label{fig:se-sop}
}
 As with the Kraus to superoperator transformation, the proof of Eq.~\eqref{eqn:se-to-sop} follows from Roth's lemma.

Note that while the vectorization wire bending duality is invertible, these transformations to the superoperator from the Kraus and system-environment representations are single directional. In both cases injectivity fails as the superoperator is unique, while both the Kraus and system-environment representations are not. Hence we have solid single directional arrows in Fig.~\ref{fig:cpreps} connecting both the Kraus and system-environment representations to the superoperator. The inverse transformation from a superoperator to the Kraus or system-environment representation requires a canonical decomposition of the operator $\2 S$ (via first reshuffling to the Choi-matrix), which is detailed in Sections~\ref{sec:to-kraus} and \ref{sec:to-se}.

%===================================================
%		TO CHOI-MATRIX
%===================================================
\subsection{Transformations to the Choi-matrix representation}
\label{sec:to-choi}

Transforming to the Choi-matrix from the Kraus and system-environment representations is accomplished via a wire-bending duality which captures the Choi-Jamio{\l}kowski isomorphism. As with the case of transforming to the superoperator, this duality transformation is surjective but not injective.

Given a set of Kraus matrices $\{ K_\alpha: \alpha=0,...,D-1\}$ where $K_\alpha\in\Lx{X,Y}$ for a CPTP-map $\2 E\in\Cx{X,Y}$, one may form the Choi-Matrix $\Lambda$ as was previously illustrated in \eqref{fig:choi-jamiolkowski} in Section~\ref{sec:choi}. In terms of both Dirac notation and tensor components we have:

\begin{eqnarray}
\Lambda	&=& \sum_{i,j}\left( \ketbra{i}{j}\otimes \sum_\alpha K_\alpha\ketbra{i}{j}K^\dagger_\alpha\right)\\
		&=& \sum_{\alpha} \dketdbra{K_\alpha}{K_\alpha}\\
\Lambda_{mn,\mu\nu}&=&\sum_\alpha (K_\alpha)_{\mu m}(\overline{K}_\alpha)_{\nu n}.
\end{eqnarray}
where $\{\ket{i}\}$ is an orthonormal basis for $\2 X$, $m,n$ index the standard basis for $\2 X$, and $\mu,\nu$ index the standard basis for $\2 Y$.

Given a system-environment representation with joint unitary $U\in \Lx{\XZ}$ and initial environment state $\ket{v_0}\in \2 Z$ we have
\begin{eqnarray}
\Lambda	&=&	\sum_{i,j}\left(\ketbra{i}{j}\otimes 
			\Tr_{\2 Z}\left[U\ketbra{i}{j}\otimes\ketbra{v_0}{v_0}U^\dagger\right]\right)
\end{eqnarray}
Graphically this is given by
% FIGURE - CHOI - STINESPRING
\figeq{\includegraphics[width=0.4\textwidth]{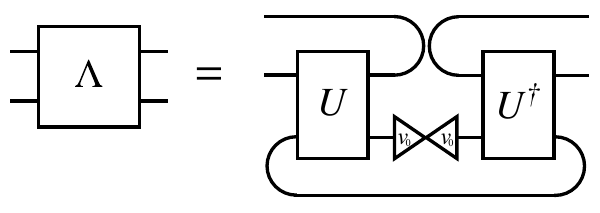} \label{fig:choi-se}}
 
The proof of these transformations follow directly from the definition of the Choi-matrix in Eq.~\eqref{eqn:col-choi}, and the tensor networks for the evolution via the Kraus or system-environment representations given in \eqref{fig:kraus-evo} and \eqref{fig:sys-env} respectively. As with the vectorization transformation to the superoperator discussed in Section~\ref{sec:to-sop}, even though the Choi-Jamio{\l}kowski isomorphism is linear these transformations are single directional as injectivity fails due to the non-uniqueness of both the Kraus and system-environment representations. Hence we have the solid single-directional arrows connecting both the Kraus and system-environment representations to the Choi-matrix in Fig.~\ref{fig:cpreps}. 
 
This completes our description of the linear transformations between the representations of CP-maps in Fig.~\ref{fig:cpreps}. We will now detail the non-linear transformations to the Kraus and system environment representations. 
 
%===================================================
%		TO KRAUS
%===================================================
\subsection{Transformations to the Kraus Representation}
\label{sec:to-kraus}

We may construct a Kraus representations from the Choi-matrix or system-environment representation by the non-linear operations of spectral-decomposition and partial trace decomposition respectively. To construct a Kraus representation from the Superoperator however, we must first reshuffle to the Choi-matrix. 

To construct Kraus matrices from a Choi matrix we first recall the graphical Spectral decomposition we introduced as an example of our color summation convention in \eqref{fig:spectral-sum}. If  $\2 E$ is CP, by Eq.~\eqref{eqn:choithm} we have $\Lambda\ge0$ and hence the spectral decomposition of the Choi-matrix is given by
\begin{equation}
\Lambda=\sum_\alpha \mu_\alpha \ketbra{\phi_\alpha}{\phi_\alpha}, 
\end{equation}
where  $\mu_\alpha \ge 0$ are the eigenvalues, and $\ket{\phi_\alpha}$ the eigenvectors of $\Lambda$. Hence we can define Kraus operators $K_\alpha= \lambda_\alpha A_\alpha$ where $\lambda_\alpha = \sqrt{\mu_\alpha}$ and $A_\alpha$ is the unique operator satisfying $\dket{A_\alpha}=\ket{\phi_\alpha}$ as illustrated: 
% FIGURE - CHOI 2 KRAUS
\figeq{\includegraphics[width=0.35\textwidth]{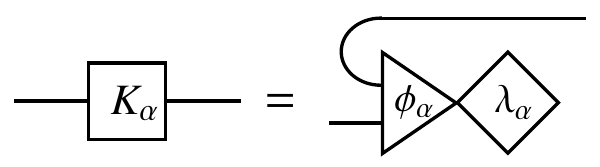}
\label{fig:choi2kraus}}
The number of Kraus operators will be equal to the rank $r$ of the Choi-matrix, where $1\le r\le \mbox{dim}(\Lx{X,Y})$. The graphical proof of this transformation is as follows:
 % FIGURE - CHOI TO KRAUS PROOF
 \figeq{\includegraphics[width=0.42\textwidth]{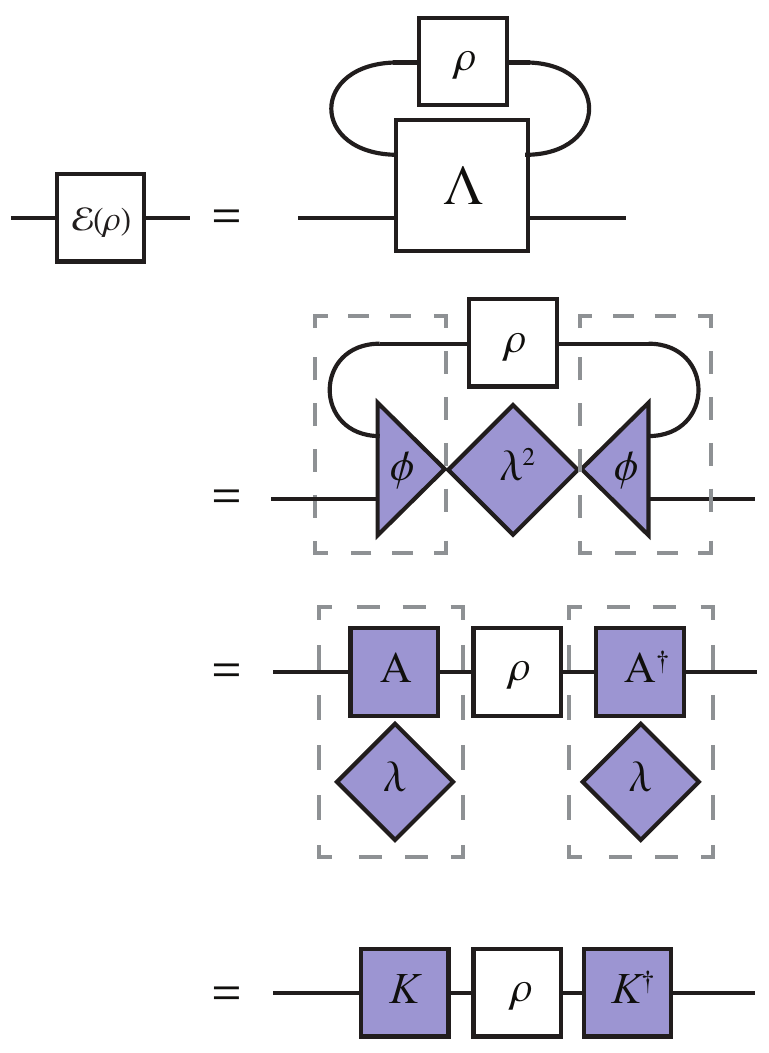} \label{fig:choi-kraus-proof}}
The proof that Kraus operators satisfy the completeness relation follows from the trace preserving property of $\Lambda$ in Eq.~\eqref{eqn:choi-TP}: 
\figeq{\includegraphics[width=0.43\textwidth]{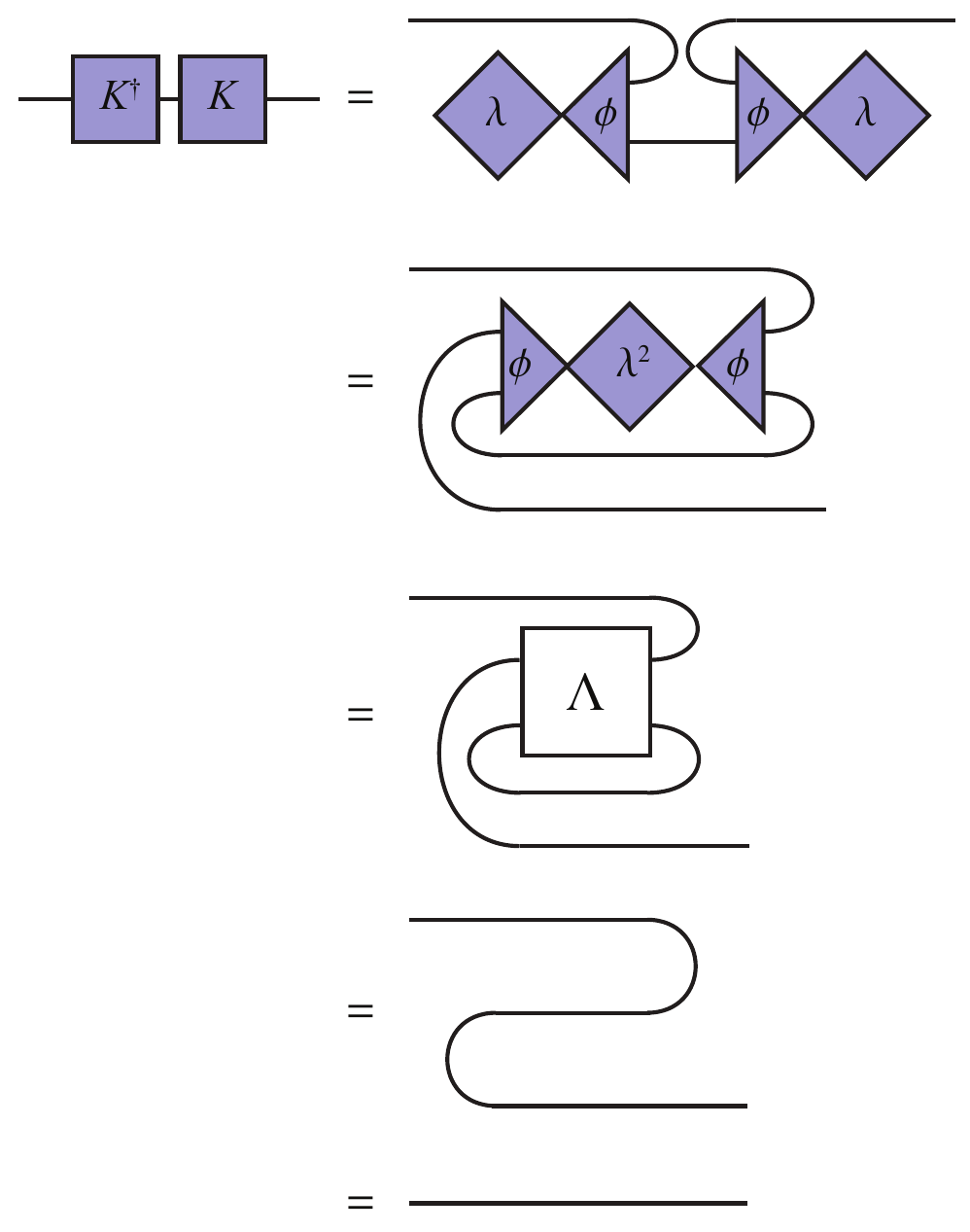}\label{fig:completeness-proof}}
  
Note that since $\Lambda$, and the $\chi$-matrix are related by a unitary change of basis, the Kraus representations constructed from their respective spectral decompositions will also be related by the same transformation. Each will give a unitarily equivalent Canonical Kraus representation of $\2 E$ since the eigen-vectors are orthogonal. Thus we have described the arrow in Fig.~\ref{fig:cpreps} connecting the Choi-matrix to the Kraus representation. It is represented as a dashed arrow as it involves a non-linear decomposition, and is single directional as this representation transformation is injective, but not surjective. Surjectivity fails as we can only construct the canonical Kraus representations for $\2 E$. The reverse transformation is given by the Jamio{\l}kowski isomorphism described in Section~\ref{sec:to-choi}.

Starting with a system-environment representation with joint unitary $U\in \Lx{\XZ}$ and initial environment state $\ket{v_0}\in\2 Z$, we first choose an orthonormal basis $\{\ket{\alpha}:\alpha=0,...,D-1\}$ for $\2 Z$. We then construct the Kraus representation by decomposing the partial trace in this basis as follows
\begin{eqnarray}
\2 E(\rho) 
	&=&\Tr_E\left[U\left(\rho\otimes\ketbra{v}{v}\right)U^\dagger\right] \\
	&=& \sum_{\alpha=0}^{D-1} \bra{\alpha}U\ket{v_0} \rho \bra{v_0}U^\dagger\ket{\alpha}\\
	&=& \sum_{\alpha=0}^{D-1} K_\alpha \rho K_\alpha^\dagger.
\end{eqnarray}
Hence we may define Kraus matrices 
\begin{equation}
K_\alpha = \bra{\alpha}U\ket{v_0}
\label{eqn:kraus-se-proof}
\end{equation}
 leading to the tensor network 
 % FIGURE - SE TO KRAUS
 \figeq{\includegraphics[width=0.35\textwidth]{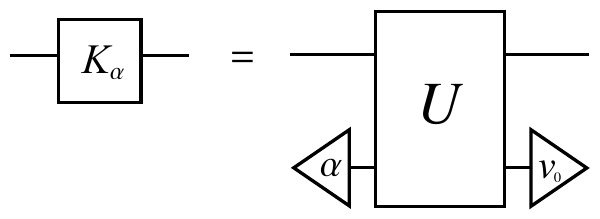} \label{fig:kraus-sys-env}
}
The graphical proof of Eq.~\eqref{eqn:kraus-se-proof} and \eqref{fig:kraus-sys-env} is as follows
 % FIGURE - SE TO KRAUS PROOF
 \figeq{\includegraphics[width=0.45\textwidth]{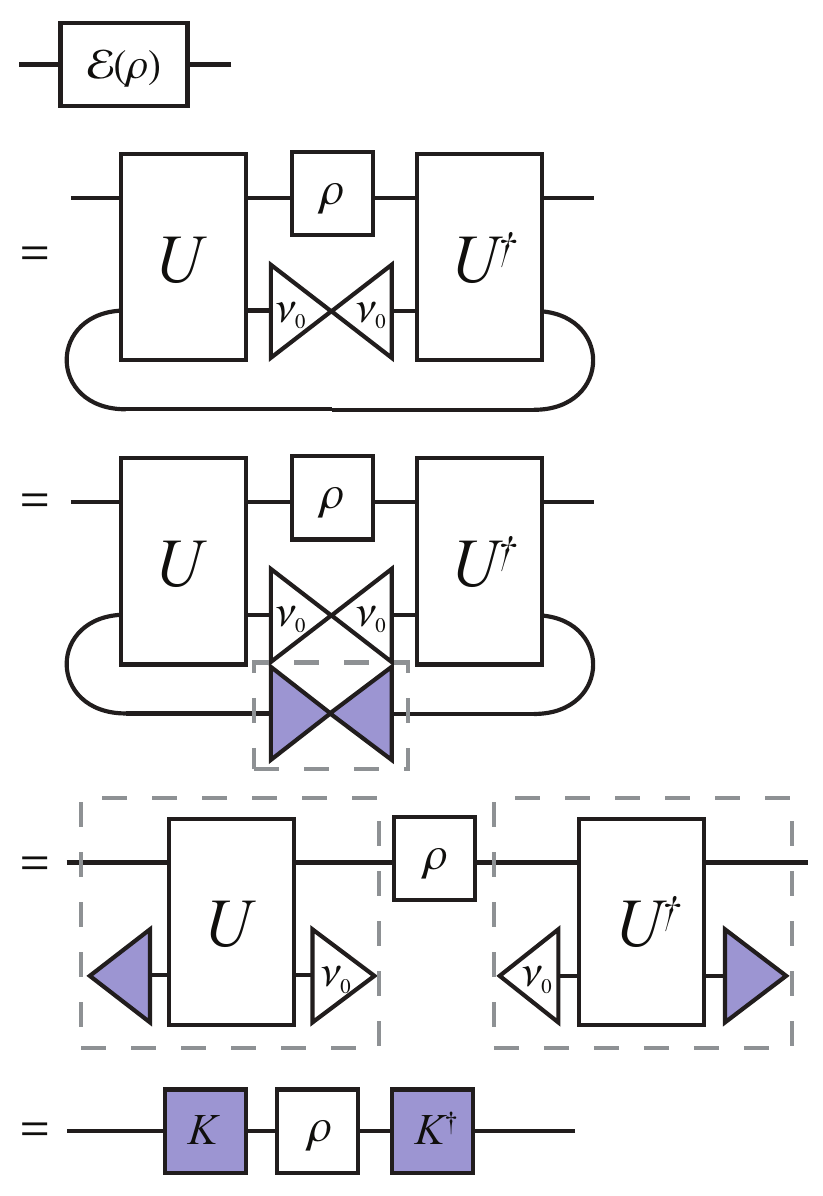} 
 \label{fig:kraus-sys-env-proof}}

Though the Kraus and system-environment representations are both non-unique, for a fixed environment basis this partial trace decomposition is an injective transformation between the Kraus and Stinespring representations (or equivalently between the Kraus and system-environment representations when the joint unitary is restricted to a fixed initial state of the environment). 
To see this  let $\{K_\alpha\}$ and $\{J_\alpha\}$ be two Kraus representations for a CPTP-map $\2 E\in\Cx{X,Y}$, constructed from Stinespring representations $A$ and $B$ respectively. We have that
\begin{eqnarray}
K_\alpha &=& J_\alpha\\
&\Leftrightarrow& (K_\alpha)_{ij}= (J_\alpha)_{ij}\\
&\Leftrightarrow& A_{i\alpha,j} = B_{i\alpha,j}\\
&\Leftrightarrow& A=B.
\end{eqnarray}
Since the Stinespring operators satisfy $A=U\ket{v_0}$ and $B=V\ket{v_0}$ for some joint unitaries $U$ and $V$, we must have that $U_0=V_0$ where $U_0$ and $V_0$ are the joint unitaries restricted to the subspace of the environment spanned by $\ket{v_0}$.

This transformation can be thought of as the reverse application of the Stinespring dilation theorem, and hence for a fixed choice of basis (and initial state of the environment) it is invertible. The inverse transformation is the Stinespring dilation, and as we will show in Section~\ref{sec:to-se}, since the inverse transformation is also injective this transformation is a bijection. However, since the partial trace decomposition involves a choice of basis for the environment it is non-linear --- hence we use a dashed bi-directional arrow to represent the transformation from the system-environment representation to the Kraus representation in Fig.~\ref{fig:cpreps}.

%===================================================
%		TO SYS-ENV
%===================================================
\subsection{Transformations to the system-environment representation}
\label{sec:to-se}

We now describe the final remaining transformation given in Fig.~\ref{fig:cpreps}, the bijective non-linear transformation from the Kraus representation to the system-environment, or Stinespring, representation. The system-environment representation is the most cumbersome to transform to as it involves the unitary competition of a Stinespring dilation of a Kraus representation. Thus starting from a superoperator one must first reshuffle to the Choi-matrix, then from the Choi-matrix description one must then spectral decompose to the canonical Kraus representation, before finally constructing the system-environment as follows.

Let $\{ K_\alpha:\alpha=0,..., D-1\}$, where $1\le D\le \mbox{dim}(\Lx{X,Y})$, be a Kraus representation for the CP-map $\2 E\in \Tx{X,Y}$. Consider an ancilla Hilbert space $\2 Z\cong \C^D$, this will model the environment. If we choose an orthonormal basis for the environment, $\{\ket{\alpha}: \alpha=0,...,D-1\}$, then by Stinesprings dilation theorem we may construct the Stinespring matrix for the CP map $\2 E$ by 
\begin{equation}
A = \sum_{\alpha=0}^{D-1} K_\alpha \otimes\ket{\alpha}.
\label{eqn:kraus2stinespring}
\end{equation}

Recall from Section~\ref{sec:se} that the Stinespring representation is essentially the system-environment representation when the joint unitary operator is restricted to the subspace spanned by the initial state of the environment. Hence if we let $\ket{v_0}\in \2 Z$ be the initial state of the environment system, then this restricted unitary is given by
\begin{equation}
U_0 = \sum_\alpha K_\alpha\otimes\ketbra{\alpha}{v_0},\label{eqn:kraus-to-U}.
\end{equation} 
The tensor networks for Eq.~\eqref{eqn:kraus2stinespring} and \eqref{eqn:kraus-to-U} are:
% FIGURE - STINESPRINE SE
\figeq{
\begin{tabular}{c|c}
\includegraphics[width=0.42\textwidth]{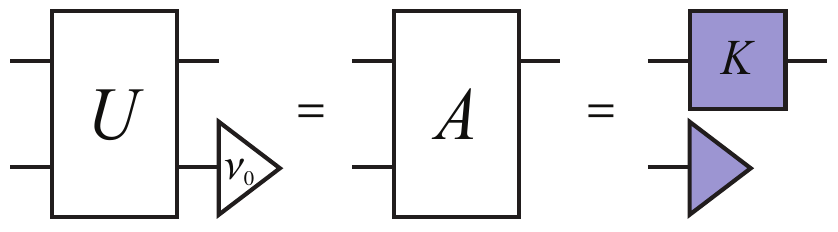}	
\quad\quad&\quad\quad
\includegraphics[width=0.25\textwidth]{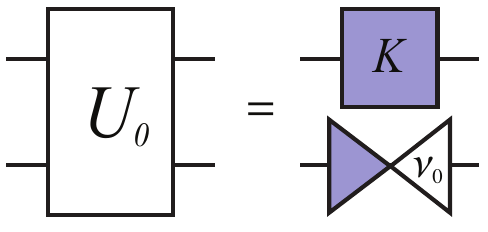}	\\
\footnotesize{(a) Stinespring operator}	&
\footnotesize{(a) Restricted unitary}
\end{tabular}
 \label{fig:sys-env-kraus}
 }
The graphical proof that this construction gives the required evolution of a state $\rho$ is as follows
% FIGURE - STINESPRINE SE PROOF
\figeq{
\includegraphics[width=0.33\textwidth]{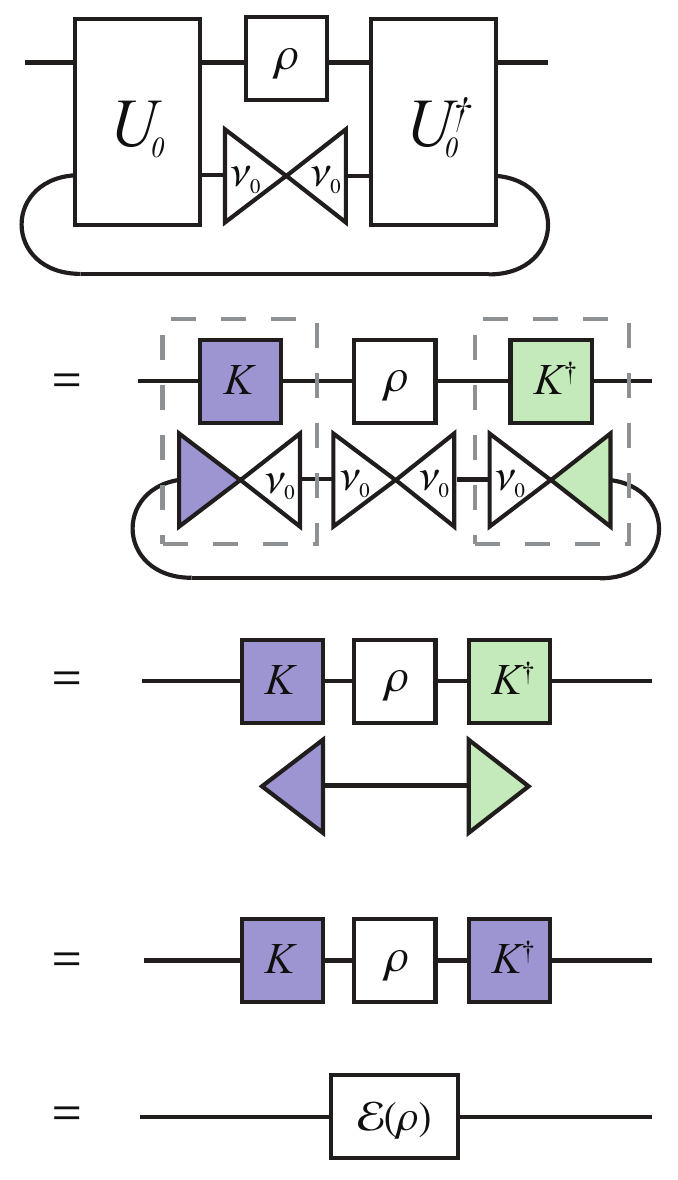}\label{fig:kraus-se-proof}}
In principle, one may complete the remaining entries of this matrix to construct the full matrix description for the unitary $U$, however such a process is cumbersome and is unnecessary to describe the evolution of the CP-map $\2 E$~\cite{Bengtsson2006}.

We have now finished characterizing the final transformations depicted in Fig.~\ref{fig:cpreps} connecting the Kraus representation to the system-environment representation by Stinespring dilation. As previously mentioned in Section~\ref{sec:to-kraus}, for a fixed choice of basis and initial state for the environment, the transformation between Kraus and Stinespring representations is bijective (and hence so is the transformation between Kraus and system-environment representations when restricted to the subspace spanned by the initial state of the environment). Though both these representations are non-unique, by fixing a basis and initial state for the environment we ensure that this transformation is injective. To see this let $U_0$ and $V_0$ be unitaries restricted to the state $\ket{v_0}$ constructed from Kraus representations, $\{K_\alpha\}$ and $\{J_\alpha\}$ respectively, for $\2 E\in \Cx{X,Y}$. Then 
\begin{eqnarray}
U_0=V_0
&\Leftrightarrow&	\sum_\alpha K_\alpha\otimes\ketbra{\alpha}{v_0} = \sum_\alpha J_\alpha \otimes\ketbra{\alpha}{v_0}\\
&\Leftrightarrow&	\sum_\alpha K_\alpha\braket{\beta}{\alpha} = \sum_\alpha J_\alpha \braket{\beta}{\alpha}\\ 
&\Leftrightarrow& K_\beta = J_\beta
\end{eqnarray}
Bijectivity then follows from the injectivity of the inverse transformation --- the previously given construction of a Kraus representation by the partial trace decomposition of a joint unitary operator in \eqref{fig:kraus-sys-env}. 

%% file: Sections/tnoqs-5-examples.tex
 %==============================================================
%==============================================================
%		APPLICATIONS
%==============================================================
%==============================================================
\section{Applications}
\label{sec:ex}

We have now introduced all the basic elements of our graphical calculus for open quantum systems, and shown how it may be used to graphically depict the various representations of CP-maps, and transformations between representations. In this section we move onto more advanced applications of the graphical calculus. We will demonstrate how to apply vectorization to composite quantum systems, and in particular how to compose multiple superoperators together, and construct effective reduced superoperators from tracing out a subsystem. We also demonstrate the superoperator representation of various linear transformations of matrices. These constructions will be necessary for the remaining examples where we derive a succinct condition for a bipartite state to be used for ancilla assisted process topography, and where we present arguably simpler derivations of the closed form expression for the average gate fidelity and entanglement fidelity of a quantum channel in terms of properties of each of the representations of CP-maps given in Section~\ref{sec:cpmaps}.
 
%===================================================
% VECTORIZING COMPOSITE SYSTEMS
%===================================================
\subsection{Vectorization of composite systems}
\label{sec:vec-comp}

We now describe how to deal with vectorization of the general case of composite system of $N$ finite dimensional Hilbert spaces. Let $\2 X_k\cong \C^{d_k}$ be a $d_k$-dimensional complex Hilbert space, and let $\{\ket{i_k}:i_k=0,...,d_k-1\}$ be the standard basis for $\2 X_k$. We are interested in the composite system of $N$ such Hilbert spaces, 
\begin{equation}
\2 X = \2 X_1\otimes...\otimes \2 X_{N}=\bigotimes_{k=1}^{N} \2 X_k
\end{equation}
 which has dimensions $D=\prod_{k=1}^N d_k$. Let $\{\ket{\alpha}: \alpha = 0,..., D-1\}$ be the computational basis for $\2 X$. We can consider vectors in $\2 X$ and the dual space $\2 X^\dagger$ as either 1st-order tensors where their single wire represents an index running over $\alpha$, or as a $N$th-order tensor where each of the $N$ wire corresponds to an individual Hilbert space $\2 X_k$. The correspondence between these two descriptions is made by the concatenation of the composite indices according to the lexicographical order 
 \begin{equation}
 \alpha =\sum_{k=1}^{N} c(k)\, i_k \quad\mbox{where}\quad c(k) := \frac{D}{\prod_{j=1}^k d_j}.
 \end{equation} 
Note that one could also consider the object as any order tensor between 1st and $N$th by the appropriate concatenation of some subset of the the wires.

We define the unnormalized Bell-state on the composite system $\XX$ to be the state formed by the column (or row) vectorization of the identity operator $\id_{\2X} \in L(\2X)$
\begin{eqnarray}
\dket{\id_{\2 X}} 
	&=& \sum_{\alpha=0}^D \ket{\alpha}\otimes\ket{\alpha}\nonumber\\
	&=& \sum_{i_1=0}^{d_1-1}....\sum_{i_N=0}^{d_N-1} \ket{i_1,..., i_N}\otimes\ket{i_1,...,i_N}.
\label{eqn:comp-bell}
\end{eqnarray}
where $\ket{i_1,...,i_N}:=\ket{i_1}\otimes...\otimes\ket{i_N}$. The tensor network for this state is
% FIGURE - MULTI BELL
\figeq{\includegraphics[width=0.25\textwidth]{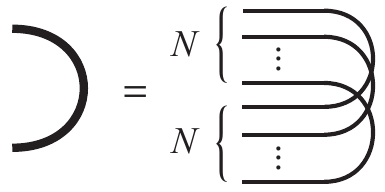}
 \label{fig:composite-bell}}
 
 As with the single system case the column vectorization of a composite linear operator $A \in L(\2X,\2Y)$, where $\2 Y=\bigotimes_{k=1}^{N} \2 Y_k$,  is given by bending all the system wires upwards, or equivalently by the identity 
 \begin{equation}
 \dket{A}\equiv (\id\otimes A)\dket{\id}.
 \end{equation}
Graphically this is given by
% FIGURE - MULTI VEC
\figeq{\includegraphics[width=0.45\textwidth]{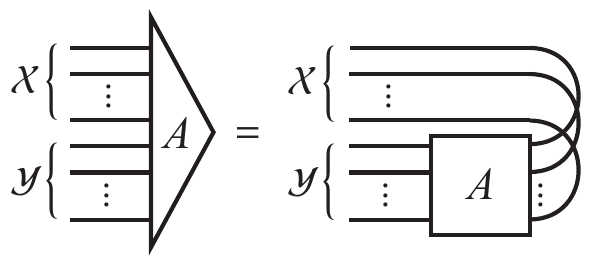}
\label{fig:composite-vec}}
 Note that the order of the subsystems for the bent wires is preserved by the vectorization operation.
 
 % UNRAVELLING
In some situations it may be preferable to consider vectorization of the composite system in terms of vectorization of the individual component systems. Transferring between this component vectorization and the joint-system vectorization can be achieved by an appropriate index permutation of vectorized operators which has a succinct graphical expression when cast in the tensor network framework. 

Suppose the operator $A\in\Lx{X,Y}$, where $\2 X=\bigotimes_{k=1}^N \2 X_k$, $\2 Y=\bigotimes_{k=1}^N \2 Y_k$, is composed of subsystem operators such that 
\begin{equation}
A= A_1\otimes...\otimes A_N
\end{equation} 
where $A_k\in L(\2 X_k,\2 Y_k)$ for $k=1,...,N$. As previously stated the vectorized composite operator $\dket{A}$ is a vector in the Hilbert space $\XY$. 

We define an operation $\2V_{N}$ called the \emph{unravelling} operation, the action of which unravels a vectorized matrix $\dket{A}= \dket{A_1\otimes\hdots\otimes A_N}$ into the tensor product of vectorized matrices on each individual subsystem $\2 X_k\otimes\2 Y_k$
\begin{equation}
\2V_{N} \dket{A_1\otimes\hdots\otimes A_N}
= \dket{A_1}\otimes\hdots\otimes\dket{A_N} \label{eqn:unravelling}. 
\end{equation}
The inverse operation then undoes the unravelling
\begin{equation}
\2V_{N}^{-1} \big(\dket{A_1}\otimes\hdots\otimes\dket{A_N}\big)
	=\dket{A_1\otimes\hdots\otimes A_n}. 
\end{equation}

More generally the unravelling operation $\2V_{N}$ is given by the map
\begin{eqnarray}
\2V_{N}: \ket{x_{\2 X}}\otimes\ket{y_{\2 Y}}
	&\longmapsto& 
	\bigotimes_{k=1}^{N}\left( \ket{x_k}\otimes\ket{y_{k}}\right)
\end{eqnarray}
where 
$\ket{x_{\2 X}} \equiv\ket{x_1}\otimes\hdots\otimes\ket{x_N},
\ket{y_{\2 Y}}\equiv\ket{y_1}\otimes\hdots\otimes\ket{y_N}$.
Hence we can write $\2 V_N$ in matrix form as
\begin{equation}
\2V_N = \sum_{i_1,\hdots,i_{N}}\sum_{j_1,\hdots,j_N} \ket{i_1,j_{1},\hdots,i_N,j_{N}}\bra{i_{\2 X},j_{\2 Y}}.
\label{eqn:unravelling}
\end{equation}
where $\ket{i_{\2 X}} \equiv\ket{i_1}\otimes\hdots\otimes\ket{i_N},
\ket{j_{\2 Y}}\equiv\ket{j_1}\otimes\hdots\otimes\ket{j_N}$, and $\ket{i_k},\ket{j_l}$ are the standard bases for $\2 X_k$ and $\2 Y_l$ respectively.

We can also express $\2V_N$ as the composition of SWAP operations between two systems. For the previously considered composite operator $A\in L(\2X,\2Y)$ we have that $\dket{A}$ has $2N$ subsystems. If we label the SWAP operation between two subsystem Hilbert spaces indexed by $k$ and $l$ by $\mbox{SWAP}_{k:l}$, where $1\le k,l\le2N$, then the unravelling operation can be composed as
\begin{equation}
\2V_{N} = W_{N-1}...W_{1}   
\label{eqn:unravelling}      
\end{equation}
where
\begin{equation}
W_{k} =\prod_{j=0}^{k-1}\,\mbox{SWAP}_{N-k+2j+1:N-k+2j+2}.
\end{equation}
For example
\begin{eqnarray}
W_1 &=& \mbox{SWAP}_{N:N+1}\\
W_2 &=& \mbox{SWAP}_{N-1:N}\mbox{SWAP}_{N+1:N+2} \nonumber\\
W_{N-1} &=& \mbox{SWAP}_{2:3}\mbox{SWAP}_{4:5}\hdots \mbox{SWAP}_{2N-2:2N-1}. \nonumber
\end{eqnarray}
While this equation looks complicated, it has a more intuitive construction when depicted graphically. The tensor networks for the unravelling operation in the $N=2,3$ and $4$ cases are shown below
% FIGURE - UNRAVELLING
 \figeq{
 \begin{tabular}{ccc}
 \includegraphics[width=0.07\textwidth]{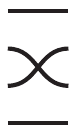}
 \quad\quad&\quad\quad
 \includegraphics[width=0.1\textwidth]{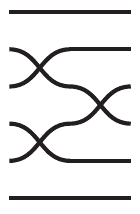}
 \quad\quad&\quad\quad
 \includegraphics[width=0.13\textwidth]{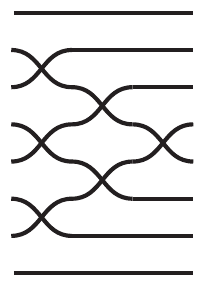}
 \\
\footnotesize{(a) $\2V_2$} 
 \quad\quad&\quad\quad
\footnotesize{(a) $\2V_3$} 
  \quad\quad&\quad\quad
  \footnotesize{(a) $\2V_4$} 
 \end{tabular}
  \label{fig:unravelling}
 }
 We also present a graphical proof of this for the $N=3$ case: 
% FIGURE - UNRAVELLING N=3 PROOF
\figeq{\includegraphics[width=0.5\textwidth]{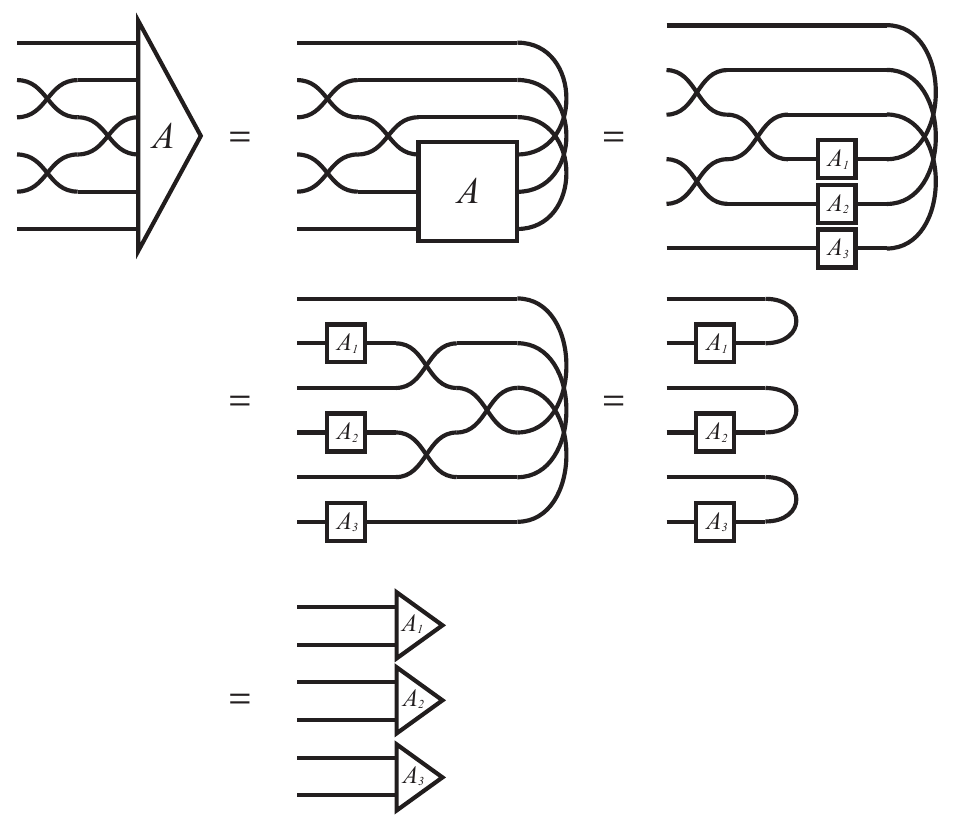}
\label{fig:unravelling-proof}}

%===================================================
% COMPOSING SUPEROPERATORS%===================================================
\subsection{Composing superopators}
\label{sec:comp-sop}

We now discuss how to compose superoperators on individual subsystems to form the correct superoperator on the composite system, and vice-versa.  Given two superoperators $\2S_1$, and $\2S_2$, if we construct a joint system superoperator via tensor product $(\2S_1\otimes \2S_2)$, this composite operator acts on the tensor product of vectorized inputs $\dket{\rho_1}\otimes\dket{\rho_2}$, rather than the the vectorization of the composite input $\dket{\rho_1\otimes\rho_2}$. To construct the correct composite superoperator for input $\dket{\rho_1\otimes\rho_2}$ we may use the unravelling operation $\2V_N$ from Eq.~\eqref{eqn:unravelling} and its inverse.

If we have a set of superoperators $\{\2S_{k} : k=1,...,N\}$ where $ \2S_k\in L(\2 X_k\otimes\2X_k,\2Y_k\otimes \2 Y_{k})$, then the joint superoperator $\2 S\in L(\2X\otimes\2X,\2Y\otimes\2Y)$, where $\2 X=\bigotimes_{k=1}^N \2 X_k$, $\2 Y=\bigotimes_{k=1}^N \2 Y_k$,  is given by
\begin{equation}
\2 S = \2 V_N^{\dagger}\left( \2 S_{1}\otimes\hdots\otimes\2 S_{N}\right)\2 V_N.
\label{eqn:comp-sop}
\end{equation} 
The tensor networks for this transformation in the $N=2$ and $N=3$ cases are shown below
% FIGURE - COMPOSE SOP
\figeq{
\begin{tabular}{c|c}
\includegraphics[width=0.3\textwidth]{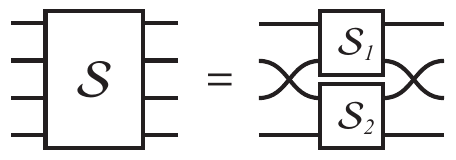}
\quad\quad&\quad\quad
\includegraphics[width=0.38\textwidth]{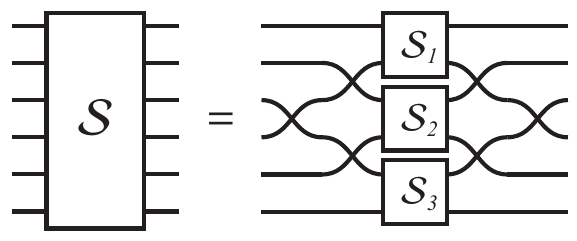}
\\
\footnotesize{$N=2$}
\quad\quad&\quad\quad
\footnotesize{$N=3$}
\end{tabular}
\label{fig:comp-sop}
 }

Composing superoperators from individual subsystem superoperators is useful when performing the same computations for multiple identical systems. For an example we consider vectorization in the Pauli-basis for an $N$-qubit system. While it is generally computationally more efficient to perform vectorization calculations in the col-vec (or row-vec) basis, as these may be implemented using structural operations on arrays, it is often convenient to express the superoperator in the Pauli basis, or the Choi-matrix in the $\chi$-matrix representation, when we are interested in determining the form of correlated errors. However, transforming from the col-vec to the Pauli-basis for multiple (and possibly arbitrary) number of qubits is inconvenient. Using our unravelling operation we can instead compute the single qubit change of basis superoperator $T_{c\rightarrow\sigma}$ from \Eqref{eqn:vec-change-1}, where $\sigma=\{\id,X,Y,Z\}/\sqrt{2}$ is the Pauli-basis for a single qubit, and use this to generate the transformation operator for multiple qubits. In the case of $N$-qubits we can construct the basis transformation matrix as
\be
T^{(N)}_{c\rightarrow \sigma} = \2 V_N^{\dagger}\cdot T_{c\rightarrow\sigma}^{\otimes N} \cdot \2 V_N.
\ee
The joint-system superoperator in the Pauli-basis is then given by 
\be
\2 S^\sigma = T^{(N)}_{c\rightarrow \sigma}\cdot\2 S \cdot T^{(N)\dagger}_{c\rightarrow \sigma}
\ee
The same transformation can be used for converting a state $\rho=\rho_1\otimes\hdots\otimes\rho_N$ to the Pauli basis: $\dket{\rho}_\sigma = T^{(N)}_{c\rightarrow \sigma}\dket{\rho}_c$. These unravelling techniques are also useful for applying operations to a limited number of subsystems in a tensor network as used in many tensor network algorithms.

%===================================================
% OPERATIONS AS SUPEROPERATORS
%===================================================
\subsection{Matrix operations as superoperators}
\label{sec:ops-as-sops}

We now show how several common matrix manipulations can be written as superoperators. These expressions are obtained by simply vectorizing the transformed operators. We begin with the trace superoperator $S_{\mbox{\scriptsize{Tr}}}$ which implements the trace of a matrix $S_{\mbox{\scriptsize{Tr}}}\dket{A} := \Tr[A]$ for a square matrix $A\in L(\2X)$. This operation is simply given by the adjoint of the unnormalized Bell-state:
\be
S_{\mbox{\scriptsize{Tr}}} := \dbra{\id}
\ee
where $\id\in L(\2X)$ is the identity operator. If $\2X$ is itself a composite system, we simply use the definition of the Bell-state for composite systems from \Eqref{eqn:comp-bell}. This is illustrated in our graphical calculus as
% FIGURE - TRACE
\figeq{
S_{\mbox{\scriptsize{Tr}}}\dket{A} = \parbox[c]{1em}{\includegraphics[width=0.1\textwidth]{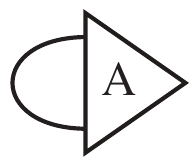}}
\label{fig:vec-tr}
}

For a rectangular matrix $B\in L(\2X,\2Y)$, the transpose superoperator $S_{T}$ which implements the transpose $S_T\dket{B}=\dket{B^T}$ is simply a swap superoperator between $\2X$ and $\2 Y$.
\begin{eqnarray}
S_{T} &=& \mbox{SWAP}	\\
\mbox{SWAP}&:& \XY \mapsto \YX  
\end{eqnarray}
The tensor network for the swap superoperator is 
% FIGURE - TRANS
\figeq{
S_{\mbox{\scriptsize{T}}}\dket{B} = \parbox[c]{0.1\textwidth}{\includegraphics[width=0.1\textwidth]{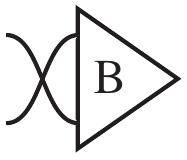}}
\label{fig:vec-trans}
}
If $\2X$ and $\2Y$ are composite vector spaces we may split the crossed wires into their respective subsystem wires. 

Next we give the superoperator representations of the bipartite matrix operations in \eqref{fig:bipartite} acting on vectorized square bipartite matrices $M\in L(\XY)$. These are the partial trace over $\2X$ ($S_{\mbox{\scriptsize{Tr}}_{\2 X}}$) (and $S_{\mbox{\scriptsize{Tr}}_{\2Y}}$ over $\2 Y$), transposition $S_T$, 
and col-reshuffling $(S_{R_c})$. 
\begin{eqnarray}
S_{\mbox{\scriptsize{Tr}}_{\2 X}}
	&:& \XY\otimes\XY \mapsto \2Y\otimes\2Y\\
S_{\mbox{\scriptsize{Tr}}_{\2 Y}}
	&:& \XY\otimes\XY \mapsto \2X\otimes\2X\\
S_{\mbox{\scriptsize{T}}}
	&:&	 \XY\otimes\XY \mapsto \XY\otimes\XY\\
S_{\scriptsize{R_c}}
	&:&	 \XY\otimes\XY \mapsto \XX\otimes\YY
\end{eqnarray}
The graphical representation of the superoperators for these operations are:
% FIGURES - BIPARTITE
\figeq{
\begin{tabular}{c|c|c|c}
$S_{\mbox{\scriptsize{Tr}}_{\2 X}}\dket{M}
=$ \parbox[c]{0.11\textwidth}{\includegraphics[width=0.09\textwidth]{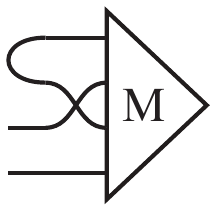}}
&\quad
$S_{\mbox{\scriptsize{Tr}}_{\2 Y}}\dket{M}
=$ \parbox[c]{0.11\textwidth}{\includegraphics[width=0.09\textwidth]{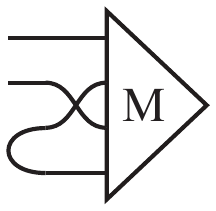}}
&\quad
$S_{\mbox{\scriptsize{T}}}\dket{M}
=$ \parbox[c]{0.11\textwidth}{\includegraphics[width=0.09\textwidth]{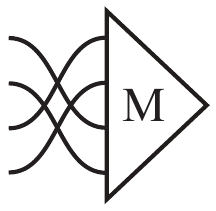}}
&\quad
$S_{\scriptsize{R_c}}\dket{M}
=$ \parbox[c]{0.11\textwidth}{\includegraphics[width=0.09\textwidth]{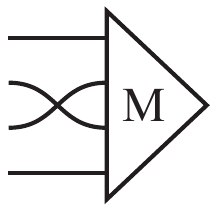}}
\end{tabular}
\label{fig:bipartite-sop}
}

Algebraically they are given by
\begin{eqnarray}
S_{\mbox{\scriptsize{Tr}}_{\2 X}}
	&=&	\big[\dbra{\id_{\2 X}}\otimes\id_{\2 Y}\otimes \id_{\2 Y}\big]\2 V_2\\
S_{\mbox{\scriptsize{Tr}}_{\2 Y}}
	&=&	\big[\id_{\2 X}\otimes \id_{\2 X}\otimes\dbra{\id_{\2 Y}}\big]\2 V_2\\
S_{\mbox{\scriptsize{T}}}
	&=&	\mbox{SWAP}_{1:3}\mbox{SWAP}_{2:4}\\
S_{\scriptsize{R_c}}
	&=&	 \2V_2
\end{eqnarray}
where $\2V_2$ is the unravelling operation in \Eqref{eqn:unravelling}. 

In the general multipartite case for a composite matrix $A\in L(\2X)$ where $\2 X=\bigotimes_{k=1}^N \2 X_k$, we can trace out or transpose a subsystem $j$ by using the unravelling operation in \Eqref{eqn:unravelling} to insert the appropriate superoperator for that subsystem with identity superoperators on the remaining subsystems:
\begin{eqnarray*}
\2S_{O_j} 
&=& 
	\2 V_{N-1}^{-1}\left[
		\left(\bigotimes_{k=1}^{j-1}\2S_{\2 I_{k}}\right)
		\otimes\2S_{O}\otimes
		\left(\bigotimes_{k=j+1}^{N}\2S_{\2 I_{k}}\right)
	\right]\2 V_N	\\
\end{eqnarray*}
where $\2S_O\in T(\2 X_j)$ is the superoperator acting on system $j$ and $\2S_{\2 I_k}\in T(\2 X_k)$ is the identity superoperator for subsystem $ L(\2 X_k)$. Similarly by inserting the appropriate operators at multiple subsystem locations we can perform the partial trace or partial transpose of any number of subsystems.

%===================================================
% REDUCED SUPEROPERATORS
%===================================================
\subsection{Reduced superoperators}
\label{sec:reduced-sop}

We now present a simple but useful example of the presented bipartite operations in the superoperator representation to show how to construct an effective reduced superoperator for a a subsystem out of a larger superoperator on a composite system.

Consider states $\rho_{XY}\in L(\XY)$ which undergo some channel $\2 F\in C(\XY)$ with superoperator representation $\2S$. Suppose system $\2Y$ is an ancilla which we initialize in some state $\tau_{0}\in L(\2Y)$, and we post-select on the output state of system $\2Y$ being in a state $\tau_{1}$. We may construct the effective reduced map $\2 F^\prime\in T(\2X)$ for this process for arbitrary input and output states of system $\2X$, given by a superoperator $\2S^\prime$, as shown:
% FIGURE - REDUCED SOP
\figeq{\includegraphics[width=0.35\textwidth]{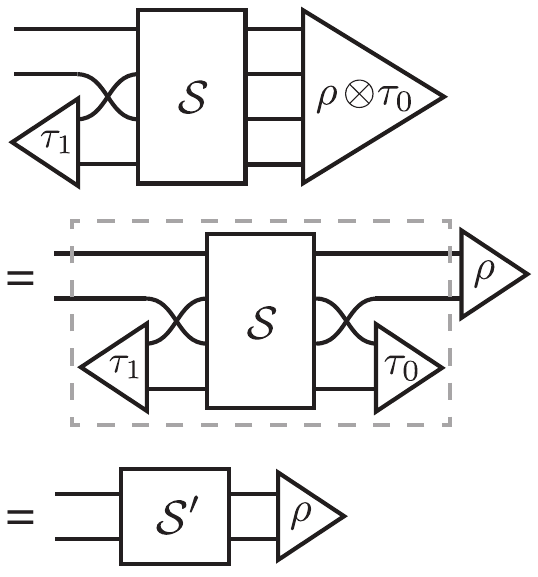}
\label{fig:reduced-sop}
}

Formally, we are defining the superoperator representation $\2S^\prime$ of the effective channel $\2 F^\prime$ as the map
\begin{eqnarray}
\2 F^\prime(\rho) &=& \Tr_{\2 Y}\left[(\id_{\2X}\otimes\tau_1)\2 F(\rho\otimes\tau_0)\right]	\\
\2 S^\prime &=& \dbra{\tau_1}\2V_2 \2S\2V_2^\dagger\dket{\tau_0}
\end{eqnarray} 
where $\2S$ is the superoperator representation of $\2F$ and $\dket{\tau_j}$ is implicitly assumed to have the identity operation on the vectorization of subsystem $\2X\otimes\2X$ $(\dket{\tau_j}:= \id_{\2X}\otimes\id_{\2X}\otimes \dket{\tau_j})$.

%==============================================================
%		AAPT
%==============================================================
\subsection{Ancilla Assisted Process Tomography}
\label{sec:aapt}

Quantum state tomography is the method of reconstructing an unknown quantum state from the measurement statistics obtained by performing a topographically complete set of measurements on many identical copies of the unknown state~\cite{Nielsen2000}. Quantum process tomography is an extension of quantum state tomography which reconstructs an unknown quantum channel $\2E\in C(\2X)$ from appropriately generated measurement statistics. One such procedure, known as \emph{standard quantum process tomography}, involves preparing many copies of each of a topographically complete set of input states, subjecting each to the unknown quantum channel, and performing state tomography on the output~\cite{DAriano2001}. 

An alternative approach is to directly measure the Choi-matrix for the channel via a method known as \emph{ancilla assisted process tomography} (AAPT)~\cite{White2003}. The simplest case of AAPT is \emph{entanglement assisted process tomography}(EAPT) which is an experimental realization of the Choi-Jamio{\l}kowski isomorphism. Here an experimenter prepares a a maximally entangled state 
\be
\rho_{\Phi}= \frac{1}{d}\dketdbra{\id}{\id}
\ee 
across the system of interest $\2X$ and an ancilla $\2Z \cong \2X$, and subjects the system to the unknown channel $\2E$, and the ancilla to an identity channel $\2I$. The output of this joint system-ancilla channel is the rescaled Choi-matrix:
\bea
\rho_{\phi}^\prime = (\2I\otimes\2E)\left(\rho_{\Phi}\right) 
	&=& \frac{\Lambda}{d}.
\eea
which can be measured directly by quantum state tomography. The tensor network for EAPT is
% FIGURE - EAPT
\figeq{
\includegraphics[width=0.4\textwidth]{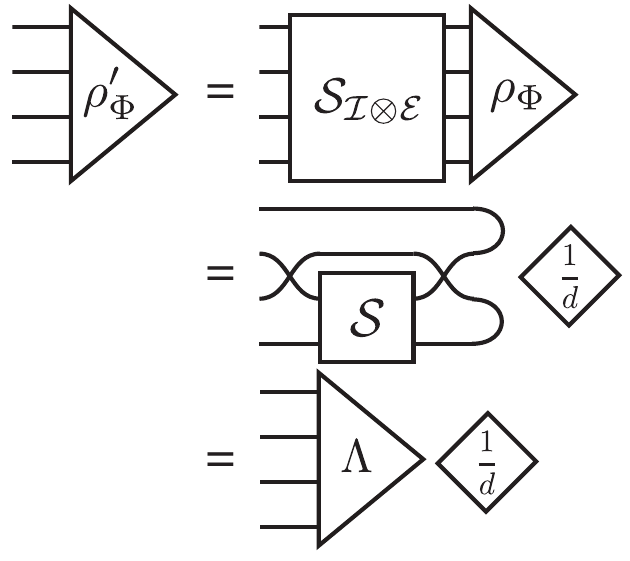}
\label{fig:aapt-choi}
}

In general AAPT does not require $\rho_{AS}$ to be maximally entangled. It has been demonstrated experimentally that AAPT may be done with a state which does not have any entanglement at all, at the expense of an increase in the estimation error of the reconstructed channel~\cite{White2003}. A necessary and sufficient condition for a general state $\rho_{AS}$ to allow recovery of the Choi-matrix of an unknown channel $\2E$ via AAPT is that it have a \emph{Schmidt number} equal to $d^2$ where $d$ is the dimension of the state space $\2X$~\cite{White2003}. This conditions has previously been called \emph{faithfulness} of the input state, and one can recover the original Choi-matrix for the unknown channel $\2E$ by applying an appropriate inverse map to the output state in post-processing~\cite{DAriano2003}. We provide an arguably simpler derivation of this condition, and the explicit construction of the inverse recovery operator. The essence of this proof is that we can consider the bipartite state $\rho_{AS}$ to be Choi-matrix for an effective channel via the Choi-Jamio{\l}kowski isomorphism (but with trace normalization of 1 instead of $d$) . We can then apply channel transformations to this initial state to convert it into an effective channel acting on the true Choi-matrix, and if this effective channel is invertible we can recover the Choi-matrix for the channel $\2E$ by applying the appropriate inverse channel.

% PROPOSITION
\begin{proposition}
\label{prop:aapt}
$(a)$ A state $\rho_{AS}\in L(\2X\otimes\2X)$ may be used for AAPT of an unknown channel $\2E\in  C(\2X)$ if and only if the reshuffled density matrix $\2S_{AS} = \rho_{AS}^{R_c}$ is invertible. 

$(b)$ The channel can be reconstructed from the measured output state by $\Lambda_{\2E} = (\2R\otimes \2I)(\rho_{AS}^\prime)$
where $\rho_{AS}^\prime = (\2 I\otimes \2E)(\rho_{AS})$ is the output state reconstructed by quantum state tomography, and $\2R$ is the recovery channel given by superoperator $\2S_{\2R} = (\2S_{AS}^T)^{-1}$.
\end{proposition}

\begin{figure}[h]
\centering
\includegraphics[width=0.4\textwidth]{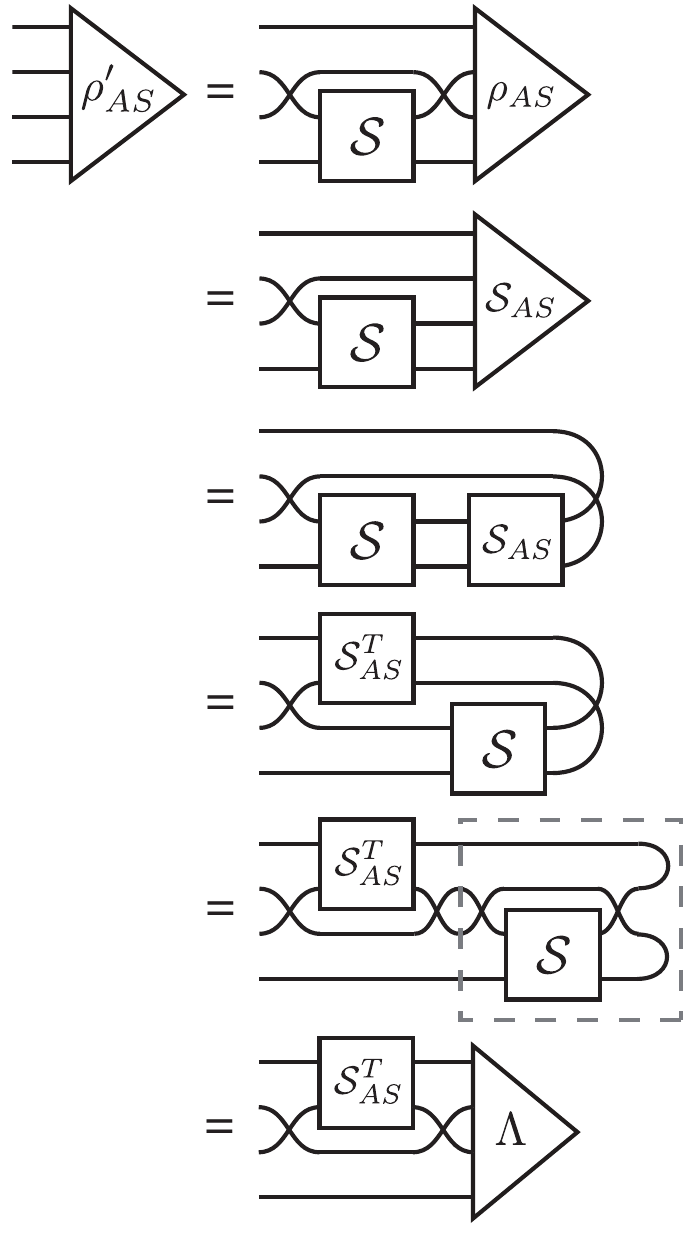}
\caption{Graphical proof of the equivalence of an initial state $\rho_{AS}$ used for performing AAPT of an unknown CPTP map $\2E$ with superoperator representation $\2S$, to a channel $(\2R\otimes\2I)$ acting on the Choi-matrix $\Lambda$ for a channel $\2E$. The Choi-matrix can be recovered if and only if the the superoperator $\2S_{\2R}=\2S_{AS}^T=(\rho_{AS}^{R_c})^T$ is invertible.}
\label{fig:aapt-proof}
\end{figure}

The graphical proof of Prop.~\ref{prop:aapt}  is illustrated in Fig.~\ref{fig:aapt-proof}. This proof demonstrates several useful features of the presented graphical calculus. In particular it applies the vectorized reshuffling transformation to a bipartite density matrix input state to obtain an effective superoperator representation of a state, and uses the unravelling operation for composition of superoperators.
From this construction we find that if the initial state $\rho_{AS}$ is maximally entangled, then it can be expressed as $\rho_{AS}= \dketdbra{V}{V}$ for some unitary $V$. In this case the reshuffled superoperator of the state corresponds to a unitary channel $\2S_{AS} = \overline{V}\otimes V$, and hence is invertible with $\2S_{AS}^{-1} = \2S_{AS}^\dagger$. If the input state is not maximally entangled, then the closer it is to a singular matrix, the larger the condition number and hence the larger the amplification in error when inverting the matrix.

% FIGURE - AAPT GRAPHICAL PROOF
%

%==============================================================
%		AVE GATE FIDELITY
%==============================================================
\subsection{Average Gate Fidelity}
\label{sec:gatefid}

When characterizing the performance of a noisy quantum channels a widely used measure of the closeness of a CPTP map $\2E\in C(\2X)$ to a desired quantum channel $\2F\in C(\2X)$ is the \emph{Gate Fidelity}. This is defined to be
\be
F_{\2E,\2F}(\rho) = F(\2F(\rho)\2E(\rho))
\ee
where
\be
F(\rho,\sigma)=\left(\Tr\left[ \sqrt{\sqrt{\rho}\sigma\sqrt{\rho}}\right]\right)^2
\ee
is the fidelity function for quantum states~\cite{Nielsen2000}.

In general we are interested in comparing a channel $\2E$ to a unitary map $\2U\in C(\2X)$ where $\2U(\rho)=U\rho U^\dagger$. In this case we have
\bea
F_{\2E,\2U}(\rho) 
	&=& \left[\Tr\sqrt{\sqrt{U\rho U^\dagger}\2E(\rho)\sqrt{U\rho U^\dagger}}\right]^2\\
	&=&  \left[\Tr\sqrt{\sqrt{\rho}U^\dagger\2E(\rho)U\sqrt{\rho}}\right]^2\\
	&=&  \left[\Tr\sqrt{\sqrt{\rho}\,\2U^\dagger(\2E(\rho))\sqrt{\rho}}\right]^2\\
	&=& F_{\2U^\dagger\2E,\2 I}(\rho)
\eea
where $\2I$ is the identity channel and $\2U^\dagger(\rho) = U^\dagger \rho U$, is the adjoint channel of the unitary channel $\2U$. Thus without loss of generality we may consider the gate fidelity $F_{\2E}(\rho)\equiv F_{\2U^\dagger\2F,\2I}(\rho)$ comparing $\2E$ to the identity channel, where we simply define $\2E \equiv \2U^\dagger\2F$ if we wish to compare $\2F$ to a target unitary channel $\2U$.

The most often used quantity derived from the gate fidelity is the \emph{average gate fidelity} taken by averaging $F_{\2E}(\rho)$ over the the Fubini-Study measure. Explicitly the average gate fidelity is defined by
\be
\overline{F}_{\2E} = \int d\,\psi\, \bra{\psi} \2E(\ketbra\psi\psi)\ket{\psi}.
\label{eq:avegf}
\ee
where due to the concavity of quantum states we need only integrate over pure states $F_{\2E}(\ketbra\psi\psi) = \bra\psi \2E(\ketbra\psi\psi)\ket\psi$. 

Average gate fidelity is a widely used figure of merit in part because it is simple to compute. The expression in Eq.~\eqref{eq:avegf} reduces to explicit expression for $\overline{F}_{\2E}$ in terms of a single parameter of the channel $\2E$ itself. This has previously been given in terms of the Kraus representation~\cite{Horodecki1999,Nielsen2002}, superoperator~\cite{Emerson2005} and Choi-matrix in~\cite{Johnston2011}. We now present an equivalent graphical derivation of the average gate fidelity in terms of the Choi-matrix which we believe is simpler than previous derivations. We start with the tensor network diagram corresponding to Eq.~\eqref{eq:avegf} and perform graphical manipulations as follows
% AVE GATE FIDELITY GRAPHICAL PROOF PT 1
\figeq{
\includegraphics[width=0.35\textwidth]{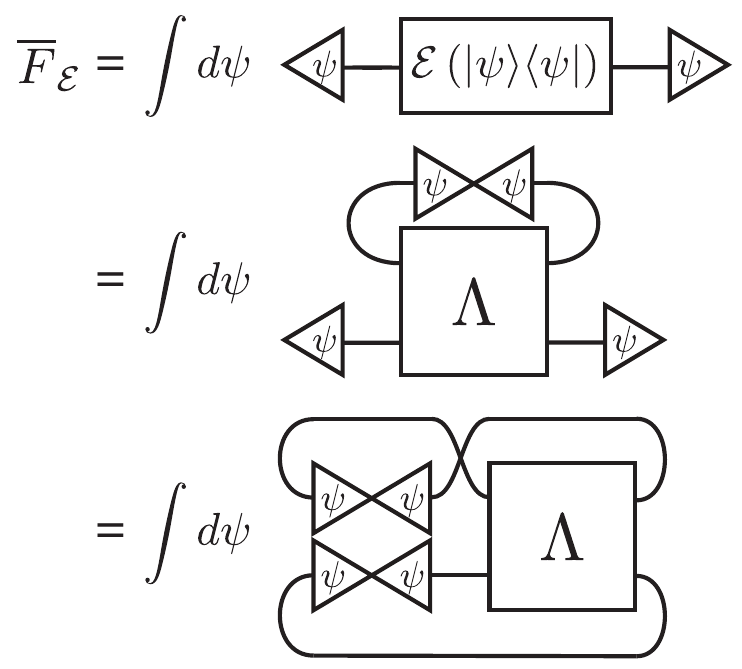}
\label{fig:ave-fid-proof-1}
}
For the next step of the proof we use the result that the average over $\psi$ of a tensor product of states $\ketbra{\psi}{\psi}^n$ is given by
\be
\int d\psi\, \ketbra\psi\psi^{\otimes n} = \frac{\Pi_{\scriptsize\mbox{sym}}(n,d)}{\Tr[\Pi_{\scriptsize\mbox{sym}}(n,d)]}
\label{eq:symsub}
\ee
where $ \Pi_{\scriptsize\mbox{sym}}(n,d)$ is the projector onto the symmetric subspace of $\2X^{\otimes n}$. This project may be written as~\cite{Magesan2011}
\be
\Pi_{\scriptsize\mbox{sym}}(n,d) = \frac{1}{n!}\sum_\sigma P_{\sigma}
\label{eq:perm-sum}
\ee
where $P_{\sigma}$ are operators for the permutation $\sigma$ of $n$-indices. These permutations may be represented as a swap type operator with $n$ tensor wires. For the case of $n=2$ we have the tensor diagram:
% FIG: SYM SUB
\figeq{
\includegraphics[width=0.45\textwidth]{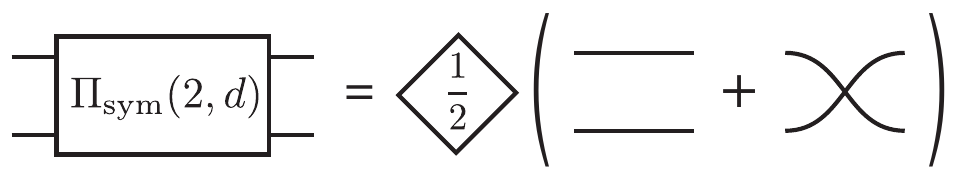}
\label{fig:symsub2}}
Here we can see that $\Tr[\Pi_{\scriptsize\mbox{sum}}(2,d)] = (d^2 + d)/2$, and hence we have that
\bea
\Pi_{\scriptsize\mbox{sym}}(2,d) &=& \frac{1}{2}\left(\id\otimes\id + \mbox{SWAP}\right)	\\
\Tr[\Pi_{\scriptsize\mbox{sym}}(2,d)] &=& \frac{d^2+d}{2}	\\
\Rightarrow
\int d\psi\, \ketbra\psi\psi^2 &=& \frac{\id\otimes\id + \mbox{SWAP}}{d(d+1)}
\eea
where $\2X\cong \C^d$, $\id\in L(\2X)$ is the identity operator, and SWAP is the SWAP operation on $\XX$. Subsituting \eqref{fig:symsub2} into \eqref{fig:ave-fid-proof-1} completes the proof:
% AVE GATE FIDELITY GRAPHICAL PROOF PT 2
\figeq{
\includegraphics[width=0.48\textwidth]{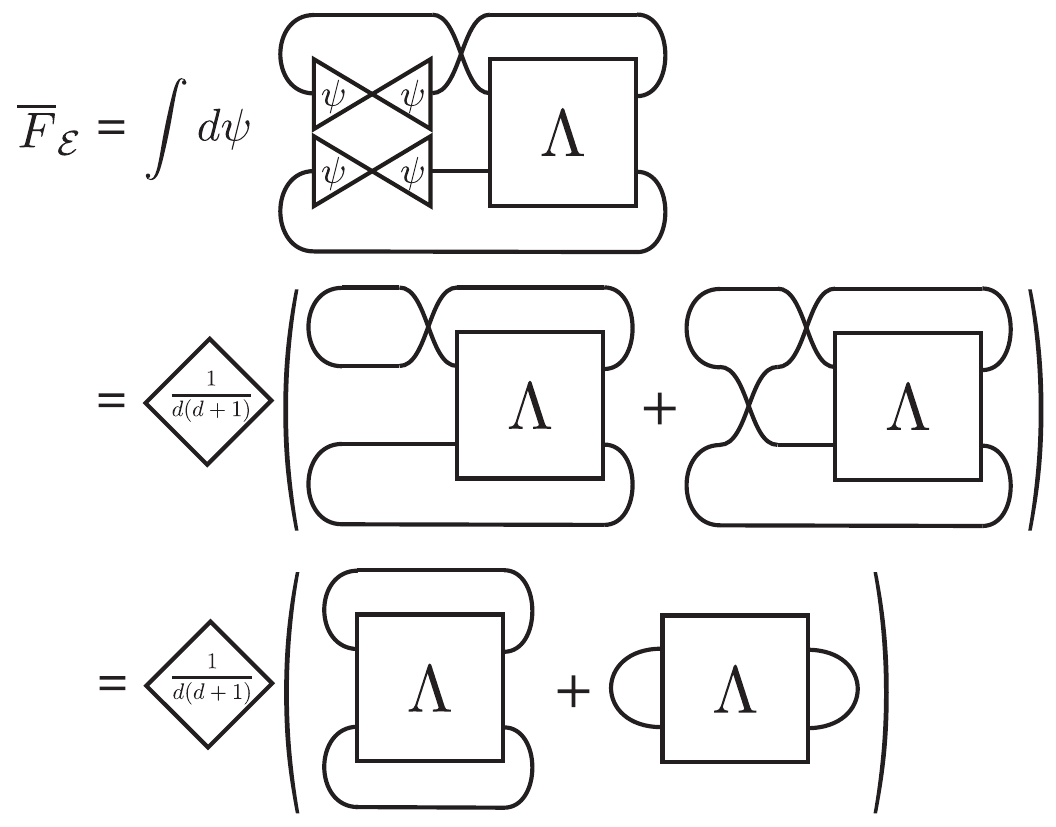}
\label{fig:ave-fid-proof-2}
}
Hence we have that the average gate fidelity in terms of the Choi-matrix is given by
\be
\overline{F}_{\2E} = \frac{d + \dbra{\id}\Lambda\dket{\id}}{d(d+1)}	
\ee
where we have used the fact that the Choi-matrix is normalized such that$\Tr[\Lambda] = d$.
From this proof one may derive expressions for the other representations using the channel transformations in Section~\ref{sec:trans} as illustrated in Appendix~\ref{app:fid-proofs}. The resulting expressions are
\bea
\overline{F}_{\2E} 
	&=& \frac{d + \Tr[\2S]}{d(d+1)}	\\
	&=& \frac{d + \dbra{\id}\Lambda\dket{\id}}{d(d+1)}	\\
	&=& \frac{d + \sum_j |\Tr[K_j]|^2}{d(d+1)}		\\
	&=& \frac{d + d\chi_{00}}{d(d+1)}		\\
	&=& \frac{d + \Tr_{\2X}[A^\dagger]\cdot\Tr_{\2X}[A]}{d(d+1)}	
\eea
where $\2S$, $\Lambda$, $\{K_j\}$, $\chi$, $A$ are the superoperator, Choi-matrix, Kraus, $\chi$-matrix and Strinespring representations for $\2E$ respectively. In the case of the $\chi$-matrix representation, $\chi$ is defined with respect to a basis $\{\sigma_j\}$ satisfying $\Tr[\sigma_j] = \sqrt{d}\delta_{j,0}$.

Similar techniques can be applied for tensor networks that may be graphically manipulated into containing a term $\int d\psi\, \ketbra\psi\psi^{\otimes n} $ for $n>2$. This could prove useful for computing higher order moments of fidelity functions and other quantities defined in terms of averages over quantum states $\ket\psi$. In this case there are $n!$ permutations of the tensor wires for the permutation operator $P_\sigma$ in Eq.~\eqref{eq:perm-sum}, and these can be decomposed as a series of SWAP gates. For example, in the case of $n=3$ we have
\bea
\Pi_{\scriptsize\mbox{sym}}(3,d) 
&=& \frac{1}{6}\big(
\id^{\otimes 3} + \mbox{SWAP}_{1:2} 
+ \mbox{SWAP}_{1:3}
+ \mbox{SWAP}_{2:3}
+ \mbox{SWAP}_{1:2}\mbox{SWAP}_{2:3}
+ \mbox{SWAP}_{2:3}\mbox{SWAP}_{1:2}\big) \\
\Tr[\Pi_{\scriptsize\mbox{sym}}(3,d)]
	&=& \frac{d^3+3d^2+2d}{6}.
\eea

%==============================================================
%		ENTANGLEMENT FIDELITY
%==============================================================
\subsection{Entanglement Fidelity}
\label{sec:entfid}

Another useful fidelity quantity is the \emph{entanglement fidelity} which quantifies how well a channel preserves entanglement with an ancilla~\cite{Schumarcher1996,Nielsen2000}. For a CPTP map $\2E\in C(\2X)$ and density matrix $\rho\in L(\2X)$ the entanglement fidelity is given by 
\bea
F_{\scriptsize\mbox{e}}(\2E,\rho) 
&=& \inf
	\big\{
	F\left(\ketbra{\psi}{\psi}, (\2I_{\2Z}\otimes\2E)(\ketbra{\psi}{\psi})\right):
\nonumber\\&&
	\Tr_{\2Z}[\ketbra{\psi}{\psi}]=\rho \big\}
\label{eq:ent-fid}
\eea
where $\ket\psi\in \2X\otimes\2Z$ is a purification of $\rho$ over an ancilla $\2Z$. Entanglement fidelity turns out to be independent of the choice of purification $\ket\psi$, and a closed form expression has been given in terms of the Kraus representation~\cite{Nielsen2000} and Choi-matrix~\cite{Fletcher2007}. Here we present a simple equivalent derivation in terms of the Choi-matrix representation of the channel $\2E$ using graphical techniques. Then by applying the channel transformations of Section~\ref{sec:trans} we obtain expressions in terms of the other representations. The resulting expressions for entanglement fidelity are:
\bea
F_{\scriptsize\mbox{e}}(\2E,\rho) 
	&=& \dbra{\rho}\Lambda\dket{\rho}	\\
	&=& \Tr\left[(\rho^T\otimes\rho)\2S\right]	\\
	&=& \sum_j |\Tr[\rho K_j]|^2 \\
	&=& \sum_{i,j}\chi_{ij} \Tr[\rho\, \sigma_i]\Tr[\rho \sigma^\dagger_j]\\
	&=&\Tr_{\2X}[\rho A^\dagger]\cdot\Tr_{\2X}[A\rho]
\eea
where $\2S$, $\Lambda$, $\{K_j\}$, $\chi$, $A$ are the superoperator, Choi-matrix, Kraus, $\chi$-matrix and Strinespring representations for $\2E$ respectively. In the case of the $\chi$-matrix representation, $\chi$ is defined with respect to a basis $\{\sigma_j\}$ satisfying $\Tr[\sigma_j] = \sqrt{d}\delta_{j,0}$. 

For the graphical proof in terms of the Choi-representation we start with Eq.~\eqref{eq:ent-fid} and perform the following tensor manipulations
% ENTANGLEMENT FIDELITY GRAPHICAL PROOF - 1 
\figeq{
\includegraphics[width=0.45\textwidth]{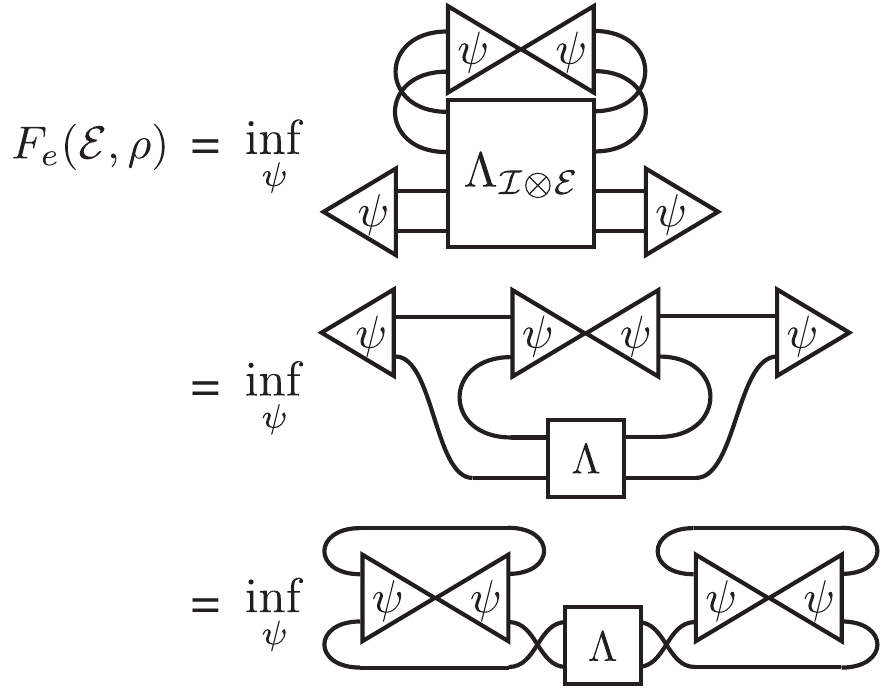}
\label{fig:ent-fid-proof-1}
}
Now since the infimum is over all $\ket{\psi}\in \2Z\otimes\2X$ satisfying $\Tr_{\2Z}[\ketbra{\psi}{\psi}]=\rho$ the result is independent of the specific purification $\psi$ and we have:
% ENTANGLEMENT FIDELITY GRAPHICAL PROOF - 2 
\figeq{
\includegraphics[width=0.45\textwidth]{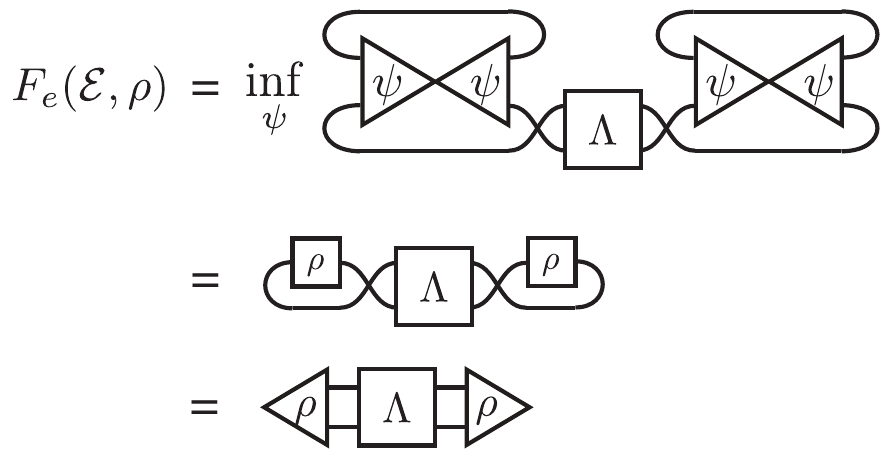}
\label{fig:ent-fid-proof-2}
}
The transformations to the  other representations are illustrated in Appendix~\ref{app:fid-proofs}.

Entanglement fidelity is equivalent to gate fidelity for pure states and hence average entanglement fidelity is equivalent to average gate fidelity. This can be shown graphically as follows
% ENTANGLEMENT FIDELITY REPS PROOF
\figeq{\includegraphics[width=0.45\textwidth]{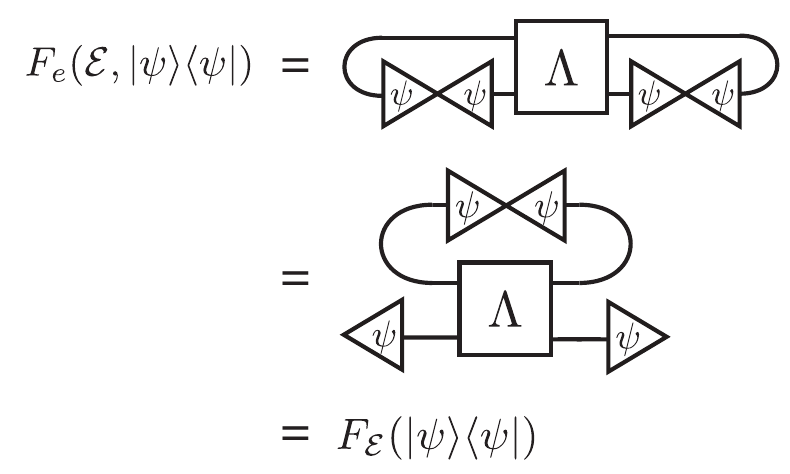}
\label{fig:ent-gate-proof}}

Alternatively we can also define the average gate fidelity in terms of the entanglement fidelity with the identity operator
\be
\overline{F}_{\2E} = \frac{d + F_e(\2E, \id)}{d(d+1)}.
\ee.

%% file: Sections/tnoqs-6-conc.tex
%========================================================================
%========================================================================
%		CONCLUSION
%========================================================================
%========================================================================
\section{Conclusion}\label{sec:sup}

The study of completely-positive trace-preserving maps is an old topic, so it is perhaps surprising that there are still new insights to be gained by investigating their structure using new techniques. Further, while the application of CPTP-maps to describing the evolution of open quantum systems is well understood, it is a surprisingly difficult task to find a concise summary of the properties of, and transformations between, the various mathematically equivalent representations used in the quantum information processing literature. The graphical calculus for open quantum systems presented in this paper has enabled us to unify, and hence transform freely between, the various common representations of CPTP-maps by performing diagrammatic manipulations of their respective tensor networks. A summary of these transformations between the different representations was given in Fig.~\ref{fig:cpreps}. 

We found that many of these transformations between representations of CPTP-maps corresponded to wire bending dualities in our graphical calculus, which have a particularly succinct tensor network description. These transformations are depicted by solid arrows between two boxes labelling representations in Fig.~\ref{fig:cpreps}. Of these duality transformations, only the reshuffling operation connecting the Choi-matrix and Liouville-superoperator is bi-directional --- the reshuffling operation is bijective and self-inverse, and hence the same transformation takes the Choi-Matrix to the superoperator as takes the superoperator to the Choi-matrix. The two other wire bending dualities are vectorization, which transforms both the Kraus and system-environment representations to the superoperator representation, and the Choi-Jamio{\l}kowski isomorphism, which transforms the same two representations to the Choi-matrix. These duality transformations are only single directional as they are not injective, and hence depicted by a one-way arrow connecting the appropriate boxes labelling these representations in Fig.~\ref{fig:cpreps}. The reason these transformations are single directional, as apposed to the bi-directional transformation between the Choi-matrix and superoperator, is due to the non-uniqueness of the Kraus and system-environment representations of a CPTP-map. The transformation is a many-to-one (surjective) mapping and not strictly invertible without first specifying some form of decomposition of the superoperator or Choi-matrix.

The transformations we presented for converting from the Choi-matrix to the Kraus representation, and between the Kraus and system-environment representations, were not based solely on wire bending dualities. These transformations are depicted by dashed arrows in Fig.~\ref{fig:cpreps}, where the dash is meant to indicate that they are non-linear transformations. This non-linearity arose from the decompositions and constructions involved, for example the spectral decomposition of a positive-semi definite operator in the Choi-matrix to Kraus representation transformation. In our case, these non-linear transformations were all also one directional due to the non-uniqueness of the representation being transformed to. There is unitary freedom in constructing them --- for the Choi-matrix to Kraus representation transformation, one could change the basis of the eigenvectors with respect to a vectorization convention and still arrive at a valid Kraus representation; for Kraus to the system environment representation one may choose any orthonormal basis in the construction of the joint system-environment unitary in Eqn~\eqref{eqn:kraus-to-U}; and for the system environment to Kraus representation one may decompose the partial trace over the environment in any orthonormal basis. 

To further demonstrate the utility of the presented graphical calculus we gave several demonstrations of more advanced constructions, and of proof techniques for various quantities used in quantum information processing. In particular we showed how one may deal with vectorization of composite quantum systems and freely transform between a description of the vectorized composite system, and the composite system of individually vectorized systems. These tools were useful for constructing composite system superoperators and effective reduced system superoperators, and for applications where we wish to update or modify a subset of a composite system. 

By vectorizing bipartite matrices and their bipartite matrix transformations we found that we could consider a bipartite density matrix over a system and ancilla as a type of Choi-matrix via the Choi-Jami{\l}kowski isomorphism. By applying the reshuffling transformation we can convert this into an effective superoperator channel description which allowed us to derive a succinct expression for a necessary sufficient condition for a bipartite state to be usable for ancilla assisted process tomography. While this is equivalent to the perviously known result we believe it is a simpler proof, and the resulting recovery channel is simpler to construct using the presented method. 

In the average gate fidelity example we were able to give a shorter proof of the average gate fidelity of a quantum channel by using the graphical representation of the average over states $\int d \psi \ketbra\psi\psi^{\otimes n}$ in terms of the projector onto the symmetric subspace of $\2X^{\otimes n}$. This projector can be expressed as the sum of  $n!$ permutation operators which have a natural representation as tensor network diagrams consisting of a series of SWAP operations corresponding to all left-to-right permutations of $n$ wires. The power of tensor network framework was to manipulate the string diagram for a given expression to form the tensor product $\ketbra\psi\psi^{\otimes n}$ irregardless of where the $n$ copies of $\ketbra\psi\psi$ appear in the original expression. After substituting in the projector onto the symmetric subspace we can contract the $n!$ resulting diagrams to arrive at the final value. Similar techniques could prove useful for calculating other quantities such as higher order moments of fidelity functions and other quantities defined in terms of averages over quantum states.  

Having new tools to investigate old problems can often lead to surprising new results, and we believe there are many more potential applications in QIP for the graphical calculus we have presented in this paper.

%===================================================
%		ACKNOWLEDGEMENTS
%===================================================
\acknowledgements

We thank Alexei Gilchrist and Daniel R. Terno for useful discussions.
This work was supported by the Canadian Excellence Research Chairs (CERC) program, the Canadian Institute for Advanced Research (CIFAR), the Natural Sciences and Engineering Research Council of Canada (NSERC), and the Collaborative Research and Training Experience (CREATE) program. JDB completed part of this work while visiting Michele Mosca at IQC, and acknowledges financial support from the Foundational Questions Institute (FQXi, under grant FQXi-RFP3-1322).

%% file: Sections/tnoqs-7-appendix.tex
%==============================================================
%==============================================================
%  APPENDIX
%==============================================================
%==============================================================
\newpage
%===================================================
% APP: TENSOR NETWORKS
%===================================================
\section{Tensor network proofs}
\label{app:tensor}

We will now prove the consistency of several of the basic tensor networks introduced in Section~\ref{sec:tensor}, and in doing so illustrate how one may use our graphical calculus for diagrammatic reasoning. 

The color summation convention we have presented represents diagrammatic summation over a tensor index by coloring the appropriate tensors in the diagram. In this convention summation over a Kronecker delta, $\sum_{i,j}\delta_{ij}=\sum_{i,j} \braket{i}{j}$, is as shown:
% FIGURE - DELTA
\figeq{\includegraphics[width=0.3\textwidth]{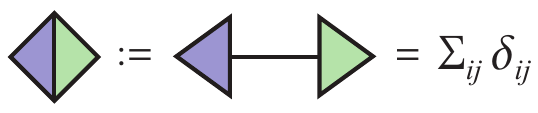}
 \label{fig:delta-sum}}
This expression is used in several of the following proofs.

We begin with the proof of the trace of an operator $A$:
% FIGURE -  TRACE PROOF
\figeq{
\includegraphics[width=0.3\textwidth]{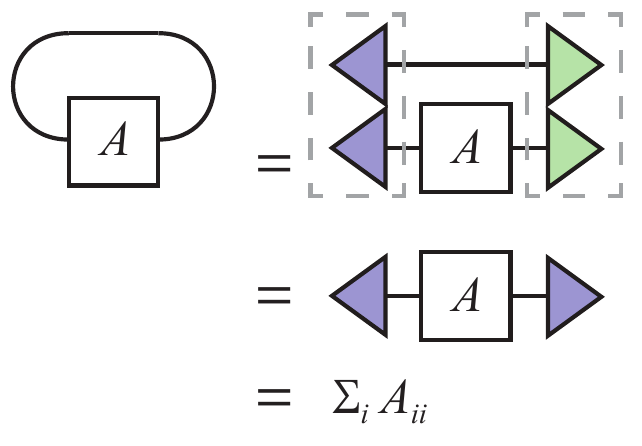}
\label{fig:trace-proof}
}
 
For illustrative purposes, to prove this algebraically we note that the tensor networks for trace correspond to the algebraic expressions $\bra{\Phi^+}A\otimes\id\ket{\Phi^+}$ and $\bra{\Phi^+}\id\otimes A\ket{\Phi^+}$, and that
\begin{eqnarray}
\bra{\Phi^+}\id\otimes A\ket{\Phi^+}
&=& \sum_{i,j}\braket{i}{j}\bra{i}A\ket{j}
= \sum_{i,j} \delta_{ij} A_{ij}\nonumber\\
&=& \sum_i A_{ii}\\
&=& \Tr[A].\nonumber
\end{eqnarray}
Similarly we get $\bra{\Phi^+}A\otimes\id\ket{\Phi^+}=\Tr[A]$.

To prove the snake equation we must first make the following equivalence for tensor products of the elements $\ket{i}$ and $\bra{j}$:
\begin{equation}
\bra{j}\otimes\ket{i}\equiv\ket{i}\otimes\bra{j}\equiv \ketbra{i}{j}
\label{eqn:tensor-product-equiv}
\end{equation}
This is illustrated diagrammatically as
% FIGURE -  TENSOR EQUIV
 \figeq{
  \includegraphics[width=0.4\textwidth]{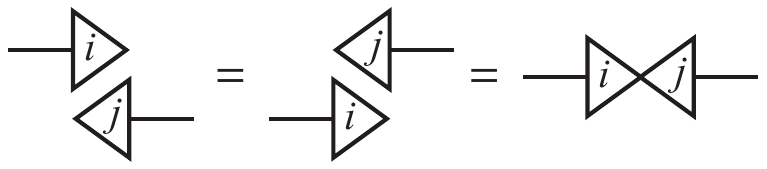}
    \label{fig:tensor-product-equiv}
 }

With this equivalence made, the proof of the snake-equation for the ``S''  bend is given by
% FIGURE -  SNAKE PROOF
\figeq{ \includegraphics[width=0.3\textwidth]{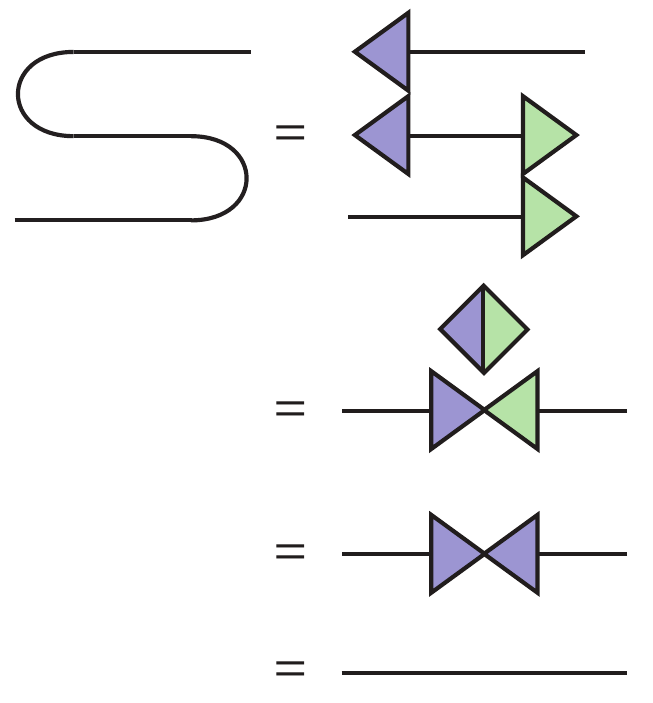}
  \label{fig:snake-proof}
}
The proof for the reflected ``S'' snake-equation follows naturally from the equivalence defined in \eqref{fig:tensor-product-equiv}.

The proof of our tensor network for the transposition of a linear operator $A$ is as follows:
% FIGURE -  TRANSPOSE PROOF 1
\figeq{ \includegraphics[width=0.35\textwidth]{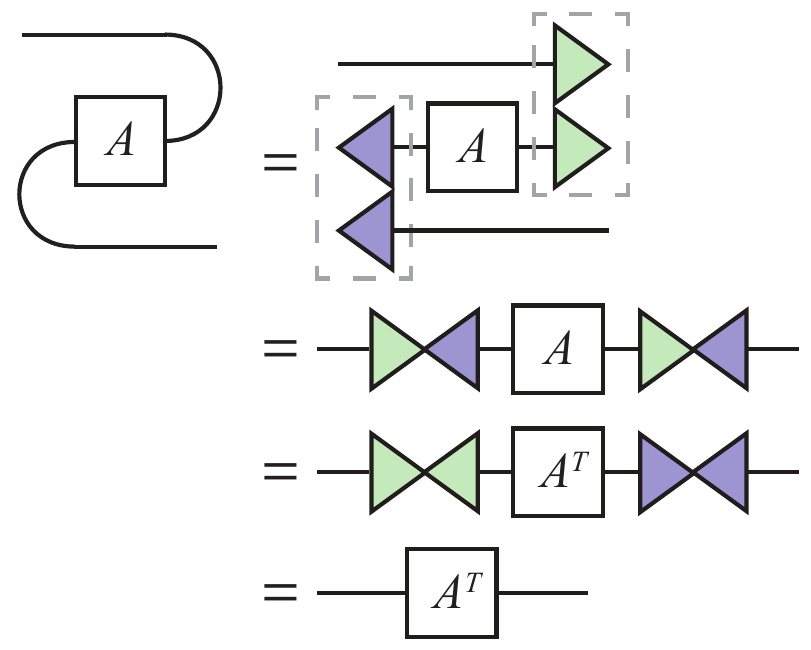}
\label{fig:tensor-transpose-proof-1}
}
To prove this algebraically we note that the corresponding algebraic equation for the transposition tensor network is
\begin{eqnarray}
\includegraphics[width=0.09\textwidth]{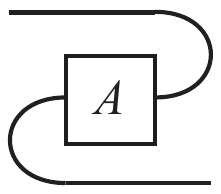}&=&\id\otimes\bra{\Phi^+}(\id\otimes A\otimes\id)\ket{\Phi^+}\otimes\id\\
&=& \sum_{i,j} \bra{j}A\ket{i}\ \ket{i}\otimes\bra{j}\\
&=& \sum_{i,j} \bra{j}A\ket{i} \ketbra{i}{j}\\
&=& \sum_{i,j} \bra{i}A^T\ket{j} \ketbra{i}{j}\\
&=& \sum_{i,j} \ketbra{i}{i}A^T\ketbra{j}{j}\\
&=& A^T.
\end{eqnarray}
The proof for transposition by counter-clockwise wire bending follows from the equivalence relation in \eqref{eqn:tensor-product-equiv} and \eqref{fig:tensor-product-equiv}. 

With the tensor network for transposition of an operator proven, the proof of transposition by contracting through a Bell-state $\ket{\Phi^+}$ is then an application of the snake equation as shown:
 % FIGURE -  TRANSPOSE PROOF 2
 \figeq{\includegraphics[width=0.3\textwidth]{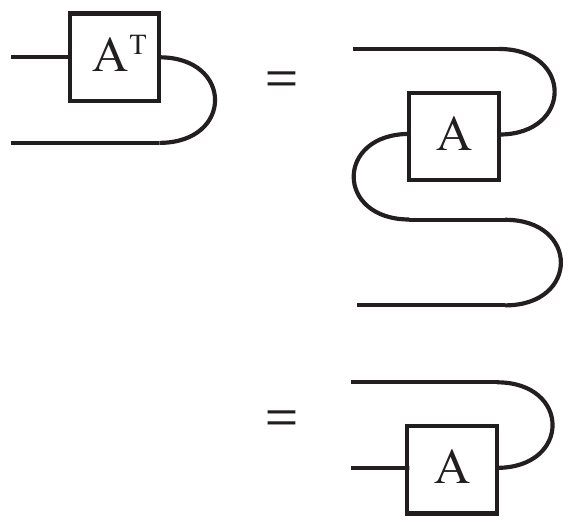}
  \label{fig:bell-tr-proof}}

%===================================================
% APP: VECTORIZATION PROOFS
%===================================================
\section{Vectorization change of basis}
\label{app:vec-proofs}

We now prove that the vectorization change of basis operator $T_{\sigma\rightarrow\omega}$ indeed functions as claimed. Let $\2 X$ be a $d$-dimensional complex Hilbert space and let $\{\sigma_\alpha: \alpha=0,...,d^2-1\}, \{\omega_\alpha:\alpha=0,...,d^2-1\}$ be orthonormal operator bases for $L(\2X)$. Define an operator $T_{\sigma\rightarrow\omega}\in \2 L(\XX)$ by
\begin{equation}
T_{\sigma\rightarrow\omega}:= \sum_\alpha \ket{\alpha}\dbra{\omega_\omega}_\sigma.
\end{equation}
where $\{\ket{\alpha}:\alpha=0,d^2-1\}$ is the computational basis for $\XX$.

We claim that for any linear operator $A\in L(\2X)$, 
\begin{equation}
T_{\sigma\rightarrow\omega}\dket{A}_\sigma=\dket{A}_\omega.
\end{equation}

The proof is as follows:
\begin{eqnarray}
T_{\sigma\rightarrow\omega}\dket{A}_\sigma
	&=& \left(\sum_\alpha \ket{\alpha}\dbra{\omega_\alpha}_\sigma\right)\dket{A}_\sigma\\
	&=& \sum_\alpha \ket{\alpha} \dbradket{\omega_\alpha}{A}_\sigma\\
	&=& \sum_\alpha \ket{\alpha}\Tr[\omega_\alpha^\dagger A]\\
	&=& \dket{A}_\omega.
\end{eqnarray}

The inverse of $T_{\sigma\rightarrow\omega}$ is given by
\begin{equation} 
T_{\sigma\rightarrow\omega}^{-1}=T_{\sigma\rightarrow\omega}^{\dagger}=T_{\omega\rightarrow\sigma}
\end{equation}
and hence $T_{\sigma\rightarrow\omega}$ is unitary.

%===================================================
% APP: GATE FIDELITY
%===================================================
\section{Fidelity proofs}
\label{app:fid-proofs}

We now derive the expressions for the average gate fidelity and entanglement of a CPTP map $\2E \in C(\2X)$ in terms of the superoperator, Kraus, $\chi$-matrix and Stinespring representations of $\2E$ given in Section~\ref{sec:gatefid}. These follow from applying the channel transformations of Section~\ref{sec:trans} to the quantity \be
F_e(\2E,\rho) = \dbra{\rho}\Lambda\dket{\rho}
\ee 
where $\Lambda$ is the Choi-matrix for $\2E$. This is illustrated below:
% ENTANGLEMENT FIDELITY REPS PROOF
\figeq{
	\includegraphics[width=0.4\textwidth]{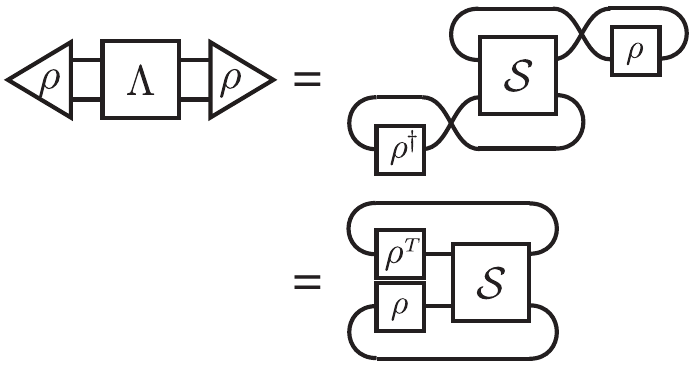}
	\label{fig:ent-fid-sop}
}
\figeq{
	\includegraphics[width=0.4\textwidth]{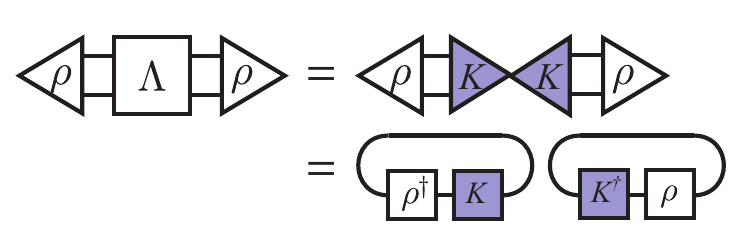}
	\label{fig:ent-fid-kraus}
}
\figeq{
	\includegraphics[width=0.4\textwidth]{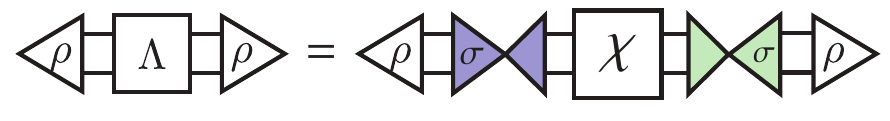}
	\label{fig:ent-fid-chi}
}\figeq{
	\includegraphics[width=0.4\textwidth]{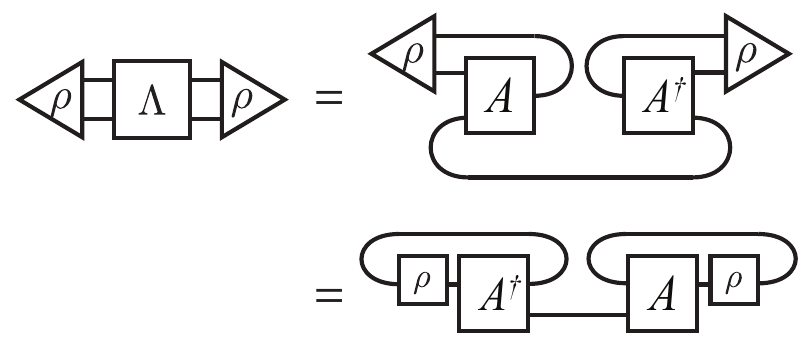}
	\label{fig:ent-fid-stinespring}
}
In the case of the $\chi$-matrix representation we assume that the $\chi$-matrix is defined with respect to an orthonomral basis $\{\sigma_\alpha\}$, $\alpha = 0,...,d^2-1$ satisfying $\Tr[\sigma_j] = \delta_{j0}\,\sqrt{d}$.

The expressions for the average gate fidelity are then obtained by from $F_e(\2E,\id),$ where $\id\in L(\2X)$ is the identity operator, via the equation:
\be
\overline{F}_{\2 E} = \frac{d+F_e(\2E,\id)}{d(d+1)}.
\ee